\documentclass[12pt]{article}
\usepackage{amsmath,amssymb,graphicx}
\usepackage{amsthm}
\usepackage{euscript}
\usepackage{eufrak}
\usepackage{slashed}

\newtheorem{defn}{Definition}
\newtheorem{lem}{Lemma}
\newtheorem{thm}{Theorem}
\newtheorem{prop}{Proposition}
\newtheorem{cor}{Corollary}

\newcommand{\pr}{\noindent{\bf Proof}. }
\newcommand{\rem}{\noindent{\bf Remark}. }
\newcommand{\rems}{\noindent{\bf Remarks}. }

\newcommand{\pa}{\partial}
\newcommand{\one}{\cO(1)}
\newcommand{\bpsi}{\bar \psi}

\newcommand{\hs}{ \hspace{1cm}}
\newcommand{\tr}{\textrm{ tr }}

\newcommand{\loc}{ \textrm{loc}}

\newcommand{\B}{\Big}
\newcommand{\blan}{\Big  \langle} 
\newcommand{\bran}{\Big  \rangle}  
\newcommand{\sgn}{\textrm{sgn}}
\newcommand{\sq}{\square}

\newcommand{\bg}{\bar g} 
\newcommand{\reg}{\textrm{reg}}

\newcommand{\be}{\begin{equation}}
\newcommand{\ee}{\end{equation}}

\newcommand{\bx}{\mathbf{x}}
\newcommand{\by}{\mathbf{y}}

\newcommand{\slp}{\slashed{p}} 
 
\newcommand{\slpa}{\slashed{\partial} } 
\newcommand{\slx}{\slashed{x}} 
\newcommand{\sly}{\slashed{y}} 

\newcommand{\sZ}{\mathsf{Z}}

\newcommand{\al}{\alpha}
\newcommand{\De}{\Delta}
\newcommand{\de}{\delta}
\newcommand{\ga}{\gamma}
\newcommand{\Ga}{\Gamma}

\newcommand{\la}{\lambda}

\newcommand{\om}{\omega}

\newcommand{\ep}{\epsilon}
\newcommand{\si}{\sigma}

\newcommand{\vep}{\varepsilon}

\newcommand{\cB}{{\cal B}}
\newcommand{\cC}{{\cal C}}

\newcommand{\cO}{{\cal O}}

\newcommand{\cH}{{\cal H}}

\newcommand{\cR}{{\cal R}}

\newcommand{\cG}{{\cal G}}

\newcommand{\cW}{{\cal W}}

\newcommand{\cL}{{\cal L}}

\newcommand{\bbR}{{\mathbb{R}}}
\newcommand{\bbZ}{{\mathbb{Z}}}
\newcommand{\bbC}{{\mathbb{C}}}

\newcommand{\bbT}{{\mathbb{T}}}

\begin{document}

\title{Structural stability of the RG flow in the Gross-Neveu model}

\author{J. Dimock\footnote{dimock@buffalo.edu}  and Cheng Yuan\footnote{chengyua@buffalo.edu}  \\
Dept. of Mathematics, SUNY at Buffalo \\
Buffalo, NY 14260, USA}

\maketitle

\begin{abstract}    We study flow of renormalization group (RG) transformations for the  massless Gross-Neveu model in a non-perturbative formulation.   The model is defined  on a two dimensional Euclidean space with  a finite volume. 
The quadratic approximation to the flow stays bounded after suitable renormalization.  We show that for weak coupling  this property
also is true for the complete flow.   As an application we prove an ultraviolet stability bound for the model.   Our treatment is an application of a method of Bauerschmidt, Brydges, and Slade.  The method was  developed for an
infrared problem, and is  now applied to an ultraviolet problem. 
\end{abstract}

%\tableofcontents

%\newpage

\section{Introduction} 

In a quantum field theory model  a renormalization group transformation replaces the given action by an effective action on a larger length scale.
When iterated  one gets a flow of the effective actions which distill  information about  the large scale structure of model.    At each stage one identifies
an effective coupling constant   and a few other parameters like field strength  and mass.   The flow of the coupling constant is particularly important
as it  drives the other parameters.     In weakly coupled models in critical dimension
it has the form 
\be \label{flowflow} 
g_{k+1}  = g_k  \pm \beta_k g_k^2  + \cO(g_k^3)  \hs  \hs \beta_k >0
\ee
Structural stability is the statement that  this quadratic flow $g_{k+1}  = g_k  \pm \beta_k g_k^2$ , together with the quadratic flow of the other parameters,  is the 
dominant  behavior  in complete  flow of the effective actions.  In particular with a minus   sign  the flow  is toward zero,  the effective actions become free,   and infrared (long distance) problems  become tractable.   With a plus sign the flow 
is away from the origin and,  after renormalization to tiny initial values,  ultraviolet  (short distance)  problems  become tractable.

Bauerschmidt, Brydges, and Slade \cite{BBS15c} have developed a method for proving structural stability in the infrared case.   They are
interested in an $n$-component  scalar field $\phi$ on a four dimensional  unit lattice with a $| \phi |^4$  interaction term \cite{BBS14} , 
as well as the self-avoiding random walk which is formally the case $n=0$ \cite{BBS15a}, \cite{BBS15b}.    
In these  models  the minus sign occurs in (\ref{flowflow}) and  they are able to use the structural stability result 
to prove results on the bulk behavior of the model.

In this paper we show that this method  works  for an  ultraviolet problem  and in the continuum.   We consider  the massless Gross-Neveu model in dimension two.    It features
 anti-commuting  fermi fields $\psi, \bar \psi$
with  $n$ internal  components.  The   action is
\be
S(\psi) = \int  \B(  - \vep + (1+z)  \ \bar \psi \slpa \psi     -  g \ ( \bar \psi \psi )^2    \B) dx
\ee
with  coupling constant $g$,   field strength $(1+z)$, and energy density $\vep$. 
We work in a finite volume and add a high  momentum cutoff  to provide a short distance  regularization of   the model.   Precise definitions to follow shortly. We scale up to a large volume and unit momentum cutoff.   We take  initial values   $g_0, z_0, \vep_0$,  depending on the cutoff, and make a sequence  of RG transformations. 
  After $k$-transformations we preserve the unit cutoff, but 
have a new effective action which will include terms $ \int ( - \vep_k + (1+z_k)  \ \bar \psi \slpa \psi     -  g_k \ ( \bar \psi \psi )^2   )  $  as well as 
psuedo-scalar terms   $p_k \int     \ ( \bar \psi \ga_5 \psi )^2  $  and vector terms    $v_k   \int   \sum_{\mu}  ( \bar \psi \ga_{\mu}  \psi )^2  $, and also 
an infinite dimensional piece $E_k$.     Each RG transformation reduces the volume,  and we continue until we reach a bounded volume  and a tame 
expression for the action.     The variables $g_k,z_k, \vep_k, p_k,v_k, E_k$  and    evolve as a nonlinear discrete dynamical system (see(\ref{stunning})).    We assume 
$g_k$  and all other variables are small.      At first we ignore terms which are   $\cO(g_k^3)$  which includes the infinite dimensional piece $E_k$. 
Then we have the quadratic equations
\be
 \begin{split}
g_{k+1}   =&   g_k + \beta_k g_k^2 \\
z_{k+1} = & z_k + \theta_k g_k^2 \\
\end{split}
\ee
with $\beta_k>0$ and bounded above and below and $\theta_k$ bounded.    It is straightforward to show that
one can choose initial values  $g_0,z_0$ such that  $g_k$ increases  and  takes any sufficiently small final value $g_f$
and $z_k = \cO( g_k)$.          Structural stability is the statement that this remains true for the full system.   That is for the
full flow  $g_k$ increases  and  takes any sufficiently small final value $g_f$  and the other variables     $z_k, \vep_k, p_k,v_k, E_k$ 
are bounded by some power of $g_k$. 
  In particular one has bounds on the effective actions uniform in the cutoff,  and this  gives a uniform 
 bound on the partition function
  \be
\sZ   =   \int    e^{-S(\psi) } D \psi    
\ee
The method could lead to a complete construction of the model which would involve   taking the limit as the cutoff is removed both in the partition function
and correlation functions,   and taking  the infinite volume limit.

Here is  some history of work on the model.  The model  was first considered by Mitter, Weisz  \cite{MiWe73}, and then taken up by 
by Gross and Neveu  \cite{GrNe74} who established dynamical mass generation for large $n$.      Rigorous treatments were given somewhat later  by Gawedzki and Kupiainen 
\cite{GaKu85b}  and by Feldman, Magnen, Rivasseau, and Seneor \cite{FMRS86}.   
In both  these cases the authors work directly in infinite volume and give a renormalization group analysis based on perturbation
theory to all orders.  They take  advantage of the fact the the perturbation theory for this model  is convergent.   Further results 
along these lines  can be found in 
   \cite{IaMa87},   \cite{IaMa88a}, \cite{IaMa88b}, \cite{KMR95}, \cite{DR00}.

Our nonperturbative method is an alternative to these perturbation theory methods.    It  is (arguably) simpler and potentially
can be used for models with bosons where perturbation theory does not converge. 
But for the Gross Neveu model it has the disadvantage that it  does not capture all the features of the model.   For example
Gawedzki and Kupiainen are able to argue in perturbation theory that in the renormalization procedure no pseudoscalar or vector quartic 
terms occur.  We  cannot make this argument and are obliged  to allow that they might occur.

%>>>>>>>>>>>>>>>>>>>>>>>>>>>>>>>>>>>>>>>>>>>>>>>>>>>>>>>>>>>>>>>>>>>>>>>>>>>>>>>
%>>>>>>>>>>>>>>>>>>>>>>>>>>>>>>>>>>>>>>>>>>>>>>>>>>>>>>>>>>>>>>>>>>>>>>>>>>>>>>>

\section{Preliminaries}

\subsection{Notation}

We start in a finite volume  by working on the torus $
\bbT_M = \bbR^2/ L^M   \bbZ^2
$
 with volume $L^{2M}$.  Here  $L$ is a fixed positive integer and $M$ is 
the long distance cutoff. The fields are elements on a  Grassmann  algebra generated by  independent. elements 
 $\bpsi^i _a(x),\bpsi^i _a(x) $ where $x \in \bbT_M$,   the spinor index  is $a  = 1,2$,   and $i=1, \dots,  n$ is the internal index.   
This  is an infinite dimensional Grassmann   algebra; we are more precise about its exact 
definition in  section \ref{Grassmann}.

The partition function is interpreted as an  integral on the Grassmann algebra and has the form (up to an overall constant)  
\be
\sZ  =
\int \exp  \B(  \int \B[   \vep   - z \bpsi \slpa \psi   + g ( \bar \psi \psi )^2  )  \B] \B)  d\mu_{G} (\psi) 
\ee
where bilinears have the form  $(\bpsi \psi )(x)  = \sum_{i=1}^{n}  \sum_{a = 1,2} \bpsi^i_{a} (x)\psi^i_a (x) $, etc.  
With self-adjoint Dirac matrices satisfying  $\{ \ga_{\mu}, \ga_{\nu} \} = \de_{\mu \nu} $, 
the Euclidean  Dirac operator  is   $\slpa = \sum_{\mu}  \ga_{\mu} \pa_{\nu}  = \ga_0 \pa_0 + \ga_1 \pa_1$. 
Further  $\int [ \cdots ] d\mu_G$ is the Grassmann Gaussian integral 
with covariance $G = \slpa ^{-1} $.
Formally the kernel $G(x-y) $ is defined by the Fourier series  \footnote{ See Appendix \ref{A} for our  Fourier series conventions} 
\be 
G (x-y) =   L^{-2M }  \sum_{ p  \in  \bbT_M^*}  e^{ip(x-y)}  \frac{-i\slp}{p^2}    
 \ee
 where $\bbT_M^* = 2 \pi   L^{-M } \bbZ^2$.    More precisely if we include internal indices it is $G (x-y) \de_ {ij}$.

     We regularize  this  by introducing  a smooth momentum cutoff  at approximately $|p| = L^N$ 
and we  also exclude $p=0$.    The modified covariance is 
\be \label{fine} 
G (x-y) =   L^{-2M }  \sum_{ p  \in  \bbT_M^*, p \neq 0}  e^{ip(x-y)}  \frac{-i\slp}{p^2}    e^{-   p^2/L^{2N} }  
 \ee
is now well defined pointwise.   Our goal is to get uniform bounds  as $N \to \infty$. 

It is convenient to scale up to the torus $\bbT_{N+M}$ with a unit momentum cutoff.   
This is accomplished    replacing $\psi$ on $\bbT_{M} $ by    $\psi_{L^{-N}  }$  for $\psi$ on $\bbT_{M+N}$ where 
\be 
  \psi_ {L^{- N }}(x)   = L^{ N/2}   \psi( L^{N} x )   \hs      \bpsi_ {L^{- N }} (x)  =   L^{ N/2}   \bpsi( L^{N} x )   
\ee 
Then we find (up to an overall constant) 
\be
\sZ =  \int e^{S_0( \psi)  } d \mu_{G_{0 } } ( \psi ) 
\ee   
The action becomes   
 \be 
 S_0 =   \int  \B( \vep_0  -     z_0 \bpsi  \slpa  \psi  +  g_0   ( \bar \psi \psi )^2   \B) 
 \ee
 with new parameters      $ \vep_0  = L^{-3N} \vep, z_0 = z,  g_0 = g$.
 The covariance is now 
\be
G_{0}  (x-y) =   L^{-2(M+N)  }  \sum_{ p  \in  \bbT_{M+N}^*, p \neq 0 }   e^{ip(x-y)} \frac{-i\slp}{p^2}   e^{- p^2 } 
 \ee
 Then $N  \to \infty$ limit is now a long distance problem.   Instead of potential singularities as   $p \to \infty$,   we have potential singularities as $p \to 0$.

%>>>>>>>>>>>>>>>>>>>>>>>>>>>>>>>>>>>>>>>>>>>>>>>>>>>>>>>>>>>>>>>>>>>>>>>>>>>>>>>

\subsection{Renormalization group} 

The renormalization group method  is a technique to find alternate  expressions for $\sZ$ on coarser momentum lattices,  thereby softening the
singularities.     The improvement comes in a sequence of steps.  After $k$ steps one has fields $\psi$  on $\bbT_{M+N -k} $   and the expression
\be 
\sZ = \int e^{S_k(\psi) } d \mu_{G_k} ( \psi) 
\ee
Here the new covariance is essentially the same as  $G_0$ and is given
by
 \be \label{gk} 
G_k(x - y)  =    {\sum}'_{p  \in \bbT^*_{N+M-k}}    e^{ip(x-y)}  \frac{-i\slp}{p^2}    e^{- p^2 }   
\ee
The  primed sum is a weighted sum with $p\neq 0$   defined by  
\be \label{prime} 
 {\sum}'_{p  \in \bbT^*_{N+M-k}}[ \cdots ] =    L^{ -2(M+N-k)}     \sum_{p  \in \bbT^*_{N+M-k }, p \neq 0} [\cdots ] 
\ee
The new    $S_k$  is yet to be specified,  but should have leading terms  similar to $S_0$.

Given such an expression on $\bbT_{M+N -k} $  we generate the expression on $\bbT_{M+N -k-1} $ as follows.
We split off the highest momentum piece of the covariance by 
\be
G_k  =   G_{k+1,L }  + C_k
\ee
where   $G_{k+1,L } (x, y)   = L^{-1} G_{k+1} (L^{-1}x, L^{-1} y )$.   Then we have 
\be 
\begin{split}  
G_{k+1,L } (x-y)  = &  {\sum}' _{ p \in  \bbT_{M+N-k }^*}  e^{ip(x-y)}  \frac{-i\slp}{p^2}   e^{- L^2p^2  }      \\ 
C_k(x-y)  = &  {\sum}' _{ p  \in  \bbT_{M+N-k }^*}  e^{ip(x-y)}  \frac{-i\slp}{p^2}  \B(  e^{- p^2 } -  e^{- L^2p^2  } \B)     \\
\end{split} 
\ee
The integral splits  as 
\be
 \sZ =   \int        \B[  \int e^{   S_{k} ( \psi + \eta)} d\mu_{C_k} (\eta) \B]   d \mu_{G_{k+1},L} ( \psi ) 
 \ee
 The expression in brackets is the fluctuation integral,  and one searches for a new action $\tilde S_{k+1}$
 such that
 \be
 e^{   \tilde S_{k+1}(\psi)    }  =   \int e^{  S_{k} ( \psi + \eta)} d\mu_{C_k} (\eta) 
 \ee
 Replace  $\psi(x)$ on   $\bbT_{M+N -k }$       by  $\psi_{L } (x) = L^{-\frac12} \psi(L^{-1} x)$ now with  $\psi$ on  $\bbT_{M+N - k -1}$.   
Under this scaling $G_{ k+1,L } $ scales to $G_{k+1} $  and  $\tilde S_{k+1} $ scales to 
$S_{k+1}(\psi)   = \tilde S_{k+1} ( \psi _L  ) $.
We have then
\be
\sZ  =    \int   e^{  S_{k+1} (\psi) }  d \mu_{  G_{k+1}  } ( \psi ) 
\ee
We are back to a unit momentum  cutoff,  but the volume has been reduced.  The issue is to keep track the effective actions $S_k$.  
\bigskip

In the main text will  use a modification of the above in which the field strength is absorbed into 
the measure before studying the fluctuation integral.

%>>>>>>>>>>>>>>>>>>>>>>>>>>>>>>>>>>>>>>>>>>>>>>>>>>>>>>>>>>>>>>>>>>>>>>>>>>>>>>>

\subsection{The Grassmann algebra} \label{Grassmann} 

We give a more precise definition of the Grassmann algebras (see also  \cite{FKT00},  \cite{BrSl15} ).  
The algebras  are formally   generated by    elements  $\psi(\xi)$ satisfying    $\psi(\xi) \psi(\xi') =  -\psi(\xi') \psi(\xi)$.    Here the index   $\xi = (x, a, i, \om)$ consists of  a point  
 $x$  in  some   torus  $\bbT_j=  \bbR^2/  L^{j} \bbZ^2$, a Dirac index   $a$, an internal index   $1 \leq i \leq n$,   and 
$\om =0,1$ which  distinguishes between $\psi, \bpsi$.   We have
\be
\psi  (x, a, i, 0 )   =  \psi_{a, i} (x)    \hs  \psi  (x, a, i, 1 )  = \bpsi_{a, i} (x) 
\ee
A general element of the Grassman algebra is given by the formal expression  
\be \label{g0} 
  F =  \sum_{\ell =0}^{\infty} \frac{1}{\ell!}  \int   F_{\ell} ( \xi_1,  \cdots ,  \xi_{\ell}   ) \psi(\xi_1) , \cdots  \psi( \xi_{\ell} )  d \xi 
\ee   
and we assume that  the kernels $ F_{\ell} $ are anti-symmetric to match the anti-symmetry of $\psi(\xi_1) , \cdots  \psi( \xi_{\ell} ) $.  So we assume 
 $\textrm{Alt} F_{\ell} = F_{\ell}$ 
where 
\be
( \textrm{Alt} F)(\xi_1, \dots, \xi_n) = \frac{1}{n!} \sum_{\pi}  \sgn \ \pi \   F( \xi_{\pi(1)} , \dots, \xi_{\pi( n)}) 
\ee
and  the sum is over a permutations $\pi$  of $1,2, \dots, n$. 

The kernels  $F_{\ell}$ are allowed to be distributions and we  assume they are in  a Banach space of distributions $\cC'$
to  be specified.    Then we define a norm  depending on a parameter $h>0$  by 
\be
 \label{g1} 
 \| F \|_h   = \sum_{\ell =0}^{\infty} \frac{h^{\ell} }{\ell!}    \|F_{\ell} \|_{\cC'} 
\ee
Let     $\cG_h$    be    Banach space  of  all formal expressions (\ref{g0}) such that   $\|F\|_h $ is finite.

A product of two elements is formally 
 \be \label{slouch} 
\begin{split} 
FG =   \frac{1} {n!m!}  \int   F_n(\xi_1, \dots, \xi_n   ) G_m(\xi_{n+1} , \dots, \xi_m  )   \psi(\xi_1) , \cdots  \psi( \xi_{n+m} )  d \xi \\ 
\end{split} 
\ee
But $ \psi(\xi_1) , \cdots  \psi( \xi_{n+m} ) $ is anti-symmetric.  So if we  take account  that
\be <  H_1, \textrm{Alt}H_2>    = < \textrm{Alt} H_1,H_2> 
\ee
we can formally transfer an $\textrm{Alt}  $ operation from the generators to the kernel.    The precise  definition of the product is  then 
\be
FG  =\sum_{n,m }  \frac{1} {n!m!}  \int   \B(   \textrm{Alt}(  F_n \otimes G_m) \B) (\xi_1, \dots, \xi_{n+m }  )  \psi(\xi_1)  \cdots  \psi( \xi_{n+m}  )  d \xi \\  
\ee
The next result shows that   under certain conditions on the space $\cC'$  the space  $\cG_h $ is a Banach algebra. 

\begin{lem} 
Suppose that 
\be
\|  \textrm{Alt}(  F_n \otimes G_m))\|_{\cC'}   \leq  \|F_n\|_{\cC'}  \|G_m\|_{\cC'}   
\ee
\label{product} 
Then if  $F \in \cG_h$  and $G \in \cG_h$ then $FG  \in \cG_h$ and $\| FG \|_h \leq \| F\|_h \|G\|_h $
\end{lem} 
\bigskip

\pr  Let $H  = FG$.
Then 
\be
H  =\sum_{\ell }  \frac{1} {\ell! }  \int   H_{ \ell}  (\xi_1, \dots, \xi_{\ell}  )  \psi(\xi_1) , \cdots  \psi( \xi_{\ell}  )  d \xi \\  
\ee
where
\be
H_{\ell}   = \sum_{n + m = \ell }  \frac{\ell!} {n!m!}       \textrm{Alt}(  F_n \otimes G_m)  
\ee
Then
\be
\begin{split} 
\|H_{\ell} \|_{\cC'} \leq &  \sum_{n + m = \ell }  \frac{\ell!} {n!m!}    \|   \textrm{Alt}(  F_n \otimes G_m) \|_{\cC'}  
\leq   \sum_{n + m = \ell }  \frac{\ell!} {n!m!}     \|    F_n \|_{\cC'} \|  G_m \|_{\cC'}  \\
\end{split} 
\ee
and  we have the announced
\be \label{pity} 
\|H \|_h   = 
\sum_{\ell} \frac{h^{\ell}}{ \ell! }  \| H_{\ell}\| _{\cC'}    \leq 
  \sum_{n,m }  \frac{h^{n+m} } {n!m!}     \|    F_n \|_{\cC'} \|  G_m \|_{\cC'}    = \| F\|_h\|G\|_h
  \ee
  \bigskip

Now we define the space  $\cC' = \cC'( \bbT_j ^{\ell } )  $.     It is the dual space to a Banach space $\cC = \cC( \bbT_j ^{\ell } )  $ defined as 
all functions   $f$   on    $\bbT_j ^{\ell } =  \bbT_j  \times \cdots \times \bbT_j  $  such that   $f \in \cC^3(\bbT_j)$  in each of the $\ell $ variables
separately.  With    $\al = (\al_1, \dots \al_{\ell} )$  and  $\xi = ( \xi_1 \cdots, \xi_{\ell} ) $ 
the  norm is 
\be
\| f \|_{\cC}   =    \sup_{ \xi, \al  }|(  \pa^{\al}   f)( \xi)   |
\ee
Here  the multi-index  $\al$  has entries $\al_i = (\al_{i,0}, \al_{i,1})$  with   $|\al_i| \equiv  \al_{i,0} +  \al_{i,1} \leq 3 $, 
and 
\be 
\pa^{\al} =   \pa^{\al_1}_{x_1}  \cdots \pa^{\al_{\ell}}_ { x_{\ell}  }   \hs \hs       \pa^{\al_i}_{x_i} = \pa^{\al_{i,0}} _{x_{i,0}} \pa^{\al_{i,1}} _{x_{i,1}} 
\ee
Then  $\cC' $ is all continuous linear functionals on $\cC$ with norm  
\be
\| F \|_{\cC'}  = \sup_{ \|f\|_{\cC }  \leq 1  } |< F , f > |
\ee
Now we show that this satisfies the conditions of lemma 1.

\begin{lem}  \label{instant} 
\be
\|  \textrm{Alt}(  F_n \otimes G_m))\|_{\cC'}   \leq  \|F_n\|_{\cC'}  \|G_m\|_{\cC'}   
\ee
\end{lem}

\pr
We need to estimate
\be
\|  \textrm{Alt}(  F_n \otimes G_m)\|_{\cC'}   = \sup_{ \|f\|_{\cC}  \leq 1}\B|  \blan  \textrm{Alt}(  F_n \otimes G_m), f \bran \B|
\ee
We have
\be
\begin{split}
& \blan  \textrm{Alt}(  F_n \otimes G_m), f \bran =    \blan   ( F_n \otimes G_m),\textrm{Alt} f \bran \\
 = &  \int F_n(\xi_1, \dots \xi_n)G_m(  \eta_1, \dots, \eta_m )  \textrm{Alt} f(\xi_1, \dots \xi_n, \eta_1, \dots, \eta_m ) d \xi d \eta \\
  = &  \int F_n(\xi_1, \dots \xi_n)G^*_m( \xi_1, \dots \xi_n) d \xi \\
 \end{split}
 \ee
where
\be 
G^*( \xi_1, \dots \xi_n) =\int G_m(  \eta_1, \dots, \eta_m )  \textrm{Alt}\  f(\xi_1, \dots \xi_n, \eta_1, \dots, \eta_m )d \eta
\ee
This gives
\be
  | \blan  \textrm{Alt}(  F_n \otimes G_m), f \bran |  \leq  \|F_n\|  \| G^* \|_{\cC} 
\ee
All this   makes sense because  $G^*$ is in fact  an element of $\cC(\bbT_k^n)$ with derivatives acting on the test function. To establish this we would need to show, say for $n=1$ and with $\textrm{Alt} f$ replaced by $f$,  the convergence of  difference quotients 
like
\be 
h^{-1} \B(  \pa^{\al} _{\eta} f( \xi + he _{\mu}, \eta ) -   \pa^{\al} _{\eta}f(\xi, \eta ) \B )  \  \to  \ 
 \pa_{\xi_{\mu} }  \pa^{\al} _{\eta}f (\xi, \eta ) 
 \ee
  uniformly in $\eta$.  This follows by the uniform continuity of $ \pa_{\xi_{\mu} }  \pa^{\al} _{\eta}G^*(\xi, \eta  )$. 
Now we have
\be
\begin{split}
\pa^{\al} _{\xi} G^*_m( \xi_1, \dots \xi_n) = & \int G_m(  \eta_1, \dots, \eta_m ) (\pa^{\al} _{\xi}  \textrm{Alt}   f)  (\xi_1, \dots \xi_n, \eta_1, \dots, \eta_m )d \eta\\
= &\blan  G_m, ( \pa^{\al} _{\xi}  \textrm{Alt}   f)  (\xi_1, \dots \xi_n,  \cdot ) \bran \\
\end{split} 
\ee
which gives for $\|f\|_{\cC} \leq 1$
\be
\begin{split}
| \pa^{\al} _{\xi} G^*_m( \xi_1, \dots \xi_n)|  \leq  &
  \|  G_m  \|_{\cC'}     \sup_{ \beta, \eta} \B| \ ( \pa^{\beta} _{\eta} \pa^{\al} _{\xi}  \textrm{Alt} f)  (\xi_1, \dots \xi_n, \eta_1, \dots, \eta_m ) \B|  \\
  \leq &   \|  G_m  \| _{\cC'}  \|  \textrm{Alt} f   \|_{\cC }   \leq \|  G_m  \|_{\cC'}   \|  f   \|_{\cC }     \leq \|G_m\|_{\cC'}            \\
\end{split} 
\ee
Hence    $  \| G^* \|_{\cC}   \leq \|  G_m  \|_{\cC'}      $ and so 
$
\|  \textrm{Alt}(  F_n \otimes G_m))\|_{\cC'}   \leq  \|F_n\| _{\cC'} \|G_m\|_{\cC'}    
$.
\bigskip

\rem   Suppose the $F$ is presented in the form (\ref{g0}), but with kernels $F_{\ell}(\xi_1, \dots , \xi_{\ell} ) $  which are not necessarily anti-symmetric.  
In this case, as in the discussion of  products,  the interpretation is that it is the element with kernels   $\textrm{Alt} F_{\ell} $ 
and  the norm is 
\be
\|F\|_h =  \sum_{\ell} \frac{h^{\ell} }{\ell!}  \|\textrm{Alt} F_{\ell} \|_{\cC'} 
 \ee
However 
\be
| < \textrm{Alt} F_{\ell}, f>| = | <  F_{\ell},   \textrm{Alt}f>|  \leq \| F_{\ell} \|_{\cC'} \| \textrm{Alt} f \|_{\cC}  \leq  \| F_{\ell} \|_{\cC'} \|  f \|_{\cC} 
\ee
says that   $\|\textrm{Alt} F_{\ell} \|_{\cC'}  \leq    \|  F_{\ell} \|_{\cC'} $.  Thus we can estimate the norm with the original kernel as 
\be
\|F\|_h \leq  \sum_{\ell} \frac{h^{\ell} }{\ell!}  \|  F_{\ell} \|_{\cC'} 
 \ee

\bigskip
 
It will also  be convenient to use a representation which distinguishes $\psi, \bpsi$.  In the following 
 Dirac and internal indices  are suppressed.  
 
\begin{lem} 
Every element of $\cG_h$ can be written in the form 
\be \label{monkey3}  
  F =   \sum_{n, m}  \frac{1}{n! m!}   \int   F_{nm} (  x_1,\dots, x_n ,  y_1 ,  \dots ,   y_m  )   
  \psi( x_1) , \cdots     \psi( x_n)         \bpsi( y_1) , \cdots     \bpsi( y _m)       \            d x dy 
\ee  
with $ F_{nm}$ anti-symmetric in  the $x_i$ and $y_i$ separately.
 The norm is
\be \label{monkey4}  
 \| F \|_h  =   \sum_{n, m}  \frac{h^{n+m} }{n! m!}   \|  F_{nm} \|_{\cC'}   
\ee  
\end{lem}
\bigskip

\pr  Start with the representation (\ref{g0}) and  write it as
\be \label{monkey1} 
  F =  \sum_{\ell =0}^{\infty} \frac{1}{\ell!}\sum_{\om_1, \dots, \om_{\ell} }  \int   F_{\ell} (\om_1, x_1,  \cdots , \om_{\ell} ,  x_{\ell}     )
   \psi( \om_1, x_1) , \cdots  \psi(   \om_{\ell},  x_{\ell}  )  d x
\ee   
Classify the terms in the sum over the $\om_i$ by the number of $0$'s that occur.  Then move all the zeros to the left both in the kernel and in the fields.  There is no net change in sign.   A term with $n$ zeros and $m$ ones 
can occur in  $\ell!/n!m!$  different way corresponding to the $m$ different positions out of the original $\ell$ .     Thus we have
\be \label{monkey2} 
\begin{split} 
  F = &   \sum_{\ell =0}^{\infty} \frac{1}{\ell!}\sum_{n+m = \ell}  \frac{\ell!}{n! m!}   \int   F_{\ell} \B( (0, x_1),\dots, (0,x_n), ( 1,y_1) ,  \cdots ,  (1,  y_{m} )     \B)\\
 &  \psi( x_1) , \cdots     \psi( x_n)         \bpsi( y_1) , \cdots     \bpsi( y _m)       \            d x dy \\
 \end{split} 
\ee   
If we now define 
\be
 F_{nm} (  x_1,\dots, x_n ,  y_1 ,  \dots ,   y_m  )   
= F_{n+m} \B( (0, x_1),\dots, (0,x_n), ( 1,y_1) ,  \cdots ,  (1,  y_{m} )     \B)
\ee
and   collapse the sum over $\ell,m,n$ to just a sum over $m,n$ we have the expression (\ref{monkey3}). 
The analysis of the norm is entirely similiar.

 \subsection{Localized  functions and norms} 

 \label{norms} 

We cover  any torus $\bbT_j = \bbR^2/ L^j \bbZ^2$  into closed  unit blocks  (squares) $\sq$  centered on the points of  $ \bbZ^2/ L^j \bbZ^2$.      We consider paved sets   $X$  defined to be  unions of such  blocks.   They are not necessarily connected so we refrain from calling them polymers,  although the terminology is sometimes used. 

We  consider elements  of the Grassmann algebra  $\cG_h$  which have the 
form 
\be \label{polymer}
E(\psi )  =  \sum_{X} E(X, \psi )   
\ee
where  $E(X, \psi)$ depends on $\psi$  only  in $X$. 
With a weight function $\Ga(X)$ we   define a norm on the family  $E = \{  E(X)  \} $    as in  \cite{BrYa90},  \cite{DiHu92}, \cite{BDH96},  by 
 \be
\| E \|_{h, \Ga}   = \sum _{X \supset \sq}   \| E(X ) \|_h   \Ga(X)  
\ee
Both $  E(X )$  and  $  \Ga(X) $ will be invariant under translations by elements of  $\bbZ^2$,  so this is independent of the unit block $\sq$.   
The space of all $E(X, \psi)$ such that $\|E\|_{h, \Ga}< \infty $ is a Banach space denoted $\cG_{h, \Ga} $.

The weight   factor $\Ga(X)$  is take to have the form
 \be \label{Gamma} 
  \Ga(X) = A^{|X| } \Theta(X) 
  \ee
 Here  $|X| $ is the number of  unit  squares in  the paved set  $X$, so $|X|$ is the volume of $X$.
In $A^{|X|} $ the $A$ is a sufficiently large constant depending on $L$,  say $A = L^{d+2} =L^4$.  Finally 
  $\Theta(X)$ has the form
  \be
  \Theta(X)  = \inf_T \prod_{\ell \in T } \theta(| \ell | )  
  \ee 
 where the infimum  is over all tree graphs  $T$ on the centers of the blocks in $X$.  The  $\ell$ are the lines in the graph 
 and   $|\ell |$ is the length in an $\ell^{\infty}$ norm. 
 The factors $\theta(|\ell|) $ have a power law rate of increase.    Furthermore $\theta$ can be chosen so that $\theta(s/L)  \leq L^{-d-1} \theta(s) = L^{-3} \theta(s)$
 and $\theta(0) = \theta(1) =1$.  
 
 The weight factor $\Ga(X) $ satisfies  $\Ga(X) \geq 1$ and 
 \be    \label{flower} 
 \Ga(X \cup Y) \leq \Ga(X) \Ga(Y) \theta ( d(X,Y) )
 \ee
 We also will use  a modification defining for positive integer $n$  
 \be
 \Ga_n(X)  = e^{n|X|} \Ga(X) 
 \ee
 This also satisfies (\ref{flower}). 
 \footnote{ $\Ga_n(X) $ was originally defined with $2^{n|X|} $ rather than  $e^{n|X|}$, an insignifecant difference}

A paved set $X$ is defined to be   \textit{small}  if it is connected and satisfies  $|X| \leq  2^d = 4$.  Otherwise it is a \textit{large} set.   let $\bar X$ be the smallest union of
$L$-blocks containing $X$.    Then  one   can show that for $p,q>0$ there are constants $c_{p,q}= \one $ such that 
\be \label{slinky} 
\Ga_q(L^{-1} \bar X ) 
 \leq  \begin{cases}
 c_{p,q}    \Ga_{q-p}  (X)  &  \hs X \textrm{ small  } \\ 
   c_{p,q}   L^{-d-1} \Ga_{q-p}  (X)  &  \hs X \textrm{ large } \\ 
   \end{cases} 
\ee 

Here are some elementary examples of localized functions

\newpage

\begin{lem} { \ } \label{nugget} 
\begin{enumerate} 
\item If   $ E =  \int (\bpsi \psi)^2 $, then $ E= \sum_X E(X) $ with  $\| E\|_{h, \Ga}  \leq \frac14 Ah^4$.
\item If    $ E =  \int  \bpsi \slpa \psi $  then $ E= \sum_X E(X) $ with  $\| E\|_{h, \Ga}  \leq   \frac12 Ah^2$.
\end{enumerate}
\end{lem}
\bigskip

\pr   Define  $E(X) = E(\sq) = \int_{\sq}  (\bpsi \psi)^2 $ if $X$ is a single block $\sq$,  and $E(X) =0$ otherwise.   Then taking account the remark after lemma \ref{instant} we have
\be
\| E_{2,2} (\sq)\|_{ \cC'} =  \sup_{\|f\|_{ \cC} \leq 1} \B|  \int_{\sq}   f(x,x,x,x) dx \B|   \leq  1
\ee
This is the only contribution to $\|E(\sq) \|_h = \frac14 h^4 \| E_{2,2} (\sq)\|_{ \cC'} = \frac14 h^4$,
and 
\be
\| E \|_{h, \Ga}  = \|E(\sq) \|_h \Ga(\sq)    =  \frac14 h^4 A
\ee
The second case is similar,  but now  
\be
\| E_{1,1} (\sq)\|_{ \cC'} =  \sup_{\|f\|_{ \cC} \leq 1} \B|  \int_{\sq}  (\slpa_x  f) (x,x) dx \B|   \leq  1
\ee

%>>>>>>>>>>>>>>>>>>>>>>>>>>>>>>>>>>>>>>>>>>>>>>>>>>>>>>>>>>>>>>>>>>>>>>>>>>>>>>>

\subsection{Symmetries}  \label{symmetry} 
Our localized functions $E(X, \psi)= E(X, \bpsi, \psi)$  expressed in the form (\ref{monkey3}) will always have equal numbers
of $\psi$'s and $\bpsi$'s.   In addition they will have the following symmetries which are all symmetries of the massless Dirac operator.

\begin{enumerate} 
\item    (torus symmetries) Let   $R$ is a    rotation by a multiple of $\frac{\pi}{2}$ or a   reflection and  let $a$ be  a  point in the torus.
Let  ${a,R} $ be  the   symmetry   $x \to Rx + a$.    Then there is  a matrix  $S$ in the group $Pin(2)$ such that    $S^{-1}  \ga_{ \mu} S = \sum_{\nu}  R_{\mu \nu} \ga_{\nu} $.  The fields   $\psi, \bpsi$ transform respectively  to    $ \psi_{R,a}  (x)  = S  \psi(R^{-1} ( x-a ) ) $  and $\bpsi_{R,a} (x)  =  \bpsi(R^{-1} ( x-a ) ) S^{-1} $.   We require  that 
\be
E(X, \bpsi, \psi ) = E(RX+a, \bpsi_{R,a}, \psi_{R,a} ) 
\ee
   \item (internal rotations)    For $u \in U(n) $  and $(u\psi)_i = \sum_j u_{ij} \psi_j$  and  $ (\bpsi u^*)_i = \sum_j \bpsi_j (u^*)_{ji} $
    \be 
  E(X,  \bpsi, \psi   ) = E(X,   \bpsi u^* , u  \psi ) 
  \ee
    \item (chiral transformations) 
    For $\ga_5 = i \ga_0 \ga_1$ 
    \be   \label{chiral} 
  E(X,  \bpsi, \psi   ) = E(X,   \bpsi (i\ga_5) , (i\ga_5) \psi ) 
  \ee

\end{enumerate}

%>>>>>>>>>>>>>>>>>>>>>>>>>>>>>>>>>>>>>>>>>>>>>>>>>>>>>>>>>>>>>>>>>>>>>>>>>>>>>>>

\subsection{Scaling} 
\label{scaling} 

We  recall some general terminology about   the scaling  operation $\psi(x) \to \psi_L(x) = L^{-\frac12} \psi(x/L)$.    Terms  that increase like   $\int \bpsi \psi $ 
 are called \textit{relevant}.   Terms  that are invariant like   $\int (\bpsi  \psi)^2 $ and   $\int \bpsi \pa \psi $ are called \textit{marginal}.    Terms that decrease like $\int  (\bpsi \psi )^6$ 
 and $\int \pa \bpsi \pa \psi $  are called
\textit{irrelevant}.

  We want to develop a similar classification for our localized functions.  
If   $E(\psi)  = \sum_X E(X, \psi )$ is expressed as   a sum over paved sets  on   a torus  $\bbT_j=\bbR^2/ L^j \bbZ^2$,  
we first reblock to  $L $-paved sets  defined to be unions of $L$-blocks centered on points of $L\bbZ^2/ L^j \bbZ^2$.  If $X'$ is
an $L$-block then 
\be
  (\cB E) (X', \psi)  = \sum_{X: \bar X =  X'} E(X, \psi ) 
\ee
Here  the sum is over unit paved sets  $X$ in $\bbT_j$
 such that the smallest  $L$-paved set  $\bar  X$ containing $X$  is equal to $X' $. 
Then we scale down to the torus $\bbT_{j-1}$  defining for unit  paved sets   $Y$
\be  \label{suds} 
 (\cL E)(Y, \psi)  =   (\cB E)( LY, \psi_L)  =   \sum_{X: \bar X =LY }   E(X, \psi_L ) 
\ee

We    want to remove local relevant and marginal parts from our localized functions and  leave irrelevant parts.  
To do this we will need a workable criterion to identify irrelevant pieces. 
We use the following condition  introduced in \cite{BDH98}. 

\begin{defn} \label{first} 
 $E(X, \psi)$ is  irrelevant (normalized)  if 
\be \label{jinx} 
\begin{split} 
E_{0}(X)  = & 0 \\
E_{1,1}(X,1) = \int E_{1,1}(X,x,y)\   dx dy  = & 0 \\
E_{1,1}(X,x-x_0)=  \int  E_{1,1}(X, x, y) (x-x_0) \ dx dy   = & 0 \\
E_{2,2}(X,1) =  \int E_{2,2}(X,x_1,x_2,y_1,y_2 )\ dx dy     = & 0 \\
\end{split}
\ee
\end{defn} 
\bigskip

The third condition is independent of the point $x_0$ which we take to  be in $X$.

\begin{lem}  \label{sync} For $\psi$ on $\bbT_{j-1} $ and $\psi_L $  on $\bbT_j$, let $E'( X, \psi) = E(X, \psi_L)$.
Then
   \be
\| E'(X) \|_h \leq   \| E(X) \|_{L^{-\frac12}h } \leq \| E(X) \|_h 
\ee
Furthermore if  $X$ is  small    and $E(X) $ is normalized then 
\be   \label{unknown} 
\| E'(X) \|_h \leq   \one L^{-3} \| E(X) \|_h
\ee
\end{lem} 
\bigskip

\pr    (as in \cite{BDH98}) 
For $f \in \cC(\bbT^{n+m} _{j-1}) $ we have 
\be
<  E'_{nm} (X ), f > = <E_{nm} (X) , f_L>  
\ee
where  
\be
f_L (x_1, \dots, x_n , y_1, \dots, y_m ) = L^{-\frac12(n+m) } f \left( \frac{x_1}{L},  \dots ,  \frac{x_n}{L},  \frac{y_1}{L},  \dots ,  \frac{y_m}{L}  \right ) 
\ee
Then  $ \| f _L\|_{\cC}  \leq  L^{-\frac12(n+m) }  \|f  \|_{\cC} $ and so 
\be  \label{idiot} 
\begin{split}
 \|E'_{nm}(X ) \|_{\cC'}    =  &  \sup_{\| f \|_{\cC}   \leq 1 } |\blan E_{nm} (X ), f_L  \bran |  \\
 \leq  &   \sup_{\| f \|_{\cC}    \leq 1 }  \|E_{nm} (X )  \|_{\cC'}    \| f_{L}\|_{\cC}  \\   
  \leq  &    L^{-\frac12(n+m) } \|E_{nm} (X ) \|_{\cC'}    \\
 \end{split} 
 \ee
 Now multiply by  $h^{n+m}/n!m! $ and sum over $n,m$ to get the first result. 

For the second result    note that for $n+m \geq 6$ we have  $ L^{-\frac12(n+m) }  \leq L^{-3}$ so (\ref{idiot}) is sufficient for these terms. 
It suffices to consider  $n+m = 2,4$.   In analyzing these terms we are going to make use of identities like
\be \label{replace} 
  \psi(x) =     \psi (  x_0 )  + (R_{1,x_0}  \psi) (  x  ) 
\ee
where 
\be
 (R_{1,x_0}  \psi) ( x)  =   \int_0^1  (x-x_0) \cdot  ( \pa  \psi )   \B( x_0 + t( x-x_0) \B) dt
\ee
The meaning of this manipulation of formal symbols  is the corresponding operation  on test functions.
For example if  $E(\psi)  = \int E(x) \psi(x) dx $  then making the replacement (\ref{replace}) means we 
are writing the distribution $<E,f>$ as the sum 
\be
<E,f> = <E, f(x_0) > + <E,  R_{1,x_0} f>
\ee

For $n+m = 4$ we consider  
\be \label{syrup} 
E'_{2,2} (X, \psi) =   E_{2,2} (X, \psi_L ) = \int E_4(X, x_1, x_2,y_1,y_2)  \psi_L(x_1)   \psi_L(x_2)   \bpsi_L(y_1)   \bpsi_L(y_2) 
\ee
Pick a point $x_0$ in $X$ and make the replacement $  \psi_L(x_i) =    \psi_L (  x_0 )  + (R_{1,x_0}  \psi_L) (  x_i  ) $ and similarly
for $\bpsi$.
The term with all   the fields at $x_0$  is proportional to   $ \int E_{2,2}(X)$ and so vanishes by the normalization condition.
The other terms have at least one remainder term. 

Consider for example
\be    \label{silver} 
   \int E_{2,2} (X,x_1,x_2,y_1,y_2 )   (R_{1,x_0}  \psi_L) (  x_1)         \psi_L(x_0)   \bpsi_L(x_0)   \bpsi_L(x_0) 
   \ee
This refers to the distribution   $<E_{2,2} (X), f^{\star}   > $  with norm 
\be \label{gold} 
 \sup_{ \| f\|_{\cC}   \leq 1} |<E_{2,2} (X), f^{\star}   > |
\ee
where
\be 
\begin{split} 
f^{\star} (x_1, x_2,y_1,y_2 )  = & ( R_{1,x_0} f_L ) (x_1, x_0,x_0 x_0) \\
=&   \int_0^1  (x_1-x_0)_{\mu}  \frac{ \pa}{ \pa x_{1, \mu} }  f_L\B( x_0 + t( x_1-x_0)  , x_0 ,  x_0,  x_0 \B) dt \\
=& L^{-3}   \int_0^1  (x_1-x_0)_{\mu}   \frac{ \pa f}{ \pa x_{1, \mu} } \B(\frac{ x_0 + t( x_1-x_0) }{L} , \frac{x_0}{L} ,  \frac{x_0}{L}, \frac{x_0}{L} \B) dt \\
\end{split}
\ee 
Here $f_L$ supplies the $L^{-2} $ and the derivative gives the extra $L^{-1} $. 
But $|x_1-x_0| \leq 4$ thanks to our restriction to $X$ small and built from unit squares.  Thus we have
$| f^{\star}|   \leq \one L^{-3} \| f\|_{\cC} $.
Similarly we claim  that  for $| \al | \leq 3$
\be
|\pa^{\al}_ {x_1}  f^{\star}  | \leq \one L^{-3} \| f\|_{\cC} 
\ee
 We use the identity $f_L (  x_1)  = f_L  ( x_0 )  + (R_{1,x_0}f_L) (  x_1  ) $
to change  
$ \pa^{\al}_ {x_1} ( R_{1,x_0} f_L ) ( x_1, ...) $ to just   $\pa^{\al}_{x_1} f_L ( x_1, \dots)$.  
Again the derivatives supply extra factors of $L^{-1}$. 
Altogether then we have
\be
\| f^{\star} \|_{\cC}  \leq   \one L^{-3} \| f\|_{\cC} 
\ee
and hence  the norm  (\ref{gold}) is bounded by 
  $\one  L^{-3} \| E_{2,2} (X) \|_{\cC'} $.  The other contributions to $E'_{2,2} (X)$ are similar and so
\be
   \| E'_{2,2} (X) \|_{ \cC'}   \leq \one   L^{-3} \| E_{2,2} (X) \|_{\cC'}  
\ee
which suffices for (\ref{unknown})

For $n+m =2$ we consider  
\be
  E_{1,1}(X, \psi_L ) = \int E_{1,1}(X, x,y  )  \psi_L(x)     \bpsi_L(y) 
\ee
Again  pick a point $x_0$ in $X$ and  now  make the replacement
\be    \label{congo2} 
 \psi_L ( x)  =  \psi_L ( x_0 )  +  (x -x_0)_{\mu} \pa_{\mu}    \psi_L ( x_0 )   +    (R_{2,x_0}  \psi_L) (  x )  
\ee
and similarly for $\bpsi$.
Here
\be
\begin{split} 
 (R_{2,x_0}  \psi_L) (  x_i )   = &    \int_0^1  (1-t) (x-x_0)_{\mu} (x-x_0)_{\nu} ( \pa_{\mu} \pa_{\nu} \psi_L )   \B( x_0 + t( x-x_0) \B) dt \\
\end{split} 
\ee
The two leading terms here generate expressions which are seen to be proportional to $\int E_{2}(X,x,y) dxdy $ and 
$\int E_{2}(X,x,y) (x-x_0) dxdy $ and so vanish by our normalization condition.    The surviving terms have at least one factor
$(R_{2,x_0}  \psi_L) (  x_i )  $. 
 
    Let us consider for example
\be
   \int E_{1,1}(X, x, y)  (R_{2,x_0}  \psi_L) (  x )        \bpsi_L(x_0) 
   \ee
 This has norm 
\be \label{goldtooth} 
 \sup_{ \| f\|_{\cC}   \leq 1} |<E_{1,1}(X), f^{\star}   > |
\ee
where now 
\be 
\begin{split} 
f^{\star} (x, y )  = & ( R_{2,x_0} f_L ) (x, x_0 ) \\
 =& \int_0^1  (1-t) (x-x_0)_{\mu} (x-x_0)_{\nu} \pa_{\mu} \pa_{\nu}  f_L  \B( x_0 + t( x-x_0),x_0 \B) dt \\
  =& L^{-3} \int_0^1  (1-t) (x-x_0)_{\mu} (x-x_0)_{\nu}\pa_{\mu} \pa_{\nu}  f  \B(\frac{ x_0 + t( x-x_0)}{L}, \frac{x_0}{L}  \B) dt \\
\end{split}
\ee 
Again we have 
$| f^{\star}|   \leq \one L^{-3} \| f\|_{\cC} $ since the $f_L$ supplies a factor $L^{-1}$ and the derivatives give a factor $L^{-2}$. 
Similarly we claim  that  for $ 1 \leq | \al| \leq 3$
\be
|\pa^{\al}_x f^{\star}  | \leq \one L^{-3} \| f\|_{\cC} 
\ee
This follows for $|\al | =1$ from the above representation.  For $|\al| = 2,3$  we use 
\be
   (\pa_{ \nu} R_{2,x_0} f_L) (  x, x_0 )  =   \pa_{\nu}   f_L ( x,x_0  )  - \pa_{\nu}  f_L  ( x_0,x_0 )  = ( R_{1,x_0} \pa_{\nu} f_L ) (x, x_0 ) 
\ee   
   and so
      \be
    (\pa_{\mu} \pa_{\nu} R_{2,x_0}  f_L ) (  x,x_0  )  =   ( \pa_{\mu}  \pa_{\nu} f_L) (x,x_0 )  
  \ee 
  Again the extra derivatives supply the $L^{-2}$. 
   Altogether  we have   $\| f^{\star} \|_{\cC}  \leq   \one L^{-3} \| f\|_{\cC} $
and hence  the norm  (\ref{goldtooth}) is bounded by   $\one  L^{-3} \| E_{1,1}(X) \|_{\cC'} $.  The other contributions to $E'_{1,1} (X)$ are similar and so
\be
   \| E'_{1,1} (X) \|_{\cC'}    \leq \one  L^{-3} \| E_{1,1}(X) \|_{\cC'}  
\ee
 which suffices for (\ref{unknown}), and  completes the proof. 
  \bigskip

 \begin{lem}  \label{fiver}  For  $p,q  \geq 0 $ and any family of localized functions $E(X)$ 
\be
\| \cL E \|_{h, \Ga_q}   \leq \one L^2 \|E\|_{L^{-1/2}  h, \Ga_{q-p} } 
\ee
If  $E(X) $ is normalized for small sets this improves to 
\be
\| \cL E \|_{h, \Ga_q}  \leq \one L^{-1} \|E\|_{h, \Ga_{q-p}  } 
\ee
\end{lem} 
\bigskip

\pr   With $  E'(X, \psi )  = E(X, \psi_L ) $ we have  
$ (\cL E)(Y) =  \sum_{X: \bar X =LY }   E' (X) $ and
\be
\| \cL E \|_{h, \Ga}  \leq \sum_{Y \supset \sq}  \sum_{X: \bar X =LY }  \| E' (X) \|_h  \Ga(Y)\\
\ee
But 
\be
 \sum_{Y\supset \sq}    \sum_{X: \bar X =LY }  = \sum_{X: \bar X \supset L \sq}  \leq  \sum_{\sq_0 \subset L\sq} \sum_{X \supset \sq_0}  =L^2  \sum_{X \supset \sq_0}
\ee
Therefore
\be \label{tort} 
\| \cL E \|_{h, \Ga_q}  \leq   L^2  \sum_{X\supset \sq_0 }  \| E' (X) \|_h  \Ga_q(L^{-1} \bar X) 
\ee
For general  $E(X)$ we have $\| E' (X) \|_h \leq \| E (X) \|_{L^{-1/2} h}$ from lemma \ref{sync}, 
 and  $\Ga_q (L^{-1} \bar X)  \leq \one  \Ga_{q-p}( X) $ from (\ref{slinky}),  so the right side is bounded by $\one  L^2\|E\|_{L^{-1/2} h, \Ga_{q-p} } $
as announced.

In the second case  if $X$ is small and $E(X)$ is normalized we use  $\| E'(X) \|_h \leq   \one L^{-3} \| E(X) \|_{h}$ from lemma \ref{sync}
 and  again   $\Ga_q (L^{-1} \bar X)  \leq \one  \Ga_{q-p}( X) $. 
If $X$ is large we use  $\| E'(X) \|_h \leq    \| E(X) \|_h$ and   $\Ga_q (L^{-1} \bar X)  \leq \one L^{-3}  \Ga_{q-p}( X) $  from (\ref{slinky}). 
In either case we have
\be
 \| E' (X) \|_h  \Ga_q(L^{-1} \bar X) \leq \one   L^{-3} \| E(X) \|_h    \Ga_{q-p} (X)
\ee
Insert this bound in (\ref{tort}) and get the announced  $\one L^{-1} \|E\|_{h, \Ga_{q-p} } $

%>>>>>>>>>>>>>>>>>>>>>>>>>>>>>>>>>>>>>>>>>>>>>>>>>>>>>>>>>>>>>>>>>>>>>>>>>>>>>>>

 \subsection{Gaussian integrals}  \label{GI}

 We study Gaussian integrals on our  Grassmann algebra.    We study particularly  the fluctuation integrals after $k$ renormalization group steps,  although
 much of what follows is more general.   The torus is   $\bbT_{N+M-k} $  and the covariance is 
 \be
 C_k (x-y)  = {\sum}' _ { p \in \bbT^*_{N+M-k} }  \tilde C_k ( p)    \hs \hs
\tilde C_k ( p)  =  \frac{ -i\slp }{ p^2}\B(  e^{-p^2} - e^{L^2p^2} \B) 
\ee 
The   Gaussian integral with   covariance $C_k(x-y)$ is a linear function on the Grassman algebra   defined to be zero if $n \neq m$, and if $n=m$ 
 \be \label{sync1} 
  \int \
  \psi_{a_1} (x_1) \bpsi _{b_1}(y_1)  \cdots  \psi_{a_n}(  x_n )  \bpsi _{b_n}(y_n )  \  d \mu_{C}  (\psi, \bpsi)  
  =       \det  \B \{ (C_k)_{a_i b_j} (x_i - y_j)\B  \}   
\ee 
where  $ \{ ( C_k)_{a_i b_j} (x_i - y_j)\  \} $ is  a  $2n \times 2n$ matrix.   More precisely this is the integral if all internal indices are
the same.   In the general case the integral factors into a product of this type.  This definition is consistent with the more familiar
finite dimensional case.

 We also have the identity 
\be
 \psi_{a_1} (x_1) \bpsi _{b_1}(y_1)  \cdots  \psi_{a_n}(  x_n )  \bpsi _{b_n}(y_n )= \ep_n \
\psi_{a_1} (x_1)  \cdots  \psi_{a_n}(  x_n )     \bpsi _{b_1}(y_1) \cdots  \bpsi _{b_n}(y_n ) 
\ee
where $\ep_n = \pm 1$  is defined by $\ep_2= -1$ and $\ep_{n+1} = (-1)^n \ep_n$.   Thus in terms of
our standard representation for $n=m$
 \be \label{sync2} 
  \int \ \psi_{a_1} (x_1)  \cdots  \psi_{a_n}(  x_n )     \bpsi _{b_1}(y_1) \cdots  \bpsi _{b_n}(y_n )  \  d \mu_{C}  (\psi, \bpsi)  
  =   \ep_n   \det  \B \{ (C_k)_{a_i b_j} (x_i - y_j)\B  \}  
\ee 
   In the following let $C=C_k$ and suppress Dirac indices.

We   split the covariance  as $\tilde C =\tilde  C_1 \tilde C_2$ with 
\be
\tilde  C_1(p) = e^{-\frac12 p^2}   \hs \hs    \tilde C_2(p) = \frac{ -i\slp }{ p^2} \B(  e^{- \frac12  p^2} - e^{-(L^2- \frac12)p^2} \B) 
\ee
  Then we have
\be  \label{cfactor} 
C(x-y) =\int  C_1(x- z) C_2(z-y) dz  = \B(\overline{C_1( x- \cdot)} , C_2(\cdot - y)\B) 
\ee
It follows by Gram's inequality \cite{FKT00}  that
\be
\B| \det  \B \{ C(x_i- y_j)\B  \}    \B| \leq \prod_{i=1}^n  \|  C_1( x_i- \cdot)\|_2 \|  C_2(\cdot - y_j )\|_2 =  \| C_1\|_2^n \|  C_2\|^n_2  
\ee
Here $  \|C_1\|_2, \|C_2\|_2 $ are both finite and $\cO(L^2)$.   
The same is true for derivatives and we define
\be \label{sequoia} 
h(C)  = \sup_{ |\al | \leq 3}\B\{  \| \pa^{\al} C_1 \|_2, \ \| \pa^{\al} C_2 \|_2 \B\} 
\ee

\bigskip

\begin{lem}  \label{concert} 
For $F \in  \cG_{ h(C)} $  
\be
| \int  F      d \mu_{C}    |
\leq   \|  F \|_{h(C) }  
\ee
\end{lem} 
\bigskip

\pr  Take  $F$ of  the form  (\ref{monkey3}), and then by (\ref{sync2}) we have 
\be
\int  F      d \mu_{C}   
=\sum_{n=m} \frac{1} {n!m!}   \int F_{nm} (x_1,   \dots, x_n, y_1,  \dots,   y_m ) \  \ep_n \det  \B \{ C(x_i- y_j) \B \}  
\ee
 The integral here  has the form  $<F_{nm}, f > $  with $n=m$ and 
 \be
 f(x_1,   \dots, x_n,y_1, \cdots   y_m)  = \ep_n   \det  \B \{ C(x_i-  y_j) \B\}  
 \ee 
 Then  with $\pa^{\al}_x = \pa^{\al_1}_{x_1}  \cdots   \pa^{\al_n}_{x_n} $ 
 \be
 \begin{split} 
 \pa^{\al}_x \pa^{\beta}_y f(x_1,   \dots, x_n,y_1, \cdots   y_n) = & \ep_n  \det  \B \{  \pa^{\al_i}_{x_i}  \pa^{\beta_j}_{y_j}    C(x_i-  y_j) \B\}   \\
 =  & \ep_n \det \left\{  \B(\overline{ \pa^{\al_i} C_1( x_i- \cdot)} , (-\pa) ^{\beta_j}C_2(\cdot - y_j)\B) \right \}  \\
 \end{split} 
   \ee
  Then by Gram's inequality 
\be
\begin{split} 
 \| f \|_{\cC}=  & \sup_{\al, \beta, x,y} | ( \pa^{\al}_x \pa^{\beta}_y f)  (x_1,   \dots, x_n,y_1, \cdots   y_m) | \\
  \leq  &  \sup_{ |\al_i|, |\beta_j| \leq 3}     \prod_{i=1}^n   \| \pa^{\al_i } C_1 \|_2   \prod_{j=1}^m   \|  \pa^{\beta_j}C_2\|_2   \leq h(C) ^{n+m} \\
 \end{split} 
\ee
Hence
 $|<F_{nm}, f >| \leq   \|F_{nm} \|_{\cC'}  h(C)  ^{n+m}   $  and the result follows. 
\bigskip

  We also  need to consider integrals in which only some of the  generators are integrated out.    
  Suppose the Grassmann algebra has two sets of generators $ \psi, \bpsi $ and $ \eta, \bar \eta $.
  We consider elements  of the form 
 \be   \label{uncle2} 
 \begin{split}
&  F(  \psi, \bpsi, \eta, \bar \eta ) \\
& =  \sum_{n= m,n' =m' }   \frac{1}{n!m!n'!m' !}  \int  F_{nm,n'm'} ( x, y,  x', y' )
 \psi(x)  \bpsi( y   )   \eta(x')  \bar \eta( y'   )    \ d  x d y  d  x' d y'  \\
 \end{split} 
\ee 
where $x =(x_1, \dots, x_n)$, $\psi(x)  =\psi(x_1) \cdots \psi(x_n) $, etc. 
The associated norm is
 \be 
\| F \|_{h,h'}  \\
 =  \sum_{n= m,n' =m' }   \frac{h^{n+m} h'^{n'+m'} }{n!m!n'!m' !}  \|  F_{nm,n'm'}   \|_{\cC'}   
\ee

\begin{lem}    \label{stud1}  The integral
\be
F^*(   \psi, \bpsi)  = \int   F (  \psi, \bpsi, \eta, \bar \eta ) \ d \mu_{C}(\eta, \bar \eta)  
\ee
satisfies
\be
\|F^*\|_{h}  \leq \one  \| F\|_{h, h(C) } 
\ee
\end{lem} 
\bigskip

\pr    The integral is evaluated as
\be  F^*( \psi ) =    \sum_{n=  m }   \frac{1}{n!m! }  \int  F^*_{nm} ( x, y )  \psi(x)  \bpsi( y   )     \ d  x d y     
\ee
where 
\be F^*_{ nm} ( x, y )   =   \sum_{n'=m' }   \frac{1}{ n'!m' !}  \int  F_{nm,n'm'} ( x, y,  x', y' )  \ep_n \det \B(   \{ C( x'_ i  - y'_j)  \}  \B)     d  x' d y' 
\ee
Then with $g( x', y' ) =  \ep_n \det \B   \{ C( x'_ i  - y'_j) \B \} $
\be
\blan  F^*_{ nm} , f \bran    =  \sum_{n'=m' }   \frac{1}{ n'!m' !}   \blan    F_{nm, n'm'} ,  f \otimes g \bran  
\ee
It is straightforward that  $  \| f \otimes g \|_{ \cC}  \leq \| f \|_{ \cC}  \|   g \|_{ \cC}  $ and
again by Gram's inequality    we have  $\|g \|_{\cC}  \leq h(C) ^{n'+m'} $. 
Therefore
\be
\|F^*\|_{\cC'}  =  \sup_{ \|f\|_{\cC} \leq 1}  |<   F^*_{ nm} , f >   |    \leq   \sum_{n'=m' }   \frac{ h(C) ^{n'+m'} }{ n'!m' !}   \one   \|   F_{nm,n'm'}  \|_{\cC'}      
\ee
Use this in 
\be
 \|F^*\|_h =   \sum_{n= m }   \frac{h^{n+m}}{n!m! }   \|   F^*_{nm} \|_{\cC'}   
 \ee
 and we have the result.  
\bigskip

   The following result  is also useful
 
 \begin{lem} \label{stud2} 
 If  $ F^+ ( \psi, \eta)  = F( \psi + \eta ) $
 then 
 \be 
 \|  F^+  \|_{h, h'}  = \| F \|_{h+h'} 
  \ee
  \end{lem} 
 \bigskip

 \pr We use the representation (\ref{g0}) which does  not distinguish $\psi, \bpsi$.  Then we have
  \be
  \begin{split} 
&   F^+ (\psi,  \eta)  =   \sum_{\ell} \frac{1} {\ell!} \int F_{\ell} (  \xi_1, \dots, \xi_{\ell}  ) \B(\psi(\xi_1) + \eta( \xi_1 ) \B)  \cdots   \B(\psi(\xi_{\ell} ) + \eta( \xi_{\ell} ) \B)d \xi \\  
&=   \sum_{\ell} \frac{1} {\ell!} \sum_{n+m = \ell} \frac{\ell!}{n!m!} \int F_{\ell} (  \xi_1, \dots, \xi_n,  \zeta_1, \dots, \zeta_m  ) 
 \psi(\xi_1)  \cdots  \psi(\xi_n)) \eta( \zeta_1 )  \cdots    \eta( \zeta_m )\ d \xi d \eta\\
 & = \sum_{n, m} \frac{1}{n!m!} \int  F_{n+m} (  \xi_1, \dots, \xi_n, \zeta_1, \dots, \zeta_m ) 
 \psi(\xi_1)  \cdots  \psi(\xi_n) \eta( \zeta_1 )  \cdots    \eta( \zeta_m ) \  d \xi d \eta\\
 \end{split} 
\ee 
Here in the first line we multiply out the product  and in each of the  $2^{\ell}$ terms  and move all the $\psi(\xi_i) $ to the left.   Make the same permutation in the  kernel and the signs cancel.  Relabel the $\eta(\xi) $ as $\eta(\zeta) $. Terms with $n$   $\psi$-fields and $m$  $\eta$-fields all give the same contribution. There are $\ell !/ n! m!$ of these
corresponding to the number of start positions.  This  gives the second line. 
From the third line we identify
\be
  F^+ _{n,m}  (  \xi_1, \dots, \xi_n, \zeta_1, \dots, \zeta_m ) =F_{n+m} (  \xi_1, \dots, \xi_n, \zeta_1, \dots, \zeta_m ) 
\ee 
Hence
\be 
\begin{split} 
\|  F^+  \|_{h,h'}  = &    \sum_{n, m} \frac{h^n(h') ^m}{n!m!}  \|  F^+_{n,m} \| = \sum_{n, m} \frac{h^n(h') ^m}{n!m!}  \|   F_{n+m} \| \\
& = \sum_{\ell} \frac{1} {\ell!} \sum_{n+m = \ell}  \frac{\ell!}{n!m!} h^n(h') ^m \|   F_{\ell } \|   = \sum_{\ell} \frac{1} {\ell!} (h+h')^{\ell} \|   F_{\ell } \| = \| F \|_{h+h'}  \\
\end{split}
\ee

% \newpage

%>>>>>>>>>>>>>>>>>>>>>>>>>>>>>>>>>>>>>>>>>>>>>>>>>>>>>>>>>>>>>>>>>>>>>>>>>>>>>>>

\section{A single RG transformation}

\subsection{Some definitions} 
After $k$  RG transformations we will have an action which contains a term 
\be
 V_k = g_k \int   ( \bpsi \psi )^2  
\ee
with a new coupling constant $g_k$,  as well as many other terms.    If this were the only term in the next step we would want to compute
the fluctuation integral
\be
\blan e^{V_k}  \bran_{C_k}  (\psi)   \equiv \int e^{V_k( \psi + \eta) } d \mu_{C_k} (\eta) 
\ee
We would like to write this as the exponential of something.   To find the leading term of this something,  one might  look at the logarithm and expand in $V_k$
by
\be
\log \blan e^{V_k}  \bran  =    \blan V_k\bran + \frac12  \blan V_k,V_k \bran^T + \dots
\ee
where $\blan V_k,V_k \bran^T= \blan V_k^2 \bran -\blan V_k \bran^2$.   This is just formal because  $V_k$ is not small (due to the large volume) 
 and so the definition of the logarithm is problematic.    Nevertheless  $ \blan V_k\bran $ and  $ \frac 12\blan V_k,V_k \bran^T $   
are the correct leading terms and will play an important roll in the more detailed analysis.

In computing  these integrals it is useful to use the following fact.   If
\be
\blan F  \bran_{C}  (\psi)   \equiv \int F( \psi + \eta)  d \mu_{C} (\eta) 
\ee
then 
$
\blan F  \bran_{C}     = e^{\De_{C} } F
$
where $\De_C$ is the 
formal differential operator 
 \be
 \De_{C}  =  \int dx dy\  C_{ab} (x-y) \  \frac { \pa}{  \pa \bpsi_b  (y ) } \frac { \pa}{  \pa \psi_a  (x  ) }
  \ee 
    In the case at hand   $ \blan V_k\bran   = e^{\De_{C} }V_k $.  Any derivatives give terms proportions to $C_k(0) $ which vanishes
  since $C_k$ is an odd function.  (This is special to the massless model.)  Thus we
  have
  \be
  \begin{split}
  \blan V_k\bran_{C_k}    = & V_k\\
  \blan V_k, V_k \bran^T_{C_k}  = &    \blan V_k^2 \bran_{C_k}  - V_k^2 = ( e^{\De_{C_k} }-1) V_k^2 \\
  \end{split}
  \ee
  Terms like this accumulate after $k$ steps and we will find that the quadratic term has the form
 \be
  Q_k \equiv  \frac12   \blan V_k, V_k \bran^T_{w_k}   = \frac12 ( e^{\De_{w_k} }-1) V_k^2 
 \ee 
 where 
 \be  \label{wk}
w_k(x) =  \sum_{j=0}^{k-1} C_{j,L^{-(k-j} }  (x) =  {\sum}'_p e^{ipx} \frac{ -i \slp }{p^2} \B(  e^{-p^2/L^{2k} } - e^{-p^2} \B) 
\ee

In subsequent  steps  we  cannot avoid allowing  local  terms not present in the original action. 
    With $\ga_5 = i \ga_0 \ga_1$  there  may  be pseudo-scalar terms
of the form   
\be 
 V^p _k = p _k \int   ( \bpsi \ga_5 \psi )^2  
\ee
with $p_k = \cO(g_k^2)$ and $p_0 =0$.  
Again we have $\blan V^p_k \bran = V^p_k$  and 
there are associated quadratic terms   $<V_k, V^p _k>^T = \cO(g_k^3) $ and  $\frac12<V^p _k, V^p _k>^T= \cO(g_k^4)$. 
 It is only the first that we need to track in  any detail.  
 After $k$ steps these will accumulate to
 \be
 Q^p_k \equiv   \blan V_k, V^p_k \bran^T_{w'_k}   = ( e^{\De_{w'_k} }-1) V_kV^p_k
 \ee
 where $w_k'$ is $w_k$ with the $j=0$ term removed. 
 
  In addition there may be vector terms
of the form 
\be 
 V^v _k = v _k \int  \sum_{\mu}  ( \bpsi \ga_{\mu}  \psi )^2  
\ee
with $v_k = \cO(g_k^2)$ and $v_0 =0$.  Again we have $\blan V^v_k \bran = V^v_k$  and 
there are associated quadratic terms   $<V_k, V^v _k>^T = \cO(g_k^3) $ and  $\frac12<V^v _k, V^v _k>^T= \cO(g_k^4)$. 
After $k$ steps the first of  these  accumulates  to
\be
Q^v_k \equiv   \blan V_k, V^v_k \bran^T_{w'_k}   = ( e^{\De_{w'_k} }-1) V_kV^v_k
\ee

In fact it is not   $Q_k,  Q^p_k,  Q^v_k $ which occur,  but normalized versions   $  Q^{\reg}_k,  Q^{p, \reg}_k,  Q^{v, \reg} _k $
 which have relevant terms removed and have good bounds.  The precise definition is yet to come.
 In the following we group together the rogue terms by
\be
V_k' = V^p_k +V^v_k \hs      \hs Q_k' = Q^p_k +Q^v_k  \hs \hs   {Q_k'}^{\reg} =  Q^{p, \reg} _k + Q^{v,\reg} _k
\ee

%>>>>>>>>>>>>>>>>>>>>>>>>>>>>>>>>>>>>>>>>>>>>>>>>>>>>>>>>>>>>>>>>>>>>>>>>>>>>>>>

\subsection{Statement of the theorem} \label{statement} 

We going to claim that after $k$ renormalization group transformation the partition function can be written
as an integral over fields  $\psi$  on  $\bbT_{N+M-k} $     of the form    $\sZ = \int e^{  S_k(\psi) } d \mu_{G_k}(\psi) $
where $G_k$ is defined in (\ref{gk})  and 
\be  \label{h1} 
\begin{split} 
S_k  = &   \int  \B(   \vep_k   - z_k  \bpsi \slpa \psi      +  g_k  (   \bpsi \psi    )^2  +  p_k  ( \bpsi  \ga_5 \psi )^2  
+  v_k \sum_{\mu}  ( \bpsi  \ga_{\mu} \psi )^2   \B) +   Q_k^{\reg}  +   {Q_k'}^{\reg}   +  E_k  \\
= &   \int  \B(   \vep_k   - z_k  \bpsi \slpa \psi  \B) 
    +  V_k +  V_k' +  Q_k^{\reg}  +   {Q_k'}^{\reg}   +   E_k  \\
\end{split} 
\ee
and where   $E_k$  has a local expansion 
$E_k = \sum_X  E_k(X  ) $

\begin{thm} \label{one} 
 Let $L,h,C_Z,C_E$ be sufficiently large and chosen in that order (i.e. $h$ is sufficiently large depending on $L$, etc.), and  let 
 $g_{\max} $ be sufficiently small  depending on these constants. 
Suppose after $k$ renormalization steps  $\sZ = \int e^{ S_k} d \mu_{G_k}$ where $S_k$ has the form   (\ref{h1}) 
with $0< g_k   <   g_{\max} $ and  
\be  \label{region} 
\begin{split} 
\| E_k \|_{h, \Ga_4}   \leq  & C_E g_k^3 \\
|z_k| \leq &  C_Z g_k  \\
   |p_k|, |v_k|    \leq & C_E g_k^2 \\
\end{split}
\ee
And suppose  $E_k(X)$ has all the symmetries of section \ref{symmetry}. 
  Then  $\sZ = \int e^{  S_{k+1}} d \mu_{G_{k+1}}$ with  $S_{k+1} $ of the same form, with the same symmetries,  but with  new parameters.   
 There are  positive constants   constants  $\beta_k, \beta'_k$  bounded above and below and  also constants  $ \theta_k, \theta'_k$   
bounded above  such that 
 \be  \label{flow1} 
\begin{split}
E_{k+1} = & \cL( \cR E_k + E_k^*) \\
g_{k+1}   =& g_k   +   \beta_k  g^2_k -2 \beta' _kg_kp_k - 4\beta'_k g_kv_k  +  g^*_k \\
z_{k+1}  =&  z_k +  \theta_k  g_k^2 -  \theta^p_k g_kp_k - \theta^v_k g_kv_k+  z^* _k  \\
\vep_{k+1} = & L^2 (\vep_k + \vep_k^*) \\
\end{split}
\ee
as well as
 \be  \label{flow2} 
\begin{split}
p_{k+1}   = &p_k -   2  \beta_k' g_kv_k+ p_k^* \\
v_{k+1}   = & v_k-    \beta'_k g_kp_k+  v_k^* \\
\end{split}
\ee
The  starred quantities  are functions of    $E_k,  g_k, z_k, p_k, v_k$  in   the  domain (\ref{region})    and  satisfy there   
\be \label{onion1} 
\begin{split} 
 \| E_k^* \|_{\frac12 h, \Ga}   \leq  &  C_E h^{-\frac12} g_k^3  \\
 |g^*_k|, h^{-2} |z^*_k|, |p^*_k|, |v^*_k|  \leq   &   C_E h^{-4}   g_k ^3 \\
\end{split}
\ee
\end{thm} 
\bigskip

\rems
\begin{enumerate} 
\item   The linear  operator $\cR$ is defined in lemma \ref{local} to follow.     For small sets $X$ it   removes the relevant parts from $E_k(X) $  and these   show up in  the higher order corrections  $\vep^*_k,  z^*_k,  g^*_k,v_k^*,p_k^*$.   Then $\cR E_k$ is normalized for small sets in the sense of definition \ref{first}. 

\item  The operator 
$\cL$ defined in (\ref{suds}) reblocks and rescales.
 By lemma \ref{fiver}  and lemma \ref{local} 
 \be  \label{bottle1} 
 \| \cL \cR E_k  \|_{h, \Ga_4}   \leq   \cO(L^{-1} ) \| \cR E_k \|_{h, \Ga_4}   \leq  \cO(L^{-1}) \|  E_k \|_{h, \Ga_4}  \leq    \cO( L^{-1} )  C_E g_k^3
 \ee
 and  
 \be \label{bottle2} 
 \|  \cL E_k^* \|_{h, \Ga_4} \leq  \cO(L^2)  \| E_k^* \|_{L^{-\frac12} h, \Ga}  \leq  \cO(L^2)  \| E_k^* \|_{\frac12  h, \Ga}    \leq   \cO(L^2)  C_E   h^{-\frac12}g_k^3  
 \ee
It follows that   
\be \label{onion2} 
\|E_{k+1} \|_{h, \Ga_4}  \leq  \B(\cO(L^{-1}) + \cO(L^2)  h^{-\frac12}  \B )   C_Eg_k^3  <    C_E g_k^3   \leq  C_E g_{k+1}^3   
\ee
which advances the bound we started with. 
So  the  theorem is   enough to control the growth of    $E_k$  as long as $ g_k$ stays small.    

But the growth of  $g_k$ itself  (and  $z_k$)  is not necessarily controlled.   In fact the initial value 
must be carefully chosen to avoid problems.   This is  renormalization and is the content of   section \ref{five}.

\item  The following proof has many points of similarity with  the analysis of Brydges, Dimock, and Hurd  \cite{BDH96}, \cite{BDH98} who consider the infrared problem for 
a scalar $\phi^4$ model  in $4- \ep$ dimensions. 
 \end{enumerate} 
\bigskip

%>>>>>>>>>>>>>>>>>>>>>>>>>>>>>>>>>>>>>>>>>>>>>>>>>>>>>>>>>>>>>>>>>>>>>>>>>>>>>>>

\subsection{Start of the proof}  Starting  with $\sZ = \int e^{ S_k} d \mu_{G_k}$
we first absorb the incremental field strength into the measure   by  
\be \label{baking} 
  e^ {-  z_k \int  \bpsi \slpa \psi } d \mu_{G_k } = \det\B(I + z_k\slpa G_k \B)    d \mu_{G^z_k } (\psi)  
= e^{    \vep'_k    | \bbT_{N+M-k}|  }   d \mu_{G^z_k } (\psi)  
\ee
where   
 \be  \label{gkz}
G^z_k(x,  y)  =   {\sum}'_p    e^{ip(x-y)}  \frac{-i\slp}{p^2} \   \frac{ 1}{ z_k + e^{p^2} }   
\ee
This is established in  Appendix \ref{B}  where we also show that $|\vep'_k|  \leq   \one |z_k|  $. 
Now 
\be
\sZ =   e^{ ( \vep_k +  \vep'_k )   | \bbT_{N+M-k}|  }  \int  \exp\B( V_k +V'_k  + \hat Q_k + \hat Q'_k +  E_k  \B)   d \mu_{G^z_k }  
\ee
 Next  we split  $G^z_k = G^z _{k+1,L}  + C^z_k $ with 
 \be \label{h0} 
 \begin{split}
G^z_{k+1,L} (x, y)  =  &   {\sum}'_p    e^{ip(x-y)}  \frac{-i\slp}{p^2}     \frac{ 1}{ z_k + e^{L^2p^2} }   \\
C^z_k(x, y)  =  &   {\sum}'_p    e^{ip(x-y)}  \frac{-i\slp}{p^2}  \left[  \frac{ 1}{ z_k + e^{p^2} }   -   \frac{ 1}{ z_k + e^{L^2p^2} }  \right] \\
\end{split} 
\ee
Then 
\be
 \sZ =   e^{\vep_k +  \vep'_k   | \bbT_{N+M-k}|  }   \int    \left[    \exp\B( ( V_k + V_k' +\hat Q_k+ \hat Q'_k+ E_k) ( \psi+ \eta )  \B)  d \mu_{C^z_k} (\eta) \right] d \mu_{G^z_{k+1,L }  } (\psi)  
\ee
But again as in (\ref{baking}) 
\be
 d \mu_{G^z_{k+ 1,L } } 
= e^{- \vep''_k   | \bbT_{N+M-k}|  -  z_k\int  \bpsi \slpa \psi } d \mu_{G_{k+1,L  } }
 \ee
We also take out the leading term in $E_k$ defining $\de E_k$ by
\be
 E_k(\psi+ \eta) = E_k(\psi)   + \de E_k ( \psi, \eta) 
\ee
Then we have
\be  \label{ottoA} 
 \sZ =   e^{  ( \vep_k +  \vep'_k -\vep_k'')   | \bbT_{N+M-k}|  }   \int    \exp \B(   -    z_k\int  \bpsi \slpa \psi   + E_k(\psi) \B)      \    \Xi_k (\psi)    \        d \mu_{G_{k+1,L }  } (\psi)  
\ee
The  fluctuation integral is 
\be  \label{ottoB} 
   \Xi_k(\psi)        =     \int      \exp \B(  (V_k + V_k' +  Q_k^{\reg}  +   {Q_k'}^{\reg}  ) ( \psi+ \eta ) + \de E_k( \psi, \eta)  \B) \  d \mu_{C^z_k} (\eta)   
\ee
which we  abbreviate  as   
\be  
 \Xi_k(\psi)   = \blan    e^{W_k}   \bran_{ C^z_k}   (\psi)  
   \ee
with
\be 
  W_k =   V_k   +  V_k' +  Q_k^{\reg}  +   {Q_k'}^{\reg}   + \de E_k 
  \ee
 
 %>>>>>>>>>>>>>>>>>>>>>>>>>>>>>>>>>>>>>>>>>>>>>>>>>>>>>>>>>>>>>>>>>>>>>>>>>>>>>>>

 \subsection{Extraction}

Consider the leading  term $E_k(\psi) = \sum_X E_k(X, \psi) $. 
 We   isolate relevant and marginal terms and leave an irrelevant remainder.    

\begin{lem}   \label{local}     $E_k $ can be written 
\be 
E_k  =  E_k^{\loc}   +\cR E_k
\ee
where
\be  \label{otto1} 
E^{\loc} _k ( \psi  )  =        \int \B(    \vep^E_k  -    z^* _k    \   \bpsi  \slpa \psi    +  g^*_k  \  (\bpsi \psi)^2 +  p_k^* ( \bpsi \ga_5 \psi )^2 
 +  v_k^*  \sum_{\mu}   (\bpsi \ga_{\mu} \psi )^2  \B)    
\ee
and $   \cR E_k ( \psi )= \sum_X \cR E_k(X, \psi) $ is irrelevant in the sense of  (\ref{jinx}). For fixed $n$ 
we have the  $\cO(g_k^3) $ bounds with constants $\one$ independent of all parameters
\be
\begin{split} 
& |\vep_k^E| \leq \one A^{-1}  \|E_k\|_{h, \Ga_n}   \hs    |z_k^*| \leq \one A^{-1}  h^{-2} \|E_k\|_{h, \Ga_n}     \\
& |g_k^*|   |v^*_k|, |p^*_k|  \leq \one A^{-1}  h^{-4} \|E_k\|_{h, \Ga_n}      \\
\end{split} 
 \ee
 and 
 \be
 \|R E_k \|_{h, \Ga_n}  \leq \one \| E_k\|_{h, \Ga_n} 
 \ee
  \end{lem} 
\bigskip

 \pr 
  Suppress the index $k$, let $|X|$ =  volume of $X$,   and at first  choose  $ E^{\loc}  =\sum_X  E^{\loc}  (X) $ as  follows (cf.  \cite{BDH98}).
   \be 
 E^{\loc}  (X)  =  \alpha_{0} (X) |X|  +  \int_X  \bpsi  \al_{2}(X) \psi  +  \sum_{\mu} \int_X  \bpsi \  \al_{2,\mu} (X)   \pa_{\mu}  \psi 
 + \int_X  \bpsi \bpsi \ \al_{4}(X)  \ \psi \psi  
 \ee
Here if  $X$ is a small set in the sense of section  \ref{norms}  then 
\be
\begin{split}
 \alpha_{0}(X)  = &  \frac{1} {|X|} E_{00} (X) \\
\alpha_{2} (X)  = &  \frac{1} {|X|} E_{11} (X,1) \\
\alpha_{2,\mu} (X)  = &  \frac{1} {|X|} \B( E_{11} (X, x_{\mu} - x^0_{\mu}  )   - \frac{\int_X (x_{\mu} - x^0_{\mu})  }{ |X| }  E_{1,1} (X, 1 ) \B)     \\
\alpha_{4}  (X)  = &  \frac{1} {|X|}   E_{22} (X,1)   \\
\end{split} 
\ee
The  definition of   $\alpha_{2,\mu} (X) $ is independent of  the  reference  point $x^0$, but we  take $x^0 \in X$.    If $X$ is not a small set   then  all the $\al(X)=0 $ and
 $ E^{\loc}  (X)  = 0$,   We define $\cR E(X) = E(X) - E^{\loc} (X) $

Then       $(\cR E)_{11} (X,1 ) =0$ for $X$ small  since 
\be
  E^{\loc} _{11}( X, 1)  = \int_X \al_2  = \al_2 |X| = E_{11}(X,1)  
\ee
Further    $(\cR E)_{11} (X,x_{ \mu} - x^0_{\mu}  ) =0$ for $X$ small  since
\be 
\begin{split} 
 E^{\loc}_{11}  (X,x_{ \mu} - x^0_{\mu}  ) = &  \int_X \al_{2, \mu} (X) dx +  \int_X \al_{2} (X) (x_{\mu} -x^0_{\mu} ) dx \\
 = &  |X|  \al_{2, \mu} (X) +    \al_{2} (X)  \int_X (x_{\mu} -x^0_{\mu} ) dx \\
=&  E_{11} (X,x_{ \mu} - x^0_{\mu} )  \\ 
\end{split}
\ee
Similarly $(\cR E)_{00}(X) =0$ and  $(\cR E)_{22} (X,1 ) =0$.  Thus  for $X$ small $\cR E(X) $ is irrelevant (normalized)  in the sense of (\ref{jinx}).

Now we sum  the various terms in $E^{\loc}(X) $ over (small)  paved sets $X$.   We have
\be
\sum_X \alpha_0(X) |X|   =  \sum_X \alpha_0(X) \sum_{\sq \subset X}  |\sq|  
=  \sum_{\sq}  \B( \sum_{X \supset \sq}   \alpha_0(X)  \B)  |\sq|  
= \vep_k^E | \bbT_{M+N-k} | 
\ee
Here we defined
\be
\vep_k^E =  \sum_{X \supset \sq}   \alpha_0(X)  = \sum_{X \supset \sq  }   \frac{1} {|X|} E_{00} (X) 
\ee
which is independent of $\sq$ by translation invariance.   We have  $|E_{00} (X)| \leq \| E(X) \|_{h} $ 
and so   since $1 \leq A^{-1}  \Ga_n(X)  $
\be
| \vep_k^E | \leq   \sum_{X \supset \sq } \| E(X)  \|_{h}     \leq   A^{-1}  \sum_{X \supset \sq } \| E(X)  \|_{h}  \Ga_n(X)  = A^{-1}   \| E \|_{h, \Ga_n} 
\ee

For the quadratic terms we have
\be  \label{qqq} 
  \sum_X  \int_X  \bpsi  \al_2  (X) \psi  =   \sum_X   \sum_{\sq \subset X}  \int_{\sq}   \bpsi  \al_2  (X) \psi  
   =    \sum_{\sq }  \int_{\sq}   \bpsi \B(  \sum_{X \supset \sq} \al_2(X) \B) \psi   =    \int  \bpsi  \al_2 \psi 
\ee
where 
$ \al_2   =  \sum_{X \supset \sq} \al_2(X) $ is independent of $\sq$.
Similarly 
\be
\begin{split} 
 & \sum_X  \int_X  \bpsi  \al_{2, \mu } (X)\pa_{\mu}  \psi    =   \int  \bpsi    \al_{2,\mu} \pa_{\mu}   \psi \\
\end{split} 
\ee
where $\al_{2,\mu}   =  \sum_{X \supset \sq} \al_{2,\mu} (X) $ is independent of $\sq$.

 Now we claim that if $S^{-1} \ga_{\mu} S = \sum_{\nu} R_{\mu \nu}\ga_{\nu} $, then
 \be 
 \begin{split} 
 S^{-1}  \al_2 S =& \al_2 \\
  S^{-1}  \al_{2, \mu}  S =& \sum_{\nu} R_{\mu \nu}\al_{2,\nu} \\
 \end{split} 
 \ee
 To see this start with $E_{1,1}(X, \bpsi, \psi)  = S^{-1} E_{1,1}(RX, \bpsi_R, \psi_R) S$.  
 Then  $E_{1,1}(X, x,y)  =S^{-1}  E_{1,1}(RX, Rx,Ry) S$, 
 and hence   $E_{1,1} (X,1) =S^{-1}  E_{1,1} (RX,1) S$.  Then  $\al_2(X) = S^{-1} \al_2(RX) S$ and summing over
 $X \supset \sq$ gives the first identity.   The other is similar. 
 
 From these transformation properties it follows that   $\al_2 $ is a multiple of the identity and $\al_{2, \mu} $ is a multiple of $\ga_{\mu}$.
  For details on this argument see the Appendix \ref{C}. 
Thus we can define scalars $m_k^*, z_k^*$ by 
\be   \label{niki1} 
\begin{split} 
 m_k^*   =  &\al_2    =  \sum_{X \supset \sq} \al_2  (X)     \\
 z_k^*  \ga_{\mu}  = & -  \al_{2,\mu}  =     -       \sum_{X \supset \sq} \al_{2,\mu} (X)  \\
\end{split} 
\ee
and then the quadratic terms  are 
\be 
  m^*_k \int   \bpsi   \psi  
-         z^* _k    \int   \bpsi  \slpa \psi  
\ee
However  chiral invariance  (\ref{chiral})  of $E$  gives  that  $i \ga_5\al_{2} i\ga_5 =  \al_{2}$  and hence that  $i\ga_5  m_k^* i\ga_5 = m_k^*$. 
If the scalar $m_k^* \neq 0$ this says   $(i\ga_5 )^2 = 1$ which is false;   it is $-1$.  Thus $m_k^* =0$.   On the other hand  $ i \ga_5 z_k^*\ga_{\mu}  i\ga_5 =z_k^*  \ga_{\mu} $
is fine.

For a bound  on $z_k^*$ we note that    $|x_{\mu} - x^0_{\mu}| \leq |X| $    and so 
\be
\frac{1}{|X|}  | E_{11} (X,  x_{\mu} - x^0_{\mu} ) | \leq   \| E_{11}(X)  \|  \leq  \frac{1}{h^2}  \|E(X)\|_{h}  
\ee
The same holds for $  |X|^{-1}  \int (x_{\mu} - x^0_{\mu})  E_{11} (X, 1 )  $   and so twice this bound for $\al_{2,\mu} (X)$.  Then we have  
\be
|z_k^*|  = |\frac12  \sum_{X \supset \sq} \tr (\al_{2,\mu} (X)\ga_{\mu} )  | 
\leq \one  h^{-2}   \sum_{X \supset \sq}    \|E(X)\|_{h}   = \one  A^{-1}   h^{-2}    \|E\|_{h, \Ga_n}   
\ee

For the quartic terms   we have  as in (\ref{qqq}) 
\be
\begin{split}
 & \sum_X  \int_X  \bpsi  \al_{4} (X) \psi  =    \int  \bpsi \bpsi \al_{4} \psi \psi \\
  \end{split} 
\ee
where 
$ \al_{4}   =  \sum_{X \supset \sq} \al_{4} (X) $ is independent of $\sq$. 

Now $\al_{4} $  can be regarded as a  linear operator on $\bbC^2 \otimes  \bbC^2 $.   A basis for this space   is  the  sixteen operators
\be
\begin{array}{lll}  
I \otimes I &  I    \otimes \ga_{\mu}  &I  \otimes \ga_5 \\
\ga_{\mu} \otimes I   &    \ga_{\mu}  \otimes \ga_{\nu}  & \ga_{\mu} \otimes \ga_5 \\
\ga_5 \otimes I   &  \ga_5 \otimes \ga_{\mu}  & \ga_5 \otimes \ga_5 \\
\end{array}
\ee 
However from the  covariance  of $E_4(X) $   we have that. 
\be
( S^{-1}  \otimes S^{-1} )\  \al_4  \  (S  \otimes S )  = \al_4 
\ee
This is violated by all off-diagonal element in the above array for some $S$.   Furthermore one can show that $\ga_0 \otimes \ga_0$ and $\ga_1 \otimes \ga_1$
have the same coefficient in this basis. 
Thus  there are constants  $g_k^*,  p_k^*, v_k^* $. 
\be
\al_4  = g_k^* ( I  \otimes  I )+  p_k^*  ( \ga_5 \otimes \ga_5)   + v_k^*\sum_{\mu}  ( \ga_{\mu}  \otimes \ga_{\mu}) 
\ee
For details of this argument see Appendix \ref{C}.

The quartic term is then 
\be \label{acme2} 
  g^*_k  \int ( \bpsi \psi )^2   
+  p_k^*  \int ( \bpsi \ga_5 \psi )^2  +   v_k^*  \int  \sum_{\mu}  (\bpsi \ga_{\mu} \psi )^2   
\ee
 where 
 \be \label{take1} 
g^* _k   = 
\al_{4}  = \sum_{ X \supset \sq}   \al_4(X)   
\ee
We  have   $| \al_4(X) | \leq    \|  E_{22}(X)  \| \leq  4h^{-4}  \|  E(X)  \|_{h } $ and so 
\be  \label{take2} 
| g^*_k  |   \leq  \one A^{-1}  h^{-4} \|E\|_{h, \Ga_n}  
\ee
We get the same bound for  $v_k^*, p_k^*$.

Combining all the above if we sum  $ E(X)  =  E^{\loc} (X) + \cR E(X) $ over $X$ we get
 $ E  =  E^{\loc}  + \cR E$ with $E^{\loc}$ given by (\ref{otto1}) .
 \bigskip
 
 We still need the bound on  $ \cR E(X) $.     We  give a new local expansion for the term in $E^{\loc} $ based on unit blocks. 
From Lemma \ref{nugget}  we have that  $\int (\bpsi \psi)^2 $ has a local expansion $\sum_X E' (X) $         with  $\| E' \|_{h, \Ga_n}  \leq \one Ah^4$.     Combine 
this with the bound  $| g^*_k  |   \leq  \one A^{-1}  h^{-4} \|E\|_{h, \Ga_n} $    and  conclude the     $g_k^* \int (\bpsi \psi)^2 $ has a local expansion $\sum_X E'' (X) $   
      with  $\| E'' \|_{h, \Ga_n}  \leq \one \|E \|_{h, \Ga_n} $. The other terms in  $E^{loc} $ are similar and  so  $\| E ^{\loc} \|_{h, \Ga_n}  \leq \one \|E \| _{h, \Ga_n} $ and hence
  $\|\cR E  \|_{h, \Ga_n}  \leq \one \|E \| _{h, \Ga_n} $.

%\newpage

 %>>>>>>>>>>>>>>>>>>>>>>>>>>>>>>>>>>>>>>>>>>>>>>>>>>>>>>>>>>>>>>>>>>>>>>>>>>>>>>>

\subsection{Fluctuation integral}   \label{flc}

As we will see the fluctuation integral can be written as an exponential 
\be  \label{fluctuation} 
\Xi_k \equiv   \blan   e^{ W_k}   \bran_{C^z_k}   = \exp \B( W_k^\# \B) 
\ee
where    $W_k^\# $ has a local expansion.   More generally we will find 
\be
 \blan  e^{t W_k}  \bran_{C^z_k}   = \exp \B( W^\#_k (t ) \B)
\ee
where   $ W^\#_k (t )  $ is analytic   in complex  $t$ for   $|t|  <  r g_k^{-1} $ with  $r$ sufficiently small.   
Matching derivatives at $t=0$  gives 
\be
\begin{split} 
1 = &  \exp \B( W^\#_k (0 ) \B)\\
<W_k> = & (W^\#_k)' (0 ) \\
<W_k,W_k>^T = &  (W^\#_k)'' (0 ) \\
\end{split}
\ee
Then   expressing  $W_k^\# = W_k^\#(1)$   as an expansion around $t=0$ we have
\be \label{central} 
\exp \B( W^\#_k  \B) =\exp  \left (   \blan W_k  \bran  + \frac12 \blan W_k, W_k \bran^T +  \frac{1}{2 \pi i}  \int_{ |t| =rg_k^{-1} } \frac{ dt}{t^3(t-1) } W^\#_k(t)    \right) 
\ee
Since  $ W_k =   V_k   +   V_k' +\  Q_k^{\reg}  +   {Q_k'}^{\reg}   + \de E_k$ and since $<V_k> = V_k$ and $<V'_k> = V'_k$
\be  \label{central+}
\begin{split} 
 \blan W_k  \bran   = &  V_k   +   V' _k  +  \blan   Q_k^{\reg}      \bran +   \blan     {Q_k'}^{\reg}    \bran  +  \blan  \de E_k \bran  \\
\frac12   \blan W_k, W_k \bran^T = &   \frac12    \blan V_k, V_k \bran^T  +     \blan V_k, V' _k \bran^T    +   [ \cdots ]  \\ 
\end{split}
\ee
Here in the second line we have isolated the terms we want to analyze in detail in the next section.
In subsequent sections we  give   the exact definition  of $W_k^\#(t) $ and   the   remainder term in (\ref{central}).  Finally  the omitted terms in (\ref{central+}) are
analyzed. 
The omitted terms are  
\be  \label{central++}
\begin{split} 
[ \cdots ] = & \   \blan V_k,  Q_k^{\reg}  \bran^T + \blan V_k ,  {Q_k'}^{\reg} \bran^T +  \blan V_k,  \de E_k  \bran^T \\
&  + \frac12 \blan V'_k, V_k'  \bran^T  + \blan V'_k,   Q_k^{\reg}  \bran^T   + \blan V'_k,  {Q_k'}^{\reg}  \bran^T +
  \blan V'_k,  \de E_k  \bran^T  \\
  &  + \frac12   \blan  Q_k^{\reg} ,   Q_k^{\reg}  \bran^T   +   \blan   Q_k^{\reg} ,    {Q_k'}^{\reg}  \bran^T  + \blan  Q_k^{\reg} ,    \de E_k \bran^T   \\
    &  + \frac12   \blan  {Q_k'}^{\reg} ,    {Q_k'}^{\reg}  \bran^T    + \blan   {Q_k'}^{\reg} ,    \de E_k \bran^T   \\
      & +\frac12 \blan   \de E_k,    \de E_k \bran^T   \\
\end{split} 
\ee

%>>>>>>>>>>>>>>>>>>>>>>>>>>>>>>>>>>>>>>>>>>>>>>>>>>>>>>>>>>>

 \subsection{The quadratic terms} 

We study the $\cO(g_k^2) $ terms  in  (\ref{central+})  which are
\be \label{dingo}  
 \frac12    < V_k, V_k> ^T_{C^z_k}  +  < Q_k^{\reg} > _{C^z_k} 
\ee
We start with a computation of $Q_k$,  then discuss the modifications to generate $Q_k^{\reg}$, and
then  analyze the above  expression.  
By evaluating the formal derivatives we compute   
\be
   Q_k = \frac12  (e^{ \De_{w_k} } -1 ) V_k^2     \equiv   g_k^2   \cH(w_k)  
  \ee
where      
 \be
  \cH(w_k)  =  \cH_6 (w_k) +  \cH_4 (w_k) + \cH_2 (w_k)  +  \cH_0(w_k) 
  \ee
and     
\be \label{wonderful} 
\begin{split} 
 &  \cH_6(w_k)  =   4 \int  (\bpsi \psi)(x) \B( \bpsi(x)   w_k (x-y)    \psi(y)  \B)  (\bpsi  \psi)(y) dx dy \\
 &  \cH_4( w_k  )
= 
 -2n  \int  (\bpsi \psi)(x)    \tr    \B(  w_k(y-x)   w_k(x-y)  \B)  
   (\bpsi  \psi)(y)dx dy \\ 
       & +   4  \int      (\bpsi \psi)(x) \B(  \bpsi(y) w_k(y-x)w_k(x-y) \psi(y ) \B)  dx dy \\ 
   & + 2 \int   \B(\bpsi (x)  w_k(x-y)    \psi(y) \B)    \B(\bpsi (x)  w_k(x-y)    \psi(y)+     \bpsi (y)   w_k(y-x)   \psi(x)          \B)  dx dy  \\
  & \cH_2(w_k)   = 
      4  \int     \bpsi(x) \B(w_k(x-y) w_k(y-x) w_k(x-y)    \B)  \psi(y)   dx dy \\
    &\ \ \  -4n \int    \bpsi(x) w_k(x-y)   \psi(y)      \tr  \B( w_k(x-y) w_k(y-x)   \B)   dx dy    \\
  & \cH_0(w_k)   
   =  - n  \int   \tr  \B(w_k(x-y)  w_k(y-x)   w_k(x-y) w_k(y-x) \B)     dx dy \\
 & \ \ \  +    n^2 \int    \B(   \tr ( w_k(x-y) w_k(y-x) ) \B) ^2 dx dy    \\
\end{split}
\ee
Here we used
\be
\begin{split}
\De_{w}\B[  \psi^i_{ a}(x)  \bpsi^j_{b}(y)\B]  =  & \de_{ij} w_{ab}(x-y)  \\
\De_{w}\B [\sum_i  \psi^i_{ a}(x)  \bpsi^i_{b}(y)\B]  =  & n\  w_{ab}(x-y)  \\
\end{split} 
\ee

This expression will not have good estimates as $k$ increases.     The trouble is that  $w_k(x-y) $ is develops  a short distance singularity of the form $\cO( |x-y|^{-1} )$.
This is  summable (i.e has bounds independent of $N, k$)   but powers $w_k(x-y)^2$ and higher are not.    Thus the  $\cH_6(w_k) $  term  will have good estimates,  but this is  not so for the  $\cH_4(w_k)  , \cH_2(w_k) , \cH_0(w_k) $
terms.       To fix it we have to remove relevant terms, and this will give us the definition of $Q_k^{\reg}$.     

 \begin{prop}   There   exist  $ \beta(w_k),    \theta(w_k), \vep(w_k) $   such that  
  \be \label{yoyo} 
   \cH(w_k)  = \int \B( \vep(w_k) + \theta(w_k)  \bpsi \slpa \psi  + \beta(w_k) (\bpsi  \psi)^2 \B) +
 \hat  \cH(w_k) 
 \ee
 and   $ \hat  \cH(w_k) $  has no short distance singularities.   We define 
 \be
 Q_k^{\reg}  =  g_k^2 \hat \cH(w_k)
 \ee 
 \end{prop}

 \pr
First in 
$ \cH_4 (w_k)  $ we    replace  $\psi(y) , \bpsi(y) $ by   $\psi(x),   \bpsi (x) $   and let $\hat  \cH_4( w_k  )$
be the difference. 
There is no contribution from the  last term in  $\cH_4( w_k  )$  since $w_k$ is an odd function.   We have   
\be \label{just1}  
\begin{split}
 \cH_4( w_k  )
 =  &  \int      (\bpsi \psi)(x) \  \bpsi(x)  \beta(w_k)  \psi(x )   dx dy   + \hat  \cH_4( w_k  )
\end{split}
\ee
where 
\be 
\beta (w_k)  =  \int \B[ -2n\tr(  w_k(y-x)   w_k(x-y) ) + 4 w_k(y-x)w_k(x-y)\B] dy
\ee
 is independent of $x$.  The integral    $\int w_k(y-x)   w_k(x -y) )dy$  is a scalar in spin space so 
\be   \label{kingly} 
\beta (w_k)  = - 4(n-1)   \int w_k(y-x)   w_k( x-y) )dy = 4(n-1)  \int w_k(x-y)^2   dy
\ee
and
\be \label{thing1}  
\begin{split}
 \cH_4( w_k  )
 =  & \beta (w_k)  \int      (\bpsi \psi)^2    + \hat  \cH_4( w_k  )
\end{split}
\ee

In $\cH_2(w_k) $ we replace  $\psi(y) $
 by  $\psi (x)   +\sum_{\mu}  (y-x)_{\mu}   \pa_{\mu}  \psi (x) $  and  let $\hat \cH_2(w_k)  $ be the difference.   
The new  $\bpsi \psi  $   terms vanish since we are integrating an odd function  and we find
\be 
\cH_2(w_k) = \sum_{\mu}   \int    \bpsi(x)   \theta_{ \mu} (w_k)     \pa_{\mu} \psi (x)    dx   +\hat \cH_2 (w_k)  
\ee
where
\be
\begin{split} 
\theta_{\mu} (w_k) & =   4    \int \B(w_k(x-y) w_k(y-x) w_k(x-y)  (y_{\mu}  - x_{\mu} )  \B) \ dy   \\
& -4n   \int    \tr  \B( w_k(x-y)w_k(y-x)\ dy   \B)   w_k(x-y)   (y_{\mu}  - x_{\mu} ) \ dy   \\ 
\end{split}
\ee
is independent of $x$. 
But 
$
 \theta_{ \mu} (w_k)   = \ga_{\mu}  \theta (w_k) $   with   $ \theta  (w_k)   = \frac12 \tr ( \ga_{\mu} \theta_{\mu}(w_k) )
 $.
Therefore 
\be  \label{thing2} 
\cH_2  (w_k) =  \theta (w_k)  \int    \bpsi  \slpa \psi      + \hat \cH_2 (w_k)
  \ee
where with  $w(y-x) = -w(x-y)$ 
\be  \label{queenly} 
  \theta(w_k)=     2   \ \int \tr  \B( w_k^3(x-y)  ( \slx-  \sly  \B) \ dy    -2n \int       \tr  \B(   w_k^2(x-y)  \B)  \tr \B(  w_k(x-y) (  \slx- \sly )  \B) \ dy   
\ee  

We now    define  $ \hat \cH_6 (w_k)     = \cH_6(w_k)    $  and $\hat \cH_0 (w_k)  =0$,    and  $ \vep(w_k) |\mathbb{T}_{N+M-k}| = \cH_0(w_k) $.
Then we have the stated result     (\ref{yoyo})  with 
\be  
  \hat \cH(w_k)
=  \B(  \hat \cH_6 (w_k) + \hat  \cH_4 (w_k) + \hat  \cH_2 (w_k)  + \hat  \cH_0(w_k) \B) 
  \ee

 \begin{prop}   There are constants   $\beta_k, \theta_k, \vep^Q_k $   and  $E_k^z(\psi) $  such that
 \be  \label{otto2} 
 \begin{split}
&   \frac12   < V_k  ,   V_k  >^T_{C^z_k}    +  <Q_k^{\reg}  >_{C^z_k} \\
& =  
 g_k^2 \int  \B(  \vep^Q_k   +  \theta_k     \   \bpsi  \slpa \psi  + \beta_k    (\bpsi \psi)^2 \B) +  g_k^2    \hat  \cH(w_k+ C_k)    + E^z_k   \\
 \end{split} 
\ee
\end{prop} 

\rem
This is the expression we are looking for.  It exhibits the corrections to the coupling constants and  preserves the general form since
 $g_k^2    \hat  \cH(w_k+ C_k) $  scales to  $g_k^2    \hat  \cH(w_{k+1} ) $.  After a change of coupling constants   $g_k = g_{k+1} + \cO(g_k^2)$ this  
  becomes   $ Q_{k+1}^{\reg} =g_{k+1}^2    \hat  \cH(w_{k+1} )$ plus higher order terms.  $E_k^z$ is also higher order.   As we will see 
  $\beta_k, \theta_k, E_k^z$   are bounded in $k$. 
\bigskip

\pr   
 At  first suppose   it is  $ <   Q_k >_{C^z_k}  $ rather than   $ < Q_k^{\reg} >_{C^z_k} $.  
  Then we have  
\be \label{bongo1} 
\begin{split} 
 & \frac12   < V_k  ,   V_k  >^T_{C^z_k}    +  <   Q_k >_{C^z_k}  \\
 =& \frac12 (e^{ \De_{C^z_k}}  -1) V_k^2  +  \frac12 e^{ \De_{C^z_k} }(e^{ \De_{w_k} } -1 ) V_k^2 \\
 = &  \frac12 (e^{ \De_{C^z_k}}e^{ \De_{w_k} }  -1) V_k^2 
   =  \frac12 ( e^{ \De_{w_k + C^z_k} }- 1) V_k^2  = g_k^2\cH(w_k + C^z_k) \\
 \end{split} 
 \ee

Now consider the 
 actual expression   which we write as 
 \be \label{bongo2} 
\begin{split} 
 & \frac12   < V_k  ,   V_k  >^T_{C^z_k}    +  < Q_k^{\reg}   >_{C^z_k}    \\
= & \frac12   < V_k  ,   V_k  >^T_{C^z_k}    +  <   Q_k   >_{C^z_k}   +     g_k^2<   \hat  \cH  (w_k)  -    \cH  (w_k)  >_{C^z_k}          \\
= &   g_k^2 \cH(w_k + C^z_k)       +     g_k^2  (\hat  \cH  (w_k)  -    \cH  (w_k) )    \\
= & g_k^2    \B(\cH(w_k + C_k) - \cH( w_k)\B)  + g_k^2 \hat  \cH  (w_k)         + E^z_k
 \end{split} 
 \ee
Here in the second step  we used  (\ref{bongo1})  and the fact  that  $ \hat  \cH  (w_k)  -    \cH  (w_k)  $ is strictly local so the fluctuation integral has no effect.   
In the last step we defined
\be  \label{bongo3} 
E^z_k    =  g_k^2    \B(\cH(w_k + C^z_k) -  \cH(w_k + C_k)\B)  
\ee
Also    from (\ref{yoyo}) 
\be
 \cH( w_k + C_k)   -\cH ( w_k  )   =  \int \B ( \vep^Q_k  +  \beta_k    (\bpsi \psi)^2  +   \theta_k     \bpsi  \slpa \psi  \B)     +     \hat  \cH( w_k + C_k)   - \hat \cH( w_k  )    
\ee
where
\be 
\begin{split}
\vep^Q_k =  & \vep   ( w_k + C_k) - \vep ( w_k )  \\
\beta_k =  & \beta  ( w_k + C_k) -  \beta ( w_k )  \\
  \theta_k =  &   \theta   ( w_k + C_k) -  \theta  (  w_k  )  \\ 
  \end{split} 
\ee
 Insert   this into (\ref{bongo2}) and  cancel the terms $ \hat  \cH (w_k) $ to obtain the stated result.

%<<<<<<<<<<<<<<<<<<<<<<<<<<<<<<<<<<<<<<<

\subsection{The lowest order rogue terms} 

We repeat the analysis of the previous section for the terms   $    < V_k, V'_k> ^T  +  <  {Q_k'}^{\reg}   > $ in (\ref{central+}) which are $\cO(g_k^3)$.  
This expression breaks into a pseudo-scalar part and a vector part.

We start with the pseudo-scalar part which is 
 $     < V_k, V^p_k> ^T  +  < Q^{p, \reg} _k >  $.
First we  define  $\cH^p(w'_k)$  by 
   \be 
 Q_k^p  = (  e^{\De_{w'_k} }-1 ) V_kV_k^p  \equiv  g_k p_k  \cH^p(w'_k) 
\ee 
 and we find that  
 $ \cH^p =  \cH^p_6 + \dots +  \cH^p_0 $ where 
\be \label{h6p} 
   \cH^p_6(w'_k)  =   8 \int  (\bpsi \psi)(x) \B( \bpsi(x)   w'_k (x-y) \ga_5   \psi(y)  \B)  (\bpsi \ga_5 \psi)(y) dx dy 
 \ee      
\be \label{h4p} 
\begin{split} 
&  \cH^p_4( w'_k  )
= 
 -4n  \int  (\bpsi \psi)(x)    \tr    \B(  w'_k(y-x)   w'_k(x-y)\ga_5  \B)  
   (\bpsi \ga_5 \psi)(y)dx dy \\ 
       & +   8  \int      (\bpsi \psi)(x) \B(  \bpsi(y)\ga_5  w'_k(y-x)w'_k(x-y)\ga_5 \psi(y ) \B)  dx dy \\ 
   & + 4  \int   \B(\bpsi (x)  w'_k(x-y) \ga_5   \psi(y) \B)  \  \B(\bpsi (x)  w'_k(x-y) \ga_5   \psi(y)+     \bpsi (y) \ga_5  w'_k(y-x)   \psi(x)          \B)  dx dy  \\
\end{split}
\ee
\be 
\begin{split}  \label{h2p} 
   \cH^p_2(w'_k)   = 
    &  8   \int     \bpsi(x) \B(w'_k(x-y)\ga_5 w'_k(y-x) w'_k(x-y) \ga_5   \B)  \psi(y)   dx dy \\
    & -8n \int    \bpsi(x) w'_k(x-y) \ga_5  \psi(y)      \tr  \B( w'_k(x-y)\ga_5 w'_k(y-x)   \B)   dx dy    \\
\end{split}
\ee
\be 
\begin{split}  \label{h0p} 
   \cH^p_0 (w'_k)   
   = & -2 n  \int   \tr  \B(w'_k(x-y)\ga_5  w'_k(y-x)   w'_k(x-y)\ga_5 w'_k(y-x) \B)     dx dy \\
   +   & 2n^2 \int    \B(   \tr  (w'_k(x-y)\ga_5 w'_k(y-x) ) \B) ^2 dx dy    \\
\end{split}
\ee

\begin{prop}  There are constants $ \beta'(w'_k),     \theta^p (w'_k),     \vep^{p,Q}(w'_k)  $ such that 
 \be \label{willing} 
  \cH^p (w'_k)  =   \int \B( \vep^{p,Q}(w'_k)  +  \theta^p (w'_k)      \bpsi  \slpa \psi   - 2  \beta'(w'_k)      (\bpsi \psi) ^2      
    -      \beta' (w'_k)  \sum_{\mu}    (\bpsi  \ga_{\mu}      \psi )^2  \B) 
    + \hat  \cH^p (w'_k) 
\ee
and    $\hat  \cH^p (w'_k) $ has no short distance singularities.        We define
\be
 Q_k^{p,\reg}   =g_kp_k  \hat  \cH^p(w'_k)   
 \ee
 \end{prop}
\bigskip

\pr 
In $\cH^p_4(w'_k) $ we again  replace $\bpsi(y) , \psi(y) $ by $\bpsi(x), \psi(x) $ and let  $\cH^p_4(w'_k) $ be the difference.    The first
term  in $ \cH^p_4( w'_k  )$ is zero since  $\int w'_k(y-x)   w'_k(x-y) dy $ is a multiple of 
the identity in spin space and $\tr \ga_5 =0$. In the second term we identify  $\ga_5^2=1$.   In the third term we use 
$w'_k(x-y) \ga_5 = \ga_5 w'_k( y-x)$.    
Then we have 
\be \label{TMI}  
\begin{split}
 \cH^p_4( w'_k  )
 =  &  \ 8  \int      (\bpsi \psi)(x) \B(  \bpsi(x)  \B[ \int w'_k(y-x)w'_k(x-y) dy \B] \psi(x ) \B)  dx  \\ 
   & + 8  \int   \B(\bpsi (x)  w'_k(x-y) \ga_5   \psi(x) \B)  \  \B(      \bpsi (x) \ga_5  w'_k(y-x)   \psi(x)          \B)  dx dy    + \hat  \cH^p_4( w'_k  )\\
\end{split}
\ee

The bracketed expression is identified as $- \frac14 (n-1)^{-1}  \beta(w'_k) $. 
Also we write    $w'_k = \sum_{\mu}  \ga_{\mu} w'_{k, \mu} $
and then 
\be
\int   w'_{k, \mu} (y) w'_{k,\nu}(y)  dy   = \frac12   \de_{\mu \nu}  \int  \sum_{\mu} ( w'_{k,\mu} ( y) )^2dy =\frac12  \de_{\mu \nu}  \int   w'_k(y) ^2 
\ee
and the latter is identified as $ \frac18 (n-1)^{-1}  \beta(w'_k) \de_{\mu \nu} $. 
We  also use  
\be
\sum_{\mu \nu}   \de_{\mu \nu}  (   \ga_{\mu}\ga_5  \otimes \ga_5 \ga_{ \nu}    ) 
= \sum_{\mu }    (  \ga_{\mu}\ga_5  \otimes \ga_5 \ga_{\mu} )   =  - \sum_{\mu }    (  \ga_{\mu}  \otimes   \ga_{\mu} )     
\ee
Then we have with $\beta'(w'_k) =   (n-1)^{-1} \beta(w'_k) $  
 \be \label{TMIa}  
\begin{split}
\cH^p_4( w'_k  )
 =   &   - 2  \beta'(w'_k) \int      (\bpsi \psi) ^2      
    -      \beta' (w'_k)   \int  \sum_{\mu}  (\bpsi  \ga_{\mu}      \psi )^2         + \hat  \cH^p_4( w'_k  ) \\
\end{split}
\ee

Now consider the term   $ \cH^p_2 (w'_k)$.  We  replace  each  $\psi(y)$ by $\psi (x) + \sum_{\mu} (y_{\mu}  - x_{\mu} ) \pa_{\mu} \psi (x) $. 
The leading terms vanish since we are integrating an odd function.   Then we have 
\be \label{oldyear} 
\cH^p_2(w'_k) =   \sum_{\mu}     \int    \bpsi(x) \theta^p_{ \mu} (w'_k)     \pa_{\mu} \psi (x)    dx   + \hat \cH^p_k(w'_k) 
\ee
where (dropping some  $\ga_5$'s) 
\be
\begin{split} 
\theta^p_{ \mu} (w'_k) 
 = &   8 \int \B(w'_k(x-y) w'_k(y-x) w'_k(x-y) (y_{\mu}  - x_{\mu} )  \B) \ dy  \\
  & - 8n   \int   w'_k(x-y)  \ga_5  (y_{\mu} - x_{\mu})      \tr  \B( w'_k(x-y)\ga_5 w'_k(y-x)   \B)    dy    \\
 \end{split} 
\ee
But  $\tr ( \theta^p_{ \mu}) = 0$ and $\tr ( \ga_5\theta^p_{ \mu}) = 0$ and  $\theta^p_{\mu} $ is independent of $\mu$.  Thus
$
 \theta^p_{ \mu} (w'_k)   = \ga_{\mu}  \theta^p (w'_k) $  where  $ \theta^p  (w'_k)   = \frac12 \tr (  \ga_{\mu} \theta_{\mu} (w'_k) )$.
  Thus 
 \be
 \cH^p_2
 (w'_k) =   \theta^p(w'_k)  \int    \bpsi  \slpa \psi     +\hat  \cH^p_2(w'_k) 
\ee
where
\be
\begin{split} 
\theta^p (w'_k) 
 = &   4\int \tr\B(w'_k(x-y) w'_k(y-x) w'_k(x-y) (\sly-\slx)  \B) \ dy  \\
  & - 4n   \int  \tr\B(   w'_k(x-y)  \ga_5  (\sly-\slx)  \B)     \tr  \B( w'_k(x-y)\ga_5 w'_k(y-x)   \B)    dy    \\
 \end{split} 
\ee
We further   define  $ \hat   \cH^p_ 6(w'_k)  =\cH^p_ 6(w'_k) $ and $\hat \cH^p_0(w'_k) =0 $ and  $ \vep^{p,Q}(w'_k) |\bbT_{N+M-k} |   = \cH^p_0(w'_k)$. 
Then we have the stated result with 
 \be
  \hat  \cH^p(w'_k)   =   \hat \cH^p_6(w'_k) + \dots +    \hat \cH^p_0(w'_k) 
 \ee

\begin{prop}  \label{hawk}  There are constants   $\beta'_k,     \theta^p_k,    \vep^{p,Q}_k $ and $ E^{z,p}_k (\psi) $   such that  
\be   \label{licorice2} 
\begin{split} 
&   < V_k, V^p_k> ^T_{C^z_k}    +  <  Q_k^{p,\reg}  >_{C^z_k}  \\
=
 &  g_kp_k  \int   \B( \vep^{p,Q}_k 
 +  \theta^p_k     \bpsi  \slpa \psi  -2  \beta'_k         (\bpsi \psi)^2  
   -  \beta'_k   \sum_{\mu}  (\bpsi  \ga_{ \mu}    \psi) ^2    \B)  +     g_kp_k \hat  \cH^{p}(w'_k + C_k  )  + E^{z,p}_k   \\   
\end{split}
\ee
\end{prop} 
\bigskip

\pr
As in (\ref{bongo2})  and (\ref{bongo3}) (without the factor $\frac12$) 
\be   \label{licorice} 
\begin{split} 
    < V_k, V^p_k> ^T_{C^z_k}    +  <  Q_k^{p,\reg}  >_{C^z_k}  
= & g_kp_k    \B(\cH^p_k(w'_k + C^z_k)  +( \hat  \cH^p  (w'_k) -    \cH^p (w'_k))  \B)     \\   
= & g_kp_k    \B(\cH^p_k(w'_k + C_k) -    \cH^p  (w'_k) \B) +g_kp_k  \hat  \cH^p  (w'_k)   + E^{z,p}_k   \\   
\end{split}
\ee
where  $E^{z,p}_k =g_kp_k( \cH^p_k(w'_k + C^z_k)-\cH^p_k(w'_k + C_k)) $.
But from (\ref{willing}) 
\be
 \begin{split} 
   \cH^{p}(w'_k + C_k  ) -    \cH^{p}(  w'_k)    =  &  \int \B( \vep^{p,Q} _k   +  \theta^p_k      \bpsi  \slpa \psi   - 2  \beta'_k      (\bpsi \psi) ^2    
    -      \beta'_k   \sum_{\mu}   (\bpsi  \ga_{\mu}      \psi )^2  \B) \\
       + &  \hat  \cH^p(w'_k + C_k  )  -\hat  \cH^p (w'_k)  \\
       \end{split} 
\ee
where
\be  \label{skunk} 
\begin{split}
\vep^{p,Q} _k =  & \vep^{p,Q}   ( w'_k + C_k) - \vep^{p,Q} ( w'_k )  \\
\beta'_k =  & \beta'  ( w'_k + C_k) -  \beta' ( w'_k )  \\
  \theta^p_k =  &   \theta^p    ( w'_k + C_k) -  \theta  (  w'_k  )  \\ 
  \end{split} 
\ee
Now insert this into (\ref{licorice}), cancel the $g_kp_k  \hat  \cH^p  (w'_k) $   get the stated result.
This completes the proof. 

%>>>>>>>>>>>>>>>>>>>>>>>>

\bigskip 

Now consider the vector part which is $     < V_k, V^v_k> ^T    +  <  Q_k^{v,\reg}  >  $.
We compute 
\be 
 Q_k^v  = (  e^{\De_{w'_k} }-1 ) V_kV_k^v  \equiv  g_k v_k  \cH^v(w'_k) 
\ee 
The  $ \cH^v(w'_k) $ is the same as $ \cH^p(w'_k) $ in (\ref{h6p}) - (\ref{h0p}) except that we replace each $\ga_5$ by $\ga_{\si}$
and sum over $\si $. 

\begin{prop}
There are constants $ \beta'(w'_k),  \theta^v  (w'_k) ,  \vep ^{v,Q}(w'_k) $ such that   
 \be \label{willing2} 
  \cH^v (w'_k)  =   \int \B( \vep_k ^{v,Q}(w'_k)  +  \theta^v  (w'_k)      \bpsi  \slpa \psi   - 4  \beta'(w'_k)      (\bpsi \psi) ^2      
    -   2   \beta' (w'_k)   (\bpsi  \ga_5     \psi )^2  \B) 
    + \hat  \cH^v (w'_k) 
\ee
and $\hat  \cH^v (w'_k) $ has no short distance singularities.   We define
\be
 Q_k^{v,\reg}  =g_kv_k  \hat  \cH^v(w'_k)  
\ee
\end{prop} 
\bigskip

\pr
Instead of  (\ref{TMI})  we have
\be \label{TMI2}  
\begin{split}
 \cH^v_4( w'_k  )
 =  &  \ 8  \int      (\bpsi \psi)(x) \B(  \bpsi(x)  \B[ \int w'_k(y-x)w'_k(x-y) dy \B] \psi(x ) \B)  dx  \\ 
   & + 8  \int  \sum_{\si}  \B(\bpsi (x)  w'_k(x-y) \ga_{\si}    \psi(x) \B)  \  \B(      \bpsi (x) \ga_{\si}   w'_k(y-x)   \psi(x)          \B)  dx dy    + \hat  \cH^v_4( w'_k  )\\
\end{split}
\ee
The bracketed expression is again identified as $- \frac14 \beta'(w'_k)$.
Again we  write  $w'_k =\sum_{\mu} \ga_{\mu} w'_{k, \mu} $  and identify  $ \int w'_{k, \mu}(y-x)  w'_{k, \nu}(x-y) dy  = -\frac18 \beta'(w'_k) \de_{\mu \nu}  $.
We also use  
\be
\begin{split} 
\sum_{\si \mu \nu}   \de_{\mu \nu}  (   \ga_{\mu}\ga_{\si }   \otimes \ga_{\si }  \ga_{ \nu}    ) 
= &  \sum_{\si  \mu }    (  \ga_{\mu}\ga_{\si}   \otimes \ga_{\si}  \ga_{\mu} )    \\
= &  \sum_{\si = \mu }    (  \ga_{\mu}\ga_{\si}   \otimes \ga_{\si}  \ga_{\mu} ) +  \sum_{\si \neq  \mu }    (  \ga_{\mu}\ga_{\si}   \otimes \ga_{\si}  \ga_{\mu} )    \\
= &  2( I \otimes I )  + 2 (\ga_5 \otimes \ga_5) 
 \end{split} 
\ee
Therefore
\be \label{snort2}  
\begin{split}
 \cH^v_4( w'_k  )
 =  &  -4 \beta'(w'_k)   \int      (\bpsi \psi)^2 -2  \beta'(w'_k)    \int      (\bpsi \ga_5 \psi)^2      + \hat  \cH^v_4( w'_k  )
\end{split}
\ee

For the quadratic terms we have  instead of (\ref{oldyear})
\be \label{oldyear2} 
\cH^v_2(w'_k) =   \sum_{\mu}      \int    \bpsi(x) \theta^v_{ \mu} (w'_k)     \pa_{\mu} \psi (x)    dx   + \hat \cH^v_2(w'_k) 
\ee
where 
\be
\begin{split} 
\theta^v_{ \mu} (w'_k) 
 = &   16 \int \B(w'_k(x-y) w'_k(y-x) w'_k(x-y) (x_{\mu}  - y_{\mu} )  \B) \ dy  \\
 \end{split} 
\ee
The second term is absent here since it involves the trace of three gammas. 
But  $\tr ( \theta^v_{ \mu}) = 0$ and $\tr ( \ga_5\theta^v_{ \mu}) = 0$ and  $\theta^v_{\mu} $ is independent of $\mu$.  Thus
$
 \theta^v_{ \mu} (w'_k)   = \ga_{\mu}  \theta^v (w'_k) $  where  $ \theta^v  (w'_k)   = \frac12 \tr (  \ga_{\mu} \theta^v_{\mu} (w'_k) )$.
  Thus 
 \be
 \cH^v_2
 (w'_k) =   \theta^v(w'_k)  \int    \bpsi  \slpa \psi     +\hat  \cH^v_2(w'_k) 
\ee
where
\be
\begin{split} 
\theta^v (w'_k) 
 = &   8\int \tr\B(w'_k(x-y) w'_k(y-x) w'_k(x-y) (\slx-\sly)  \B) \ dy  \\
 \end{split} 
\ee

We also define  $ \hat   \cH^v_ 6(w'_k)  =\cH^v_ 6(w'_k) $ and $\hat \cH^v_0(w'_k)  =0$ and    $ \vep_k^{ v,Q}(w'_k)| \bbT_{M+nN-k} |    = \cH^v_0(w'_k)$. 
Then the result holds with
 \be
\hat \cH^v(w_k')  =    \hat \cH^v_6(w'_k) + \dots +    \hat \cH^v_0(w'_k) 
\ee

\begin{prop}
There are constants  $ \beta'_k,   \theta^v_k,  \vep^{v,Q}_k$  and $E^{z,v}_k (\psi) $   such that 
\be   \label{licorice3} 
\begin{split} 
&   < V_k, V^v_k> ^T_{C^z_k}    +  < Q_k^{v,\reg}  >_{C^z_k}  \\
=
 &  g_kv_k \int \B( \vep^{v,Q}_k  +  \theta^v_k      \bpsi  \slpa \psi   - 4  \beta'_k       (\bpsi \psi) ^2      
    -   2   \beta'_k  (\bpsi  \ga_5     \psi )^2  \B)+    g_kv_k \hat  \cH^v(w'_k + C_k  )  + E^{z,v}_k   \\   
\end{split}
\ee
\end{prop} 
\bigskip
  
  \pr We have from (\ref{willing2}) 
\be \label{esther} 
\begin{split} 
  \cH^v (w'_k+C_k ) -   \cH^v (w'_k)  =   &  \int \B( \vep^{v,Q}_k  +  \theta^v_k      \bpsi  \slpa \psi   - 4  \beta'_k       (\bpsi \psi) ^2      
    -   2   \beta'_k   (\bpsi  \ga_5     \psi )^2  \B) \\
    + &  \hat  \cH^v (w'_k+C_k ) -  \hat  \cH^v (w'_k)   \\
\end{split} 
\ee
where  $\beta'_k,  \theta^v_k, \vep^{v,Q} _k $ are defined as in (\ref{skunk}). 
The rest of the analysis follows the pseudoscalar  case Proposition \ref{hawk}   and gives the stated result.  
\bigskip

Finally  $   < V_k, V'_k> ^T   +  < {Q'_k}^{\reg}  >$ is the sum of (\ref{licorice2}) and (\ref{licorice3}).  When summing we define   $ {E^z_k}'  = E^{z,p}_k  + E^{z,v}_k$.

%>>>>>>>>>>>>>>>>>>>>>>>>>>>>>>>>>>>>>>>>>>>>>>>>>>>>>>>>>>>>>>>>>>>>>>>>>>>>>>>

\subsection{Estimates} 

Before  proceeding we develop estimates on some of the objects we have created.  
Here and throughout the paper $\cO(1)$ stands for a constant independent of all parameters.   We regard the number of internal components $n$ as fixed and $\one$.  
The letter  $C$ stands for a constant that may depend on $L$, and it  may change from line to line. 

 We start with estimates on  $\beta_k, \theta_k$.  ($\vep_k^Q$ is finite but is not bounded uniform in $k$.) 

\begin{lem}  { \ } \label{eleven} 
\begin{enumerate} 
\item There exist  positive constants $C_{\pm}$ such that
\be
 ( n  -1) C_-    \leq \   \beta_k  \   \leq  ( n  -1)C_+    \ee
 and $\beta'_k$ satisfies the same bound without the $(n-1)$. 
 \item There exists a constant $C$  such that $ |\theta_k|,   |\theta^p_k|,  |\theta^v_k| \leq C  $ 
 \end{enumerate} 
\end{lem} 
\bigskip

\pr
The coefficient $\beta_{k} =  \beta( w_k+ C_k ) - \beta(w_k) $  is from (\ref{kingly})      
\be 
   \beta_{k}   
    =  4 (n-1)  \int  [(w_k +C_k)^2   -    w_k^2  ]  
    =  4 (n-1)  \int  [2w_kC_k  + C_k^2   ]  
\ee
We first consider $\int     C_k^2  $
and compute 
\be
\int     C_k^2  =   {\sum}'_{p  \in \bbT^*_{M+N-k}}      \frac{1 }{p^2}  \ \B( e^{-p^2 }-  e^{-L^2p^2} \B)^2=   {\sum}'_{p  \in \bbT^*_{M+N-k}}    \frac{e^{-2p^2}  }{p^2}  \ \B( 1-  e^{-( L^2-1) p^2} \B)^2 
\ee
 We split this  into  two pieces.  First consider  the region   
$(L^2-1) p^2 \leq 1$. We use the bound
$
x e^{-1}      \leq   1-e^{-x}   \leq   x  $ for $0 \leq  x \leq  1   $.   
Then
\be                     
e^{-1}  ( L^2-1)^2   {\sum}'_{(L^2-1) p^2 \leq 1}   p^2 e^{-2p^2}       \leq     \int     C_k^2   \leq   ( L^2-1)^2     {\sum}'_{(L^2-1) p^2 \leq 1}  p^2 e^{-2p^2}    
\ee 
which suffices.     
In the region $(L^2-1) p^2 \geq 1$ we use the   bound 
$
1-    e^{-1}  \leq   1-e^{-x}  \leq  1$ for    $x \geq 1$ 
  Then   
 \be                     
(1-e^{-1} )^2   {\sum}'_{(L^2-1) p^2 \geq 1}     \frac{e^{-2p^2}  }{p^2}    \leq     \int     C_k^2   \leq       {\sum}'_{(L^2-1) p^2 \geq 1}    \frac{e^{-2p^2}  }{p^2}    
\ee 
 which also suffices.

Now let's look at the other term
which is 
\be 
\int  2 w_k C_k   =   2 {\sum}'_{p  \in \bbT^*_{N+M}}     \frac{1 }{p^2}  \ \B( e^{-p^2/L^{2k} }-  e^{- p^2} \B) \B( e^{-p^2 }-  e^{-L^2p^2} \B) 
\ee
Since this is positive and since we already have a lower bound from the first term, it suffice to get an upper bound on this. 
But    $( e^{-p^2- } - e^{-L^2p^2}) $ is $ \cO(p^2)$ as $p \to 0$ and $\cO(e^{-p^2})$ as $p \to \infty$.  So this term is bounded by a constant. 
This completes the  bound on $\beta_k$

The coefficient $\theta_k$ is  $    \theta( w_k+ C_k ) - \theta(w_k)$ where   $\theta(w_k)$ is defined in  (\ref{queenly}).
We  use the bounds from Appendix \ref{D} which are
\be  \label{bdds}
| C_k(x-y) |  \leq  \one e^{ - |x-y|/L}    \hs \hs 
 |w_k(x- y) | \leq      \one |x-y| ^{- 1} e^{-|x-y| }  
 \ee
Referring to (\ref{queenly})  we see that the integral defining $\theta  (w_k) $   is exponentially decaying as $x-y \to \infty$
and has an $\cO(|x-y|^{- 2})$  singularity as $x \to y$.  The latter is not summable.
But in $ \theta_k =     \theta( w_k+ C_k ) - \theta(w_k)$  at least one of the factors $w_k $  is replaced by   $C_k$.  
This reduces the short distance singularity to   $\cO(|x- y|^{- 1})$ which is summable.  The bound on $\theta_k$ follows and 
the others are similar.  This completes the proof. 
\bigskip

Next we   develop local expansions for  $V_k, V'_k,  Q_k^{\reg} ,  {Q'_k}^{\reg} $ and give associated estimates.  
In fact as in  Lemma \ref{nugget}  we already have $V_k = \sum_X V_k(X)$ and $V' _k = \sum_X V' _k(X)$  
with 
\be \label{lunar} 
  \| V_k \|_{h, \Ga_n}     \leq C  h^4  g_k    \hs \hs     \| V'_k  \|_{h, \Ga_n}     \leq C h^4 (|p_k|+|v_k| )  
  \ee
  
For the  quadratic terms we have:  
\begin{lem} \label{qlem} 
$ Q_k^{\reg} =  \sum_X   Q_k^{\reg}(X) $ and ${Q'_k}^{\reg}  =\sum_X  {Q'_k}^{\reg}(X) $  with 
\be \label{lunar1} 
\begin{split} 
  \|  Q_k^{\reg} \|_{h, \Ga_n}     \leq   & C   h^6    g^2_k   \hs \hs  \| {Q'_k}^{\reg} \|_{h, \Ga_n}     \leq     C  h^6  (|p_k|+|v_k| )   g_k  \\
  \end{split} 
  \ee
\end{lem} 
\bigskip

\pr We have    $Q_k^{\reg}  = g_k^2   \hat  \cH(w_k) $, but  before extracting the local pieces we have $Q_k =  g_k^2 \cH(w_k)$.    We first consider the quartic contribution  $ g_k^2 \cH_4 (w_k)$
which from (\ref{wonderful} )   is
\be
  \cH_4( w_k  )
= 
  2n  \int  (\bpsi \psi)(x) n   \tr  (   w^2_k(x-y) )   
   (\bpsi \psi)(y)dx dy  + \textrm{ 2 more terms }   
\ee
We study this first term in detail.  The    $\hat  \cH _4 (w_k) $ is the remainder when we   replace each
$\psi(y) , \bpsi(y) $ by    $\psi (x), \bpsi(x)$. It is given by    
\be
\begin{split} 
&\hat \cH_4( w_k  )
=   2 n \int  (\bpsi \psi)(x)    \tr  (  w^2_k(x-y)) \\
& \B( \bpsi(x)  (\psi(y)- \psi(x))    + (\bpsi(y)- \bpsi(x))  \psi(x) +    (\bpsi(y)- \bpsi(x))  (\psi(y)- \psi(x))    \B) 
 dx dy + \dots\\
\end{split}  
\ee

We consider the first term here in detail,  call it $K $.   We localize the integral  writing  
$K = \sum_{\sq, \sq'}  K(\sq \times \sq')  $ where $\sq, \sq'$ are unit blocks and
\be
 K(\sq \times \sq')  
=  2n  \int_{\sq \times  \sq'}     \tr  (  w^2_k(x-y))     (\bpsi \psi)(x)    \bpsi(x)    (\psi(y)- \psi(x))   dx dy 
\ee
This has norm 
\be  \label{quonset1} 
\| K(\sq \times \sq')  \|_h
\leq  \one h^4     \sup_{\|f \|_{\cC} \leq 1} \B| \int_{\sq \times  \sq'}   \tr  (  w^2_k(x-y)) \
\B (   f(x,x,x,y) - f(x,x,x,x) \B)   dx dy \B| 
\ee
We again use the bound  (\ref{bdds}) 
$
|w_k( x-y) | \leq \one |x-y|^{-1} e^{-|x-y|} 
$.
If $x \in \sq, y\in \sq'$ and                     $d(\sq,\sq') \geq 1$ this is less than  $e^{- d(\sq,\sq') }$  and (\ref{quonset1}) is bounded by $\one h^4e^{- d(\sq,\sq')/2 }$.
On the other hand if $d(\sq, \sq') =0$  we  use
\be
   f(x,x,x,y) - f(x,x,x,x) 
   =   \int^1_0 (y-x) \cdot  \pa  f  \B(x,x,x, x + t(y-x) \B) dt
   \ee
This gives the estimate $|  f(x,x,x,y) - f(x,x,x,x) | \leq |x-y| \|f \|_{\cC} $.   The $ |x-y|^{-2}$ singularity in $ \tr  (  w^2_k(x-y)) $
is reduced to an integrable $ |x-y|^{-1}$ singularity.   So in this case  (\ref{quonset1}) is bounded by $\one h^4$.   Altogether
then we have 
\be
 \| K(\sq \times \sq')   \|_h \leq \one h^4e^{- d(\sq,\sq') }
\ee
The other terms in $\hat \cH_4( w_k  )$ are estimated similarly.

Now we consider the quadratic term which has the form $g_k^2 \hat  \cH_2(w_k)$.
First without the extraction we have from (\ref{wonderful}) 
\be
  \cH_2(w_k )   =   -4  \int     \bpsi(x)  w_k^3  (x-y)     \psi(y)   dx dy   + \dots 
\ee  
This modified to  $\hat   \cH_2(w_k )   $ by  replacing   $\psi(y)$   by      $ \psi (x)   + (y-x)\cdot  \pa \psi (x)$.
The difference is 
\be
\hat   \cH_2(w_k )   =   -4  \int     \bpsi(x)   w^3_k  (x-y)   \B( \psi(y) - \psi (x)   - (y-x)\cdot  \pa \psi (x)\B)    dx dy   + \dots 
\ee  
Call this first term $H$  and write
$H = \sum_{\sq, \sq'} H(\sq \times \sq')  $
where now 
\be
 H(\sq \times \sq') 
=   -4  \int_{\sq \times  \sq'} \bpsi (x) w^3_k  (x-y)  \B( \psi(y) - \psi (x)   - (y-x)\cdot  \pa \psi (x)\B)    dx dy  
\ee
This has  norm
\be
\begin{split}
& \| H(\sq \times \sq') \|_h \\
& \leq \one h^2 \sup_{ \|f\|_{\cC}  \leq 1}\B|\int_{\sq \times  \sq'}  w^3_k  (x-y)  
\B(  f(x,y)  -f(x,x) -  (y-x) \cdot  ( \pa_yf)(x,x)  \B) dx dy \B|  \\
\end{split} 
\ee
By the bound on $w_k$, if  $d(\sq,\sq') \geq 1$ this is bounded by $\one h^2 e^{-d(\sq,\sq')} $.   For $d(\sq,\sq') =0$ we use the identity
\be
 f(x,y)  -f(x,x) -  (y-x) \cdot  ( \pa_yf)(x,x)   = \int_0^1 (1-t)  (y-x) (y-x)(  \pa   \pa  f) \B(x,   x + t(y-x) \B)  dt
\ee
This is $\cO( |x-y|^2) \|f\|_{\cC} $.
 So the  $ |x-y|^{-3}$ singularity in $  w^3_k(x-y)$
is reduced to an integrable $ |x-y|^{-1}$ singularity, and  the norm is bounded by  $ \one h^2$.  Altogether then
and we have
\be
 \| H(\sq \times \sq')   \|_h \leq \one h^2e^{- d(\sq,\sq') }
 \ee
The  other term in  $\hat \cH_2( w_k  )$ is estimated similarly.  
 
 The same bound   holds for  $\hat \cH_6( w_k  )= \cH_6( w_k  )$  (now with $h^6$ )  since we have only one $w_k(x-y) $ which is integrable.

Altogether we have   
\be
Q_k^{\reg}  =  \sum_{\sq, \sq'} Q_k^{\reg} (\sq \times \sq' ) 
\ee
 with
\be 
  \| Q_k^{\reg}  (\sq \times \sq) \|_h    \leq   \one   g^2_k   h^6 e^{- d(\sq, \sq' )  }
  \ee
Now define   $Q_k^{\reg} (X)  =  Q_k^{\reg}  (\sq \times \sq)$ if $X= \sq \cup \sq'$ and zero otherwise.
Then we have
\be \label{core} 
\begin{split} 
\| Q_k^{\reg}  \|_{h, \Ga_n}    = &  \sum_{X\supset \sq  }  \| Q_k^{\reg} (X)  \|_h  \Ga_n(X)  = \sum_{ \sq'}  \|  Q_k^{\reg} (\sq \times \sq' )  \|_h  \Ga_n(\sq \cup \sq')  \\
& \leq   \one   g^2_k   h^6    \sum_{\sq'}   e^{- d(\sq, \sq' )  } \Ga_n(\sq \cup  \sq')  
\leq   C g^2_kh^6  \\
 \end{split}
\ee
The last line follows since the exponential decay $ e^{- d(\sq, \sq' )  } $ beats the polynomial growth of $ \Ga_n(\sq \cup  \sq')  $. 
 This completes the proof.
 \bigskip

\begin{lem}  \label{easy} 
$E^z_k    = \vep_k^z|\mathbb{T}_{N+M-k} | + 
 \sum_X  E^z_k(X) $   and    ${E'}^z _k   = { \vep'} _k^z| \mathbb{T}_{N+M-k} | +  \sum_X {E^z_k}'(X)  $ with 
\be \label{lunar2} 
  \|   E^z_k  \|_{h, \Ga_n}     \leq C   h^6 g^2_k  |z_k|    \hs  \hs  \|  {E^z_k}' \|_{h, \Ga_n}     \leq C h^6 g_k  (|p_k|+|v_k| )|z_k|   
  \ee
\end{lem} 
\bigskip

\pr  Recall  $E^z_k    =  g_k^2    \B(\cH(w_k + C^z_k) -  \cH(w_k + C_k)\B)$.  The term $ \vep_k^z|\mathbb{T}_{N+M-k} |$ is the contribution of the constant term coming from $\cH_0$.    For the
rest  we follow the analysis of the previous lemma,  but now   we use the estimate from  Appendix \ref{D} 
\be
| C^z_k(x-y) - C_k(x-y) |  \leq \one |z_k|  e^{-|x-y|/L  }
\ee

In the quartic term $\cH_4$ we have terms of the form 
\be 
 (w_k + C^z_k)^2 - (w_k + C_k)^2   =   2w_k(C^z_k-C_k)  +  ((C^z_k)^2 - C_k^2  ) 
\ee
The only short distance singularity is the integrable   $|x-y|^{-1} $ from the $w_k$.   Thus we
can estimate the term as in the previous lemma and 
get a bound  $C   g^2_k  |z_k|  h^4 $.

In the quadratic term   $\cH_2$ we have terms of the form   
\be 
 (w_k + C^z_k)^3 - (w_k + C_k)^3   =  3w_k^2(C^z_k-C_k)  + 3 w_k ((C^z_k)^2 - C_k^2  )  +  ((C^z_k)^3 - C_k^3  ) 
\ee
Here the $w_k^2(x-y)  = \cO( |x-y|^{-2} )  $ term is not integrable and is potentially worrisome. 
This occurs in a term like 
\be
\int   \bpsi(x)  w_k^2 (x-y) \B(C^z_k(x-y) -C_k(x-y) \B)  \psi(y)\  dx dy
\ee
 However we again replace $\psi(y)$
by $\psi(x)   + (\psi(y) - \psi (x) ) $.  The $\psi(x)$ term gives zero since the integrand is odd.      The $ \psi(y) - \psi(x)  $  term  has an estimate with an extra   $|x-y|$  and we are
again reduced to an integrable singularity and a bound  $ C   g^2_k  |z_k|  h^2$.  

  The terms in $\cH_6$ only have $C_k^z - C_k$  and an easy   $C   g^2_k  |z_k|  h^6$  estimate.
This completes the proof for $  E^z_k(X) $ and the proof for  $  {E^z_k}'(X) $ is the same. 
  \bigskip

\begin{lem}   \label{Elem} 
 $\de E_k(X, \psi,  \eta) = E_k( \psi + \eta ) - E_k(\psi) $
satisfies   
\be \label{ergo} 
 \| \de E_k \|_{\frac12 h, \frac 14 h , \Ga_n  }  \leq    \one  \| E_k\|_{h, \Ga_n }   
 \ee
 Furthermore if   $\frac14 h>   h(C) $
 \be 
 \| < \de E_k >_{C} \|_{ \frac12  h, \Ga_n}  \leq   \one  h(C) h^{-1}  \  \|E_k\|_{h, \Ga_n}  
\ee 
\end{lem} 
\bigskip

\pr   Let $
  E^+_k (X, u,  \psi, \eta )  \equiv   E_k (X,  \psi + u\eta ) 
$. 
Then we have the representation  for $r >  1$
\be  \label{xxx} 
\de E_k (X, \psi, \eta) = \frac{1}{2 \pi i} \int_{|u|  = r } \frac{du}{ u(u-1) }  E^+ _k (X, u,  \psi, \eta ) 
\ee
Indeed $  E^+_k (X, u,  \psi, \eta ) $ is analytic in $u $ (i.e. the kernel is an analytic Banach space valued function)  
and  for   $|u|  \leq r$ it  satisfies by lemma  \ref{stud2} 
\be
\|  E^+ _k (X, u) \|_{\frac12h,h'}  \leq   \| E_k(X)  \|_{\frac12 h + rh'} 
\ee
which is finite provided  $\frac12 h + rh' \leq   h$.   
This gives the  bound  
\be 
 \| \de E_k(X) \|_{\frac12 h  ,h '  }  \leq   \one r^{-1}  \| E_k(X) \|_{\frac 12  h +rh' }   
 \ee
For  the first result we take $h'    = \frac14 h $ and  $r$ slightly less than 2.    

 For  second result  integrate   $\int [\cdots] d\mu_{C} (\eta)$  and
 get 
 \be  \label{xxxx} 
\blan  \de E_k (X) \bran_C(\psi)      = \frac{1}{2 \pi i} \int_{|u|  = r } \frac{du}{ u(u-1) }    \blan E^+ _k (X, u )  \bran_C (\psi) 
\ee
We have from  lemma \ref{stud1}   for $|u| \leq  r$
\be
\|  \blan   E^+ _k (X, u )  \bran_C  \|_{\frac12 h}  \leq   \|   E^+ _k (X, u )  \|_{\frac12 h, r h(C) }  \leq \|E_k(X) \|_{\frac12 h + rh(C) }
\ee  
Then 
\be \|  \blan  \de E_k (X) \bran_C \|_{\frac12 h}   \leq \one r^{ - 1}   \|E_k(X) \|_{\frac12 h + rh(C) }
\ee
The result follows by taking  $r$ slightly less than $ \frac12h h(C)^{-1}$.

 % \newpage

%>>>>>>>>>>>>>>>>>>>>>>>>>>>>>>>>>>>>>>>>>>>>>>>>>>>>>>>>>>>>>>>>>>>>>>>>>>>>>>>
 
 \subsection{Cluster expansion} 
To complete the analysis of  the terms in the fluctuation integral (\ref{fluctuation}) we need a cluster expansion.   First we present it  in a more general context. 
We still work on the torus $\mathbb{T}_{N+M-k}$, but consider general localized functions $E(\psi) = \sum_X E(X, \psi) $.     The covariance $C$ is either $C_k$  or  $C_k^z$. 

The following   is a  version of the cluster expansion was   developed by  Gawedski and Kupiainen \cite{GaKu85a} for bosons, and    
extended to fermions by    Feldman, Magnen, Rivasseau, Seneor  \cite{FMRS86}. 
 Technical details related to the norm we are using  go back to   Brydges and Yau \cite{BrYa90} (although they have a rather different  treatment of fluctuation integrals).
 Earlier references can be found in the cited papers.

\bigskip 

\begin{thm}  \label{nahnah1}  Let   $h$ be sufficiently large depending on $L$,  and   $ \| E \|_{h,\Ga_4} \leq c_0    $ with $c_0 = \one$   sufficiently small. Then there are complex numbers  
$E^{\#} (X) $ such that 
\be
\int  \exp \B( \sum_X E(X, \psi ) \B) d \mu_{C} (\psi)   = \exp \B( \sum_X  E^\#(X) \B) 
 \ee
with the estimate
 \be 
|E^\#|_{\Ga}   \equiv \sum_{X \supset \sq} |  E^\#(X) |\Ga(X)   \leq  \one  \|E\| _{h, \Ga_4  }  
\ee
\end{thm}  
\bigskip

\pr  \textbf{part I:} 
First we make a Mayer expansion.  The    $\exp ( \sum_XE(X) ) $  is written as     $\prod_X\B(  ( e^{E(X) } -1)  +1\B) $.
The product is  expanded and intersecting terms    are grouped together.
We have
\be     \exp \B( \sum_X E(X)  \B) = \sum_{ \{ Y_{\beta}\} }  \prod_{\beta}  K(Y_{\beta} )       
\ee
where the sum is over disjoint paved sets $\{ Y_{\beta} \}$, and where we  have   defined  
\be \label{interest} 
K(Y) =  \sum_{ \{X_i\}  \to Y  }  \prod_i   \B( e^{ E(X_i) } - 1 ) \B)  
\ee
Here  the sum is over distinct   overlap connected paved sets  $ \{X_i\} $ whose union is $Y$.
\footnote{overlap connected means the graph on the  $ \{X_i\} $ consisting of all pairs $ \{ X_i, X_j \} $ such that 
$X_i \cap X_j  \neq \emptyset$ is connected.  The $X_i $ themselves may not be connected. } 
To estimate this we use 
\be
  \| e^{ E(X_i) } - 1 \|_{h}  \leq  2 \|E(X_i) \|_{h}   
\ee
and so 
\be
\| K(Y) \|_{h}   \leq \sum_{ \{X_i  \}     }  \prod_i  2 \| E(X_i)   \|_{h}    
\ee
We also have from (\ref{flower}) $\Ga_3 (Y) \leq  \prod_i  \Ga_3(X_i) $ and then 
\be
 \| K \|_{h, \Ga_3} =
\sum_{Y  \supset \sq} \| K(Y) \|_{h} \Ga_3(Y)   \leq   \sum_{  \{X_i\} : \cup_i X_i  \supset \sq  }  \prod_i  2 \| E(X_i)   \|_{h}   \Ga_3(X_i) 
\ee
We replace the sum over unordered paved sets  by a sum over ordered paved sets  and have 
\be
 \| K \|_{h, \Ga_3}  \leq    \sum_{n=1}^{\infty}  \frac{1}{n!} \sum_{ (X_1, \dots, X_n)     }  \prod_i  2 \| E(X_i)   \|_{h}   \Ga_3(X_i) 
\ee
with the restrictions that  $ (X_1, \dots, X_n)   $ are overlap connected  and that  $\cup_i X_i  \supset \sq  $.

But 
\be
1( \cup_i X_i \supset \sq ) \leq \sum_{j=1}^n   1(  X_j \supset \sq )
\ee
and when summed over $X_i$ this gives a contribution independent of $j$.  Thus we can assume $j=1$.  Then $\sum_{j=1}^n = n$  and we have
 \be
\| K \|_{h, \Ga_3}   \leq   \sum_{n=1}^{\infty}  \frac{1}{(n-1) !} \sum_{ (X_1, \dots, X_n)    }  1(  X_1 \supset \sq )  \prod_i  2 \| E(X_i)   \|_{h}        \Ga_3(X_i) 
\ee
still with the overlap connected condition.   But if $ (X_1, \dots, X_n)  $ are overlap connected there is also a tree graph connecting  these elements.  
  Thus we have
\be
1(  (X_1, \dots, X_n)  \textrm{ overlap connected } ) \leq  \sum_{\tau } 1(  \tau  \textrm{ connects } (X_1, \dots, X_n) ) 
\ee
We can regard the tree graph $\tau$ as rooted on $X_1$ and specified by  
 a mapping on $(1,2, \dots, n)$ satisfying  $\tau(j) < j $.  The  statement  that $\tau$ connects means $X_j \cap X_{\tau(j) } \neq \emptyset$
for $j =2, \dots,  n$. 
Now we have.  
 \be
\| K \|_{h, \Ga_3}   \leq   \sum_{n=1}^{\infty}  \frac{2^n}{(n-1) !} \sum_{\tau} \sum_{  (X_1, \dots, X_n) : X_j \cap X_{\tau(j) } \neq \emptyset  }  1(  X_1 \supset \sq )  \prod^n_{i=1}    \| E(X_i)   \|_{h}      \Ga_3(X_i) 
\ee

We estimate successively the sums over $X_n, X_{n-1} , \dots$.  In the first step we have
\be
  \sum_{  X_n \cap X_{\tau(n) } \neq \emptyset  }    \| E(X_n)   \|_{h}      \Ga_3(X_n)  \leq \sum_{\sq \in X_{ \tau(n)} }  \sum_{X_n \supset \sq  }    \| E(X_n)   \|_{h}      \Ga_3(X_n)
=    |X_{\tau(n)} | \|E \|_{h, \Ga_3 } 
\ee
In  the $j^{th} $ step   we will have  built up a factor $|X_j|^{d_j-1} $ where the incidence number  $d_j$ is  the number of elements in $   \tau^{-1} (j)  $.    We estimate this
by   $ |X_j|^{d_j-1} \leq (d_j-1)! e^{ |X_j| }  $ and  use    $ e^{ |X_j| } \Ga_3(X) = \Ga_4(X) $     to obtain
\be
\begin{split} 
  \sum_{  X_j \cap X_{\tau(j) } \neq \emptyset  }  |X_j|^{d_j-1}   \| E(X_j)   \|_{h}      \Ga_3(X_j)  \leq  &
  (d_j-1)!   \sum_{  X_j \cap X_{\tau(j) } \neq \emptyset  }   \| E(X_j)   \|_{h}    \Ga_4 (X_j) \\
 \leq &   (d_j-1)!     |X_{\tau(j)} | \|E \|_{h, \Ga_4} \\
 \end{split} 
\ee
The last step is just
\be
\begin{split} 
  \sum_{  X_1 \cap \sq \neq \emptyset  }  |X_1|^{d_1-1}   \| E(X_1)   \|_{h}      \Ga_3(X_1)  \leq &
  (d_1-1)!   \sum_{  X_1 \cap \sq \neq \emptyset  }   \| E(X_1)   \|_{h}  \Ga_4(X_1)  \\
  =  &   (d_1-1)!   \|E \|_{h, \Ga_4} \\
  \end{split} 
\ee
 Altogether then we have
  \be
\| K \|_{h, \Ga_3}   \leq   \sum_{n=1}^{\infty}  \frac{2^n \|E \|^n_{h, \Ga_4} }{(n-1) !} \sum_{\tau}  \prod_{j=1} ^{n-1}   (d_j-1)!    
\ee
By Cayley's theorem the number of trees on $n$ vertices  with incidence numbers $d_j$ is 
\be 
 | \tau| \leq  \frac{ (n-2)!}{  \prod_{j=1}^n  (d_j-1)!}
   \ee
From  this  the sum over $\tau$ is bounded by $(n-2)! 4^{n-1} $ (for details see \cite{BrYa90} or  \cite{Dim13}).   Thus
 \be  \label{oneone} 
\| K \|_{h, \Ga_3}   \leq   \sum_{n=1}^{\infty}  8^n \|E \|^n_{h, \Ga_4}  \leq \one   \|E \|_{h, \Ga_4}
\ee

 \bigskip

\noindent
\textbf{Part II}:  
Now we have
\be \label{gong} 
\int  \exp \B( \sum_X E(X, \psi ) \B) d \mu_C(\psi)  
= \sum_{ \{ Y_{\beta}\} } \left[  \int    \prod_{\beta}  K(Y_{\beta} )        d \mu_C(\psi)  \right] 
\ee
We are assuming that $C (x,y) =(\overline{   C_1(x,\cdot)},  C_2(\cdot, y) ) $ as in (\ref{cfactor}). 
Let  $\{U_{\al} \} $ be a partition of the torus  consisting of the $Y_{\beta} $ and all the unit blocks $\sq $ not in any $Y_{\beta} $.
Correspondingly we  introduce weakening parameters $0 \leq  s_{\al,\al'} \leq 1$  in each factor separately  and  we define
 for  $i = 1,2$
\be
 C_i(s,x,y )  =   \sum_{\al}  1_{U_{\al} }(x) C_i(x,y)  1_{U_{\al} }(y)   + \sum_{\al \neq \al'}    s_{\al,\al'}   1_{U_{\al} }(x)  C_i(x,y)  1_{U_{\al'}}(y)   
\ee
and
\be
 C (s, x,y) = ( \overline{ C_1(s, x,\cdot )} ,  C_2(s, \cdot, y) ) 
 \ee

If all the $ s_{\al,\al'}  =1 $  this is equal to $C$.  In the bracketed expression we express the value  $s_{\al,\al'}  =1 $
by expanding around $  s_{\al,\al'}  =0 $.  Then  (\ref{gong}) can be written 
 \be \label{sister} 
 \sum_{ \{ Y_{\beta}\}  } \sum_{\ga}  \int ds_{ \ga}  \frac{\pa }{ \pa s_{\ga } }  \left[  \int    \prod_{\beta}  K(Y_{\beta} )        d \mu_{ C(s ) }   \right] _{s_{\ga^c} =0}  
 \ee
 Here    $\ga$ be  a  subset of the set   of all  pairs  $\{ U_{\al} , U_{\al'} \}$,   that is a graph on vertices
$\{ U_{\al} \}$.  
Also   $\int ds_{\ga} = \prod_{{\al,\al'} \in \ga} \int_0^1ds_{\al,\al'} $ and $\pa/ \pa s_{\ga} = \prod_{{\al,\al'} \in \ga} \pa /\pa s_{\al,\al'} $.
  The sum includes $\ga = \emptyset$ which is the totally decoupled term $ \int    \prod_{\beta}  K(Y_{\beta} )        d \mu_{ C(0 ) } $.

  Note   that not all $\ga$ contribute in  (\ref{sister}).   Indeed  we can replace the covariance   $C(s)$ by  $ 1_Y C(s) 1_Y$ where $Y = \cup_{\beta} Y_{\beta} $.  Equivalently we can
  replace $C_1(s)$ by $1_Y C_1(s) $ and    $C_2(s)$ by $C_2(s)1_Y $.    Then $C(s)$ is independent of $s_{\al, \al'} $ if      $\{ U_{\al} , U_{\al'} \}= \{\sq , \sq' \}  $   or if     
   $\{ U_{\al} , U_{\al'} \}= \{ Y_{\beta}, \sq\} $ and there is no matching pair $ \{\sq,  Y_{\beta'} \}   $.   In  either case if $\ga$ is a graph with such a line then  $\pa/ \pa s_{\ga} $ gives zero  and $\ga$  does not contribute. 
  \bigskip

  The  graph   $\ga$     splits into connected components $\{ \ga_j \}$.     The covariance   $C(s)$
  only connects points in the same  component  since  lines $\{ U_{\al} , U_{\al'} \}$  connecting different components are not in $\ga$ and so  have $s_{\al, \al'} =0$. 
Thus we have the factorization  
\be
   \left[  \int    \prod_{\beta}  K(Y_{\beta} )        d \mu_{ C(s ) }   \right] _{s_{\ga^c} =0}   
    = \prod_j  \left[  \int    \prod_{Y_{\beta}   \in  \ga_j }  K(Y_{\beta} )        d \mu_{ C(s ) } \right]_{s_{\ga^c} =0}   
\ee
The sum over $\ga$ is regarded as a sum  over disjoint $\{\ga_j\} $
and 
we   can write (\ref{sister})
as
 \be
 \sum_{ \{ Y_{\beta}\}  }   \sum_{ \{ \ga_j  \} }   \prod_j   \int ds_{ \ga_j}  \frac{\pa }{ \pa s_{\ga_j  } }  \left[  \int    \prod_{Y_{\beta}   \in  \ga_j }  K(Y_{\beta} )        d \mu_{ C(s ) } \right]_{s_{\ga^c} =0}   
 \ee
In this expression   classify the terms in the sum over $ \{ Y_{\beta} \}  , \{ \ga_j  \} $ by the paved sets they generate and get 
 \be \label{gong2} 
\int  \exp \B( \sum_X E(X, \psi ) \B) d \mu_C(\psi)  
= \sum_{ \{ X_j \} }
\prod_{j}  K^{\#} (X_j  )  
\ee
Here the sum is over disjoint paved sets  $\{ X_j \}$
and for   a paved set $X$   
\be
 K^{\#} (X)  =   \sum_{ \{ Y_{\beta}\}, \ga \to X }  \int ds_{ \ga}  \frac{\pa }{ \pa s_{\ga} }  \left[  \int    \prod_{\beta}  K(Y_{\beta} )   
      d \mu_{C(s) }   \right] _{s_{\ga^c} =0}  
 \ee
 The sum is over  disjoint  $  \{ Y_{\beta}\}$ and additional disjoint  blocks $\sq$   whose union is $X$,   and  a graph $\ga$
 connecting these sets.
 In the following we make a slight change in notation explicitly recording the extra blocks $\{\sq_i\} $ which were implicit in $\ga$.
 Thus we write 
 \be
  \sum_{ \{ Y_{\beta}\}, \ga \to X } =  \sum_{ \{ Y_{\beta}\}, \{\sq_i\}  \to X }   \sum_{\ga}
 \ee
 where $X$ is the union of the indicated sets and $\ga$ connects them. 
 
  \bigskip

\noindent
\textbf{Part III} 
We want to estimate $K^{\#} (X)$ and we start with the bracketed integrals in the last equation.    We write
\be
  K(Y_{\beta},   \psi, \bpsi  ) 
=  \sum_{n_{\beta} = m_{\beta} }    \frac{1}{n_{\beta}!m_{\beta}!}  \int     K_{n_{\beta}m_{\beta}} (Y_{\beta}, \bx_{\beta},  \by_{\beta}     )
 \psi(\bx_{\beta})  \bpsi(  \by_{{\beta}}  )\ d  \bx d \by  
\ee 
where $ \bx_{\beta} =  (x_{{\beta},1},  \cdots ,  x_{{\beta},n_{\beta}} )$ and $\psi(\bx_{\beta})  =  \psi(x_{{\beta},1} ) , \cdots  \psi( x_{{\beta},n_{\beta}} ) $, etc. 
Inserting this and carrying out the integral we have
\be \label{29} 
\begin{split}
&  \int    \prod_{\beta}  K(Y_{\beta} )        d \mu_{ C(s ) }  
\\
&=  \   \sum_{n_{\beta} = m_{\beta} }   \prod_{\beta}  \frac{h^{n_{\beta} + m_{\beta} }}{n_{\beta} ! m_{\beta}!}  \int   \prod_{\beta}  K_{n_{\beta}m_{\beta}} (Y_{\beta}, \bx_{\beta},  \by_{\beta}     )
 \det \B(  \{ h^{-1} C(s, x_{\beta, i} , y_{\beta,j})  \}   \B)    d  \bx d \by  \\
\end{split} 
 \ee  
Here  $  \{ C(s, x_{\beta, i} , y_{\beta,j})  \}   $ are the entries of an $2N \times 2N$ matrix where $N = \sum_{\beta}  n_{\beta} =\sum_{\beta} m_{\beta}  $.  Since 
$\det ( h^{-1} C)  =
h^{-2N} \det C$   the factors of $h$ in the above expression
cancel out.

We have
\be \label{iggy}
 \{ C(s, x_{\beta, i} , y_{\beta,j})   =\B( \overline{  C_1(s,  x_{\beta, i} , \cdot)},      C(s,  \cdot  ,  y_{\beta,j}) \B)  
\ee
and then by Gram's inequality 
 \be  \label{31} 
 \begin{split}
|   \det \B(   \{ h^{-1} C(s, x_{\beta, i} , y_{\beta,j})  \B)  | & \leq  \prod_{\beta,i}h^{-\frac12}  \|  C_1(s, x_{\beta,i} , \cdot)\|_2 
 \prod_{ \beta,j} h^{-\frac12}   \|  C_2(s,  \cdot  ,y_{\beta,j})\|_2 \\
\end{split} 
\ee
From Appendix \ref{D} we have  $|C(x,y)| \leq \one e^{-|x-y|/L}$.    The same bound holds separately for $C_1,C_2$.
Hence    $\| C_i(s,x,\cdot) \|_2 \leq  \one L^2 $ and  we   chose $h$ large   so  $h^{-\frac12}  \| C_i(s, x, \cdot) \|_2 \leq  \one h^{-\frac12}   L^2 \leq 1 $. 
Thus the determinant is less than one. Similarly  for  spatial derivatives 
\be \label{32}
| \pa^{\al}   \det \B(   \{ h^{-1} C(s, x_{\beta, i} , y_{\beta,j}) \}  \B)  | \leq 1
\ee
Thus as in lemma \ref{concert} 
\be   \label{33} 
\B|  \int    \prod_{\beta}  K(Y_{\beta} )        d \mu_{ C(s ) }  \B| \leq 
\sum_{ \{ n_{\beta}, m_{\beta} \} }  \prod_{\beta}  \frac{h^{n_{\beta} + m_{\beta} }}{n_{\beta}!m_{\beta}!}      \prod_{\beta} \|\  K_{n_{\beta}m_{\beta}} (Y_{\beta}) \|_{\cC'}   
= \prod_{\beta }\|K(Y_{\beta})  \|_h 
 \ee  
 
 The  covariance $C(s)$  is actually  analytic  in  complex  $s_{\al,\al'}$ (it is the complex inner product in (\ref{iggy})).  If  we take   $| s_{\al,\al'}|  \leq h^{\frac14} 
 \exp( d(U_{\al} ,U_{\al'} )/2L ) $
 we have  $|  C_i(s,x,y)|  \leq  \one  h^{\frac14}   e^{-  |x-y| /2L}$  and therefore  for $h$ large   $  h ^{-\frac12} \| C_i(s,x,\cdot) \|_2 \leq  \one h^{-\frac14 }  L^2 \leq 1$.
The bounds (\ref{32})  still hold  in this complex domain.    
The integral $ \int    \prod_{\beta}  K(Y_{\beta} )  d \mu_{ C(s ) } $ is analytic in the same domain and satisfies the bound (\ref{33}).   
  By  Cauchy bounds for $|s_{\al, \al'}| \leq 1$
 \be
  \frac{\pa }{ \pa s_{\ga } }  \left[  \int    \prod_{\beta}  K(Y_{\beta} )     d \mu_{C(s) }   \right] _{s_{\ga^c} =0}  
\leq  \prod_{ \{\al,\al'\}  \in \ga} h^{-\frac14}  e^{ -  d(U_{\al},U_{\al'} )/2L }  \prod_{\beta }\|K(Y_{\beta})  \|_h 
\ee
where $ \{\al,\al'\}  \in \ga$ means $ \{ U_{\al},U_{\al'} \}  \in \ga$. 
Thus
 \be \label{twotwo} 
|  K^{\#} (X) |   \leq     \sum_{ \{ Y_{\beta}\},\{ \sq_i\} \to X } \sum_{\ga}   \prod_{\{ \al,\al'\} \in \ga} h^{-\frac14}  e^{ -   d(U_{\al},U_{\al'} )/2L }  \prod_{\beta }\|K(Y_{\beta})  \|_h
 \ee
 where $\{U_{\al} \} =  \{\{ Y_{\beta}\},\{ \sq_i\} \}$ and $\cup_{\al} U_{\al} =X$. 
 \bigskip

  \noindent
 \textbf{Part IV: } 
 Next we estimate the norm  of $K^{\#}$.   
 First note that every connected graph $\ga$ has a  tree subgraph $\tau$ spanning the same vertices.  So  $\sum_{\ga} \leq \sum_{\tau } \sum_{\ga>\tau} $.
 We split the factor $h^{-\frac14} e^{ -   d(U_{\al},U_{\al'} )/2L } $ into two factors $h^{-\frac18}   e^{ -   d(U_{\al},U_{\al'} )/4L } $ and in one factor estimate the
contribution of lines not in the tree $\tau$ by 1.
  This gives
\be
\begin{split} 
  |  K^{\#} (X)  |  \leq     &   \sum_{ \{ Y_{\beta}\},\{ \sq_i\} \to X}   \prod_{\beta }\|K(Y_{\beta})  \|_h  \sum_{\tau} 
   \prod_{\{\al,\al'\} \in \tau}  h^{-\frac18}  e^{ -   d(U_{\al},U_{\al'} )/4L }\\
& \left[ \sum_{\ga> \tau  }     \prod_{ \{\al,\al'\} \in \ga}  Ch^{-\frac18}  e^{ -   d(U_{\al},U_{\al'} )/4L } \right]    \\ 
 \end{split} 
\ee
We estimate the bracketed expression by enlarging the  sum over connected $\ga > \tau$  to a sum over all graphs $\ga$ connected or not.
We have
\be
\begin{split} 
  \sum_{\ga }     \prod_{\{\al,\al'\} \in \ga}  Ch^{-\frac18}  e^{ -   d(U_{\al},U_{\al'} )/4L }  
   \leq  &    \prod_{\{\al,\al'\} }  \B( 1 +C h^{-\frac18}  
 e^{ -   d(U_{\al},U_{\al'} )/4L}  \B)  \\
    \leq  &  \exp \left(Ch^{-\frac18} \sum_{\{\al,\al'\}    }  
 e^{ -   d(U_{\al},U_{\al'} )/4L }\right)  \\
  \leq & \exp  \B(C h^{-\frac18} \sum_{\al}| U_{\al} |\B)   \leq   \prod_{\al} e^{|U_{\al} |}\\
\end{split}     
\ee 
Here we used   
\be
  \sum_{\{\al,\al'\} } e^{- d(U_{\al},U_{\al'} )/4L}  
\leq   \sum_{\{\al,\al'\}  }  \sum_{\sq \in U_{\al'} }e^{- d(U_{\al},\sq )/4L} 
\leq  \sum_{\al}   \sum_{\sq}  e^{- d(U_{\al},\sq )/4L} 
\leq C \sum_{\al} |U_{\al} |
\ee
Now  we have 
\be
 |  K^{\#} (X)  |  \leq        \sum_{ \{ Y_{\beta}\},\{ \sq_i\} \to X}   \prod_{\beta }\|K(Y_{\beta})  \|_h \prod_{\al} e^{|U_{\al} |}  
  \sum_{\tau}  \prod_{\{\al,\al'\} \in \tau}  h^{-\frac18}  e^{ -   d(U_{\al},U_{\al'} )/4L } 
\ee

Next we pass to an estimate on 
 $ |  K^{\#}  |_{\Ga_{1}  }  = \sum_{X\supset   \sq}  |  K^{\#} (X) |\Ga_1(X) $.  
 Repeatedly using the inequality (\ref{flower}) we have 
 \be
   \Ga_1(X) \leq    \prod_{\beta} \Ga_1(Y_{\beta} ) \prod_i \Ga_1(\sq_i) 
   \prod_{ \{ \al,\al'\} \in \tau } \theta( d(U_{\al} , U_{\al'} )   ) 
\ee
The factor  $\prod_{\al} e^{|U_{\al} |}  $ changes the $\Ga_1$ here to $\Ga_2$. 
We can kill the factor $\Ga_2(\sq_i) $ if we borrow a factor $h^{-1/16} $ from the product over lines.
Since $\theta$ is polynomially bounded      $ \theta( d( U_{\al}, U_{\al'} ) ) \leq C\exp(  d(U_{\al}, U_{\al'})/4L  )$
for some constant $C$.  Also we identify   the sum over  $X \supset \sq$ and $\{ Y_{\beta}\},\{ \sq_i\} \to X$ as just a sum over  $\{ Y_{\beta}\},\{ \sq_i\} $  
restricted only  by the condition that one of them contain $\sq$, a fact which we temporarily leave out of the notation.
Thus we have 
\be  \label{sharp} 
\begin{split} 
 |  K^{\#}  |_{\Ga_1  }    \leq     &  \sum_{ \{ Y_{\beta}\},\{ \sq_i\} }       \sum_{\tau} 
   \prod_{ \{\al,\al'\} \in \tau}  h^{-1/16}  e^{ -   d(U_{\al},U_{\al'} )/8L }
 \prod_{\beta }\|K(Y_{\beta})  \|_h \Ga_2(Y_{\beta})   \\ 
 \end{split} 
\ee

The sum over  $\{ Y_{\beta}\},\{ \sq_i\} $ is a sum over $\{U_{\al} \}$. 
We classify the terms in the sum by the number of $Y_{\beta}$ and the number of $\sq_i$. . 
So now 
\be \label{people} 
\begin{split} 
& |  K^{\#}  |_{\Ga_{1}}       \leq       \sum_{\ell} \sum_{n+m = \ell}   \left[   \sum_{ \{ U_{\al} \} }     \sum_{\tau} 
   \prod_{ \{\al,\al'\} \in \tau}  h^{-1/16}  e^{ -   d(U_{\al},U_{\al'} )/8L }
  \prod_{\beta }\|K(Y_{\beta})  \|_h \Ga_2(Y_{\beta})    \right]  \\ 
 \end{split} 
\ee
where the sum over $ \{ U_{\al} \}$ is now restricted  to have   $n$ elements $Y_{\beta} $ and $m$  elements $\sq_i$. 

We work on the bracketed expression.     The sum over unordered   $ \{ U_{\al} \}$ with at least one element containing $\sq$  is written as a sum over ordered sets
  $( U_1, \dots,  U_{\ell})  $ with $U_1 \supset \sq$.
The   tree is identified as a map $\tau$ on $(1, \dots,  \ell)$ with $\tau(j) <j$.  Still with the   restriction on the numbers of $Y_{\beta}$ and $\sq_i$  we have 
\be
[ \cdots ] = \frac{1}{\ell !} \sum_{ ( U_1, \dots U_{\ell} )  } 
\sum_{\tau} 
 \prod_{j=2}^{\ell}    h^{-1/16}  e^{ -   d(U_j,U_{\tau(j)}  )/8L } \prod_{\beta }\|K(Y_{\beta})  \|_h \Ga_2(Y_{\beta})  
  \ee
 We successively do the sums over $U_{\ell},   U_{\ell -1}, \dots$.  In the first step if   $U_{\ell} $  is some $Y_{\beta} $ then the sum over $U_{\ell} $ is  
 \be
\begin{split} 
&{\sum}_{Y_{\beta}  } e^{ -  d(Y_{\beta} ,U_{\tau(\ell ) } ) /8L } \| K(Y_{\beta} ) \|_h \Ga_2 (Y_{\beta} )     \\
 \leq  &   \sum_{Y_{\beta} } \sum_{\sq \subset Y_{\beta}  }   e^{ -  d(\sq,U_{\tau(\ell ) } ) /8L } \|  K(Y_{\beta} ) \|_h \Ga_2 (Y_{\beta} )     \\
  \leq    & \sum_{ \sq}   e^{ -  d(\sq,U_{\tau(\ell ) } ) /8L } \|K \|_{h, \Ga_2 }
   \leq       C | U_{\tau(\ell )}|  \|K \|_{h, \Ga_2 }   \\
 \end{split} 
  \ee 
 Here in the second step we used $ \sum_{Y_{\beta} } \sum_{\sq \subset Y_{\beta}  } = \sum_{\sq}  \sum_{Y_{\beta} \supset \sq  }$. 
 If $U_{\ell} $ is some $\sq_i $  then just
 \be
\sum_{\sq_i  }  e^{ -  d(\sq_i  ,U_{\tau(\ell ) } ) /8L }    \leq   C  |U_{\tau(\ell)}|  
 \ee

The proof now proceeds as in part I.   In the $j^{th}$ step we again accumulate  a factor $|U_j|^{d_j-1} $.    If 
$U_j = Y_{\beta}$ then  this is $|Y_{\beta}|^{d_j-1} \leq  (d_j-1)! e^{|Y_{\beta}|}$ and $e^{|Y_{\beta}|}\Ga_2(Y_{\beta}) =\Ga_3(Y_{\beta})$.
If $U_j = \sq_i$ then this is  $|\sq_i|^{d_j-1}=1$ and can be ignored.  Except for these changes the $j^{th} $ step is just like the first step.
The last step uses the pin $U_1 \supset \sq$.

      We   find the bracketed expression in (\ref{people}) is bounded by $   \|K   \| ^n _{h, \Ga_3 }   (  h^{-1/16})^{\ell-1}   C ^{\ell}     $.  If  $\ell \geq 2$ then
$ (  h^{-1/16})^{\ell-1}  C ^{\ell}   \leq  (  h^{-1/32}C)^{\ell}  \leq (\frac12)^{\ell} $. Now the sum in (\ref{people}) for $\ell \geq 2$  is bounded by  a sum of      $ \|K   \| ^n _{h, \Ga_3 }  (\frac 12 ) ^{n+m} $  
with the restrictions $n+m \geq 2$ and $n \geq 1$.    This gives has a  bound    $\one    \|K   \| _{h, \Ga_3 } $.  The case $\ell =1$ is special.  In this case
$K^{\#}(X)  = \int K(X) d \mu_{C} $ and by lemma \ref{concert} if $h \geq h(C)$ then   $|K^{\#} |_{\Ga_1} \leq \|K \|_{h, \Ga_1}  \leq \|K \|_{h, \Ga_3} $. Altogether then  
 \be \label{bistro3}     
|  K^{\#}  |_{\Ga_{1} }  \leq \one   \|K  \|  _{h, \Ga_3 }  
\ee

 \bigskip

\noindent
\textbf{Part V} 
Finally we claim that 
\be
\sum_{ \{ X_{\al }\} }
\prod_{\al}  K^{\#} (X_{\al } )   =  \exp \B( \sum_Y E^{\#} (Y)  \B) 
\ee
where 
\be
E^{\#} (Y)  =
  \sum_{n=1}^{\infty}  \frac{1}{n!}  \sum_{(X_1, \dots X_n) : \cup_iX_i = Y}  \rho^T(X_1, \dots, X_n) \prod_i K^{\#} (X_i) 
\ee
Here $ \rho^T(X_1, \dots, X_n) $  is a certain function which. vanishes unless $(X_1, \dots, X_n)$ are overlap connected,  and is bounded by the number of 
tree graphs on $(1,2, \dots, n)$.     This is a standard combinatoric argument, see for example \cite{Dim13}. 
This is again estimated by a spanning tree argument as in Part I.   The result is an estimate   $|E^\#|_{ \Ga}    \leq \one | K^{\#} | _{ \Ga_1} $.
Combining this with    (\ref{oneone}) and (\ref{bistro3}) we have the desired result
\be
|E^\#|_{\Ga }    \leq \one | K^{\#} | _{\Ga_1}   \leq \one \| K \| _{h, \Ga_3}   \leq \one \|E \| _{h, \Ga_4 }
\ee
 This completes the proof
 \bigskip

\rem 
$E^\#$  is actually an  analytic function of  $E$  on the ball $\|E\|_{h,\Ga} \leq c_0$  in  the complex Banach space $\cG_{h, \Ga_4}$, i.e.  it is continuously differentiable.  
To see this it is sufficient to establish the analyticity  for each of the steps  $E \to K \to K^\# \to E^\#$.    We discuss the first step in more detail.   First define
$F(X) = e^{E(X) } -1$.  This is analytic since it is the composition of two analytic functions.   Then as in (\ref{interest}) 
\be \label{interest2} 
K(Y) =  \sum_{ \{X_i\}  \to Y  }  \prod_i F(X_i)    = \sum_{n=0} ^{\infty} \frac{1}{n!}  \sum_{(X_1, \dots X_n) \to Y }\prod^n_{i=1} F(X_i) 
\ee
and it suffices to show this is an analytic function of $F$. 

Define a   bounded  multi-linear function $K_n$  from $ \cG_{h, \Ga_4} \times \cdots \times \cG_{h, \Ga_4}$ to complex-valued localized functions
by 
\be 
\B[K_n(F_1, \dots F_n) \B](Y)  =  \sum_{(X_1, \dots X_n) \to Y } F_1(X_1) \cdots F_n(X_n) 
\ee 
and let $[K_n(F)](Y)   = [K_n(F, \dots,  F) ](Y) $ be the associated homogeneous polynomial.    Then we have
 \be \label{interest3} 
K(Y)  = \sum_{n=0} ^{\infty} \frac{1}{n!}  [ K_n(F) ](Y ) 
\ee
The  proof of the theorem shows  this is an absolutely convergent power series.  As such it  is an     analytic function.  The other 
steps are similar.     See  Appendix A in  \cite{PoTr87} for more on 
analytic functions on a complex Banach space. 
\bigskip

 We now quote  a variation of this  cluster expansion.   We suppose  that there are additional fermi fields present that are not being integrated out. 
 So our function $E(X,\psi, \eta) $ is assumed to has the form (\ref{uncle2}) 
\bigskip

\begin{thm} \label{nahnah2}   Let $h$ be sufficiently large depending on $L$,  and let $\|E \|_{h',h, \Ga_4} \leq c_0$ be sufficiently small.   Then  there are paved functions
$E^{\#} (X) $ such that 
\be
\int  \exp \B( \sum_X E(X, \psi, \eta ) \B) d \mu_{C} (\eta)   = \exp \B( \sum_X  E^\#(X, \psi ) \B) 
 \ee
 and
 \be 
 \| E^{\#}  \|_{h', \Ga}   \leq  \one  \|E\| _{h',h, \Ga_4 }    
\ee
\end{thm}  
\bigskip

 The proof follows the proof of theorem \ref{nahnah1}  with the fields $\psi$ as spectators.   The analyticity remarks hold in this case as well.   We omit the details. 
 
%>>>>>>>>>>>>>>>>>>>>>>>>>>>>>>>>>>>>>>>>>>>>>>>>>>>>>>>>>>>>>>>>>>>>>>>>>>>>>>>

 \subsection{More estimates} 
 \label{moreestimates}

With the cluster expansion as a tool, we are now in a position to complete our analysis of the expansion of the fluctuation integral  starting with the remainder term
in (\ref{central}).

\begin{lem}  \label{Flem}     Let  $h$ be  sufficiently large and  $g_k$ sufficiently small.
\begin{enumerate}
\item    
 $W_k(X, \psi, \eta)   =  \B( V_k+V_k'+ Q_k^{\reg}  + {Q'_k}^{\reg} \B) (X,  \psi + \eta)    + \de E_k(X, \psi, \eta) $ satisfies
 \be \label{firstfirst} 
\| W_k \|_{ \frac12 h, \frac14  h, \Ga_4    }  \leq   C h^4 g_k
\ee
\item For $t$ complex with $|t| \leq c_0 (Ch^4 g_k) ^{-1} $ there is $W_k^{\#}(t)$
such that    
  \be
  \blan e^{-tW_k }   \bran_{C^z_k}   =   \exp \B( - \sum_XW^{\#}_k(t,X)  \B) 
\ee
which is analytic in $t$ and satisfies 
 \be  \label{concert2} 
\| W^\#_k(t)  \|_{\frac12h , \Ga }  \leq   \one 
\ee
\item  
The remainder term  
 \be \label{otto3} 
E^\#_k(X)  \equiv  \frac{1}{2 \pi i}   \int_{ |t| = c_0 (Ch^4 g_k) ^{-1} } \frac{ dt}{t^3(t-1) }  W^{\#}_k(t,X)  
 \ee
 satisfies
 \be \label{ottobd} 
 \|   E_k^\# \|_{\frac12 h, \Ga}   \leq     C   h^{12}    g_k^3
 \ee
 \end{enumerate} 
 \end{lem} 
 \bigskip

 \pr \begin{enumerate}
 \item
 By   lemma \ref{stud2} and (\ref{lunar}),   
 $ V_k^+ (X,  \psi,  \eta)    \equiv V_k (X,  \psi + \eta) $ and $ {V'}^+  _k (X,  \psi,  \eta) $ 
 satisfy
 \be
 \begin{split} 
 \|  V^+ _k \|_{ h, h,\Ga_4  }   \leq    &  \|  V_k \|_{2h,\Ga_4}    \leq           C  h^4 g_k   \\
  \|  {V'}^+  _k  \|_{ h,h, \Ga_4    }   \leq    &  \|  V'_k  \|_{2h, \Ga_4}     \leq    Ch^4 (|p_k| + |v_k|)   \
  \leq      C C_E h^4 g_k^2 \leq Ch^4g_k    \\
  \end{split} 
\ee 
The bounds on the other terms in $W_k$ are even smaller.   However the bound 
 $ \| \de E_k \|_{\frac 12 h, \frac 14  h, \Ga_4   }   \leq   \one  \|E_k\|_{h, \Ga_4 } \leq \one C_Eg_k^3$ from lemma \ref{Elem}
 requires that we weaken the $h$ parameter. 
 Altogether we have  the announced (\ref{firstfirst}). 
 
 \item  For  $|t| \leq    c_0 (Ch^4 g_k) ^{-1}  $  we have
 $
\| t  W_k \|_{\frac12 ,\frac 14 h, \Ga_4  }  \leq   c_0
$ 
which is  small.   The result follows by the cluster expansion theorem \ref{nahnah2}.  

\item This   follows from     $\| W^\#_k(t)  \|_{ \frac 12 h, \Ga }  \leq   \one  $ and  $|t|^{-3}=c_0^{-3} (Ch^4 g_k )^3  $.

\end{enumerate} 
\bigskip

 Another   $\cO(g_k^3) $  term is   $ < V_k,  Q_k^{\reg} >^T _{C_k^z} $  which we now estimate: 
 
   \begin{lem} $< V_k,   Q_k^{\reg} >^T_{C_k^z}$ has a local expansion which satisfies 
   \be
   \| < V_k,   Q_k^{\reg} >_{C_k^z} ^T \|_{\frac12 h, \Ga}  \leq  C h^{10}  g_k^3
\ee
\end{lem}  
\bigskip

\pr 
We could compute the integral and give a direct bound.   But since we have the cluster expansion at our disposal
it is quicker to use it. 
Let 
\be
W(t,s ) = tV_k + s  Q_k^{\reg}  
\ee
We have the bounds   $\|V_k\|_{h, \Ga_4} \leq Ch^4g_k$ and $\| Q_k^{\reg} \|_{h, \Ga_4}  \leq Ch^6g^2_k$ (lemma  \ref{qlem}).  
So if we take   $|t| \leq  \frac 12  c_0 ( Ch^4g_k)^{- 1} $ and $|s| \leq  \frac 12  c_0 ( Ch^6g^2_k)^{- 1} $
then 
\be 
\|W^+ (t,s) \|_{\frac12h, \frac12h,  \Ga}  \leq  \|W(t,s) \|_{h, \Ga_4} \leq c_0  
\ee
  By the cluster expansion (theorem \ref{nahnah2})  there  is a $W^{ \#}(t,s, \psi ) = \sum_X W^{ \#}(t,s,X, \psi ) $ such that 
\be 
  \blan   e^{ W(t,s)  } \bran_C (\psi) 
  =    \int   (e^{  W^+ (t,s)  }  )(\psi , \eta) d\mu_{C}  (\eta)             = e^{  W^{ \#}(t,s,X, \psi )  } 
\ee
with  $\|  W^{ \#}(t,s) \|_{\frac12 h, \Ga}  \leq \one   $.
Taking derivatives at $s=t=0$ we find that 
\be 
 < V_k,   Q_k^{\reg} >_{C} ^T =     \frac{ \pa^2 W^{\# } } { \pa t \pa s} (0,0) 
\ee
   So   $< V_k,  Q_k^{\reg}>^T  = \sum_X E(X) $ with  $E(X) = ( \pa^2 W^{\# } /\pa t \pa s ) (0,0,X)$.  Since $W^\# (t,s, X ) $ is analytic in $(t,s)$   
we can write this as 
   \be
  E(X) =    \frac{1}{ ( 2 \pi i)^2}  \int_{ |t| =  \frac 12  c_0 ( Ch^4g_k)^{- 1} } \int_{|s| =  \frac 12  c_0 ( Ch^6g^2_k)^{- 1} } \frac{ dt ds}{ t^2 s^2}W^\# (t,s, X ) 
  \ee
This  has the bound    $\| E\|_{\frac 12h, \Ga}  \leq   C h^{10} g_k^3$ as claimed.
 This completes the proof.
 \bigskip

\rem  The  higher order terms in (\ref{central++})  can be handled in the same way and are even smaller.  For example  the fourth order terms are
   \be \label{fourth} 
   \begin{split} 
   \| < V_k, \de E_k  >^T\|_{\frac12h, \Ga }  \leq   &   Ch^4g_k  \|E_k\|_{h, \Ga_4}  \leq CC_E h^4 g_k^4  \\
  \| < V'_k, Q_k^{\reg} >^T\|_{\frac12h, \Ga}, \  \| < V_k,  {Q'_k}^{\reg} >^T\|_{\frac12h, \Ga}   \leq   &   C h^{10} (|p_k|+|v_k| )g^2_k   \leq  CC_E h^{10} g_k^4           \\
    \| < V'_k, V'_k >^T\|_{\frac12h, \Ga}  \leq   &   C h^8 (|p_k|+|v_k| )^2   \leq  CC^2_E h^8 g_k^4    
   \end{split} 
\ee

%>>>>>>>>>>>>>>>>>>>>>>>>>>>>>>>>>>>>>>>>>>>>>>>>>>>>>>>>>>>>>>>>>>>>>>>>>>>>>>>

\subsection{Completion of the proof} 

We collect the results and finish the proof of theorem \ref{one}.   
From (\ref{ottoA})     the partition function is   
$ \sZ =   \int    \exp (   \tilde S_{k+1} )    \        d \mu_{G_{k+1,L } }    $
where  
\be
 \tilde S_{k+1}  =    ( \vep_k +   \vep'_k -   \vep''_k )|\bbT_{N+M-k}  | -z_k    \bpsi  \slpa \psi + E_k  + W_k^{\#}
\ee
The $E_k$ is inherited from $S_k$, but 
in  lemma   \ref{local}  the relevant parts   are extracted  by $E_k = E_k^{\loc} +  \cR E_k $.  
The $ W_k^\#$ is the contribution of the fluctuation integral  $\Xi_k$ as defined in section    \ref{flc}, and further  developed in subsequent sections. 
Collecting terms from   (\ref{otto1}), (\ref{central})-(\ref{central++}),  (\ref{otto2}), (\ref{licorice2}), (\ref{licorice3}), (\ref{lunar2}),  (\ref{otto3}) we find
\be  \label{composite1} 
\begin{split} 
 \tilde S_{k+1}  (\psi)  &=  
 (  \vep_k   + \vep^*_k) | \bbT_{M+N -k} | \\
  &  + \B(g_k   +   \beta_k  g^2_k - 2\beta' _kg_kp_k - 4\beta'_k g_kv_k  +  g^*_k \B )  \int  (\bpsi \psi)^2    \\
 &  - \B ( z_k +  \theta_k  g_k^2  -\theta_k^p g_kp_k- \theta_k^v g_kv_k  +  z^* _k  \B )   \int  \bpsi  \slpa \psi  \\
&+  \B(p_k -  2\beta_k' g_kv_k+ p_k^*\B)\int (\bpsi  \ga_5 \psi)^2 + \B(v_k -  \beta'_k g_kp_k+  v_k^*\B) \sum_{\mu} \int (\bpsi  \ga_{\mu} \psi)^2     \\
&   +  g_k^2 \hat  \cH(w_k + C_k  )   +   g_kp_k \hat  \cH^{p}(w'_k + C_k  ) +  g_kv_k \hat  \cH^{v}(w'_k + C_k  )   +   \cR E_k   + E^*_k     \\
\end{split}
\ee
Here    
\be \label{ekstar} 
\begin{split} 
 E_k^* = &   E_k^\#+  E_k^z+ {E^z_k}' + < \de E_k >    + \textrm{ the terms (\ref{central++})} \\
 \vep_k^*  = &   \vep'_k -   \vep''_k  +   \vep^E_k +  g_k^2 \vep^Q_k +  g_kp_k\vep_k^{p,Q}  +   g_kv_k\vep_k^{v,Q}         +   \vep^z_k +   {  \vep^z_k}'    \\
 \end{split} 
 \ee

We now feed in the bounds we have established.  We use   (\ref{ottobd}),  (\ref{lunar2}),  the bound  $\|< \de E_k> \|_{\frac12  h,  \Ga} \leq CC_E h^{-1} g_k^3 $ from lemma \ref{Elem},  and (\ref{fourth}) 
 and find  
\be \label{goodbound} 
 \|  E_k^* \|_{\frac12 h , \Ga}  \leq    C\B(  h^{12}   + C_Z h^6   + C_Eh^{-1}+C_Eh^{10} g_k\B) g_k^3 
\leq    C_E h^{-\frac12}   g_k^3  
\ee
Here we take $h$ large enough so $Ch^{-\frac12} \leq 1$ and   $C_E$ sufficiently large  depending on $h, C_Z$. 
We also have from lemma \ref{local} 
\be
|g_k^*|, h^{-2}  | z_k^*|, | p_k^*|, | v_k^*|  \leq  C_E h^{-4} g_k^3 
\ee
These  are the claimed bounds.

 Now we scale replacing $\psi$ on $\bbT_{M+N-k}$  by $\psi_L$ with $\psi$ on $\bbT_{M+N-k-1}$.  Then   $d \mu_{G_{k+1,L } }$ scales to   
 $ d \mu_{G_{k+1 }}$,    
 and  $\tilde   S_{k+1} $ scales to   $S_{k+1} (\psi) = \tilde S_{k+1} ( \psi_L) $ and $\sZ = \int e^{S_{k+1} } d \mu_{G_{k+1}}$.  
 
We identify the new values  $\vep_{k+1},   z_{k+1}, g_{k+1}$ as given by (\ref{flow1}),(\ref{flow2}).
Once $g_{k+1}$ is identified we can replace  $ g_{k}^2  \hat \cH(w_k + C_k  ) $ by  $g_{k+1}^2  \hat \cH( w_k + C_k   )$ with
a small $\cO(g_k^3)$ error which can be absorbed  into $E_k^*$ and $\vep_k^*$. 
Then
\be
g^2_{k+1} \hat  \cH( w_k + C_k  , \psi_L  ) =g^2_{k+1} \hat \cH( (w_k + C_k)_{L^{-1} }  , \psi ) =g^2_{k+1}\hat \cH( w_{k+1}   , \psi   ) =  Q_{k+1}^{\reg}(\psi) 
\ee
Similarly we identify  $ {Q'_{k+1}}^{\reg} $.   
Reblocking and scaling changes $ \cR E_k  + E^*_k $ to  $  E_{k+1} =    \cL (  \cR E_k  +   E^*_k )$.   (We  estimated $E_{k+1} $ in section \ref{statement}.)

Altogether we have    the announced  
\be
\begin{split} 
 S_{k+1}(\psi)    =&   \int   \B( \vep_{k+1}  -  z_{k+1} \bpsi \slpa \psi      +   g_{k+1}   ( \bar \psi \psi    )^2 +   p_{k+1} (\bpsi  \ga_5 \psi)^2    + v_{k+1}  (\bpsi  \ga_{\mu} \psi)^2   \B)  \\
 & + Q_k^{\reg} + {Q'_{k+1}}^{\reg}  +  E_{k+ 1}    \\
 \end{split} 
\ee
This completes the proof. 

\subsection{Some derivatives} 

As a corollary to theorem \ref{one}  we   estimate some derivatives of $E^*( E_k, g_k, z_k, p_k, v_k) $ as defined in (\ref{ekstar}).   The theorem
gives estimates  in the domain  (\ref{region}). For our estimates we shrink the domain by a factor $\frac12$.   The new domain is 
\be \label{onehalf}
 |g_k|   <  \frac12 g_{\max} \hs
\| E_k \|_{h, \Ga_4}   \leq  \frac12 C_E g_k^3 \hs
|z_k| \leq   \frac12 C_Z g_k  \hs
  |p_k|, |v_k|,   \leq   \frac12 C_E g_k ^2  
\ee

\bigskip

\begin{cor} \label{cor}  
In the domain (\ref{onehalf})
\be
\begin{split} 
\|D_{E_k}  E_k^* \|_{\cL( \cG_{h, \Ga_4}, \cG_{ h/2, \Ga}) } \leq & \one h^{-\frac12}  \\
\| \pa E_k^* / \pa g_k \|_{h/2, \Ga},\  \| \pa E_k^* / \pa z_k \|_{h/2, \Ga}  \leq &  C_E g_k^2 \\
\| \pa E_k^* / \pa p_k \|_{h/2, \Ga},\  \| \pa E_k^* / \pa v_k \|_{h/2, \Ga}   \leq & 
CC_E h^8 g_k^2 \\
\end{split}
\ee
\end{cor} 
\bigskip

\pr  
 $ E_k^* $ is an analytic function of $E_k$   in  the full domain  $\| E_k \|_{h, \Ga_4}   \leq   C_E g_k ^3$.  Indeed in the sequence of maps  $E_k \to W_k \to W_k^\# \to E_k^\#$ in lemma  \ref{Flem},  the maps $ W_k \to W_k^\# $ is analytic by the remarks following theorem \ref{nahnah1}.  The other steps are linear and hence analytic.  So $E_k^\#$ is analytic  and the other contributions to $E_k^*$ are
 either linear or quadratic and hence analytic.  
 
   For     $\| E_k \|_{h, \Ga_4}   \leq  \frac12 C_E g_k^3$   
and $\| \dot{E_k}\|_{h, \Ga_4}  =  1$  and $|t| \leq    \frac12 C_E g_k^3$       we have   $\| E_k+ t \dot{E_k} \|_{h, \Ga_4}   \leq   C_E g_k^3$
and we are in the domain of analyticity.
 The derivative $D_{E_k}  E_k^*$ is a linear function on $\cG_{h, \Ga_4}$
 and we can write  
\be
 \blan ( D_{E_k} E_k^* ) ( E_k),   \dot{E_k} \bran  = \frac{d}{dt}  E_k^*( E_k + t \dot{E_k}) |_{t=0} 
 = \frac{1}{2 \pi i } \int_{|t| = \frac12 C_E g_k^3} \frac{dt}{t^2}     E_k^*(E_k + t \dot{E_k} )
 \ee   
 The theorem says   $\| E_k^*\|_{h/2, \Ga}  \leq C_Eh^{-\frac12} g_k^3$ which yields 
\be
\|  \blan ( D_{E_k} E_k^* ) ( E_k),   \dot{E_k} \bran \|_{h/2, \Ga}  \leq  ( \frac12 C_E g_k^3)^{-1}   C_Eh^{-\frac12}  g_k ^3 \leq 2h^{-\frac12} 
\ee
Since $\|D_{E_k}  E_k^* \|_{\cL( \cG_{h, \Ga_4}, \cG_{ h/2, \Ga}) } $ is the supremum of this over $\| \dot{E_k}\|_{h, \Ga_4}  =  1$ we have the stated bound. 

For the second bound   note that  $ E_k^* $ is analytic  in  $|g_k| \leq g_{\max} $. Indeed inspection of the sequence of maps   $W_k \to W_k^\# \to E_k^\# $     shows that each 
preserves analyticity and the other contributions to   $ E_k^* $ are analytic as well.   With complex $g_k$ we have the basic bound $\|E_k^*\|_{h/2, \Ga} \leq C_E h^{- \frac12}  |g_k|^3$ as in the theorem.

Now  for real $g_k < \frac12 g_{\max} $ consider complex $g'_k$ with $|g_k'-g_k| < g_k $.  In this domain $|g'_k| < 2 g_k < g_{\max} $.  So $E_k^*$ is 
analytic here with the bound   $\|E_k^*\|_{h/2, \Ga} \leq C_E h^{- \frac12}  |g'_k|^3   \leq 8 C_E h^{- \frac12}  g_k^3 $.  Then by a Cauchy inequality   
the derivative at $g_k$ satisfies
\be
\| \pa E_k^* / \pa g_k \|_{h/2, \Ga}  \leq  g_k^{-1}  ( 8C_E h^{- \frac12}  g_k^3 )  \leq   C_Eg_k^2
\ee    

Similarly for   the third bound  $E^*_k$    is analytic  in  $|z_k| \leq C_Z g_k  $.  Then in the smaller domain
  $|z_k| \leq \frac12  C_Z g_k $   by a Cauchy inequality 
 \be
\| \pa E_k^* / \pa z_k \|_{h/2, \Ga}  \leq  ( \frac12  C_Z g_k  )^{-1} ( C_E h^{- \frac12}  g_k ^3)   \leq   C_E   g_k^2
\ee    
   
   For the derivative in $p_k$   we  look at  $E_k^*$ more carefully.      It has the form  $ E_k^* =    E_k^\#+ {E^z_k}' $ plus  the terms (\ref{central++})  plus terms
   that do not depend on $p_k$.     The first  term  $E_k^\#$  is analytic  in  $|p_k| \leq   g_k $   (just as  for  $g_k$) and satisfies the bound (\ref{ottobd}) there.  Then by a Cauchy inequality 
   in the smaller domain $ |p_k| \leq   \frac12  g_k $  (and hence for $|p_k| \leq  \frac12 C_E g_k ^2 $)  
   \be
\| \pa E_k^\# / \pa p_k \|_{h/2, \Ga}  \leq  ( \frac12   g_k  ) ^{-1}   ( C   h^{12}    g_k^3 ) 
 \leq   C h^{12}   g_k^2 \leq CC_Eg_k^2
\ee 
The  other terms are either linear or quadratic in $p_k$, and it is better to evaluate derivatives directly.    The term    ${E^z_k}'  $ is proportional to $g_k p_k z_k$,  the derivative reduces this to $g_k z_k$, and as in lemma \ref{easy} it is bounded by $CC_Vh^6g_k^2 \leq CC_Eg_k^2$. 
 Fourth order terms in   (\ref{central++}) are estimated in (\ref{fourth}).   The derivative in $p_k$ reduces the  bounds by $C_Eg_k^2$ 
and they are all less than $CC_Eh^8g_k^2$.  
  Altogether the   bound is the announced  $\| \pa E_k^* / \pa p_k \|_{h/2, \Ga} \leq CC_Eh^8g^2_k$.   The bound on   $ \pa E_k^* / \pa v_k  $ is just the same.

%\newpage

\section{The  flow} \label{five}

We  study the flow of the  renormalization group equations (\ref{flow1}),(\ref{flow2}).
Our basic structural stability  result is the following.
  \begin{thm} \label{another}  Let $ L,h,C_v, C_E$ be chosen as in theorem \ref{one}  and then let   $g_f<  \frac12g_{\max} $ be sufficiently small.  Then the flow equations 
  \be  \label{stunning} 
\begin{split}
E_{k+1} = & \cL( \cR E_k + E_k^*) \\
g_{k+1}   =& g_k   +   \beta_k  g^2_k - 2\beta' _kg_kp_k - 4\beta'_k g_kv_k  +  g^*_k \\
z_{k+1}  =&  z_k +  \theta_k  g_k^2- \theta^p_k g_kp_k - \theta^v_k g_kv_k+  z^* _k  \\
p_{k+1}   = &p_k -    2 \beta_k' g_kv_k+ p_k^* \\
v_{k+1}   = & v_k-     \beta'_k g_kp_k+  v_k^* \\
\vep_{k+1} = & L^2 (\vep_k + \vep_k^*) \\
\end{split}
\ee
 have a unique solution   $\{E_k, g_k,  z_k, p_k, v_k, \vep_k  \}_{0  \leq k \leq N} $
with  boundary conditions
\be \label{BC} 
  g_N = g_f,      \  \vep_N =0   \ \ \textrm{ and } \ \   E_0 =0, \     z_0=0, \   p_0=0, \  v_0 =0 
 \ee
\end{thm}

The problem is an analysis of the flow around
the origin.  As noted the condition  $\beta_k>0$ means   the coupling constant grows which it what we want for an ultraviolet problem.
Our analysis follows the work of Bauerschmidt, Brydges, and Slade \cite{BBS15c}  who study the case   $\beta_k  <0$ 
for an infrared problem.   

The proof will occupy the remainder of this section, and includes bounds of the various quantities. 
 For most of the analysis we can ignore  the energy density $\vep_k$  since it does not affect the other variables.  The final condition $\vep_N =0 $
 on the energy density is quite arbitrary.    One might contemplate tuning the  field strength  so the final  $z_N$   takes some pre-assigned value,  but for now it 
 is simpler to work with the initial  condition $z_0 =0$. 
 
 \subsection{The quadratic flow} 
 
We start by analyzing the quadratic flow of $g_k,z_k$, dropping terms of order $\cO(g_k^3)$. 
Recall that from  lemma \ref{eleven} there are positive  constants $C_{\pm}, C $ so that $C_- \leq \beta_k \leq C_+ $ and $|\theta_k| \leq C$.
 (We have absorbed the factors $n-1$ into the constants.)

 \begin{lem} { \ }  \label{english} 
 \begin{enumerate} 
 \item    Define   $\bar g_k$  for $0 \leq k \leq N$  by   
$
 \bg_{k+1}  =    \bg_k  + \beta_{k}  \bg_k^2 
$
and   $ \bg_N = g_f>0 $.   For $g_f$ sufficiently small it is bounded by
\be \label{candy}
 \frac{ g_f} { 1  +    C_+ g_f(N-k) } \leq \bar g_k \leq   \frac{ g_f}{  1 + \frac12 C_- g_f(N-k)  }   
\ee
\item  Define    $\bar z_k$ for  $ 0 \leq k \leq N$  by  $\bar  z_{k+1}  =\bar  z_k + \theta_k    \bg_k^2  $
and  $z_0 = 0$.    Then   $|\bar z_k| \leq  C \bg_k$
\end{enumerate} 
\end{lem} 
\bigskip

\pr  
$ \bar g_{k+1}>0 $  determines $ \bar g_k> 0 $   as the unique positive root of  $\beta_k   \bar g_k^2 +   \bar g_k -  \bar g_{k+1}=0 $
which  is 
\be 
  \bar g_k   = \frac{1}{2 \beta_k } \B( -1 + \sqrt{ 1 + 4 \beta_k  \bar g_{k+1}  } \B) 
  \ee 
    So  starting with $\bar g_N = g_f$ we get $\bar g_{N-1} $,then $\bar g_{N-2}$, etc. and 
  $ \bar g_N >\bar g_{N-1}  >  \cdots > \bar g_0>0$.

For the bound  it is convenient to work with $\bar g_k^{-1} $ rather than $\bar g_k$.    We
have
\be
\begin{split} 
\bar g_k^{-1}   - \bar g_{k+1}^{-1}  =  &  \bar g_k^{-1}  - (  \bar g_k  +  \beta_k \bar g_k^2)^{-1} 
=   \frac{   \beta_k \bar g_k^2} { \bar g_k(\bar g_k + \beta_k \bar g_k^2)}   
=   \frac{   \beta_k } {1 + \beta_k \bar g_k }   \\
\end{split}
\ee
It follows that  
\be
 \frac{C_-}{ 1 + C_+ \bar g_k }   \leq  ( \bar g_k^{-1}   - \bar g_{k+1}^{-1} )  \leq    \frac{C_+}{ 1 + C_- \bar g_k }
\ee
We  can assume $C_+ \bar g_k  \leq C_+  g_f \leq 1$ and so
\be
   \frac12 C_- 
\leq  (\bar g_k^{-1}   - \bar g_{k+1}^{-1} ) \leq  C_+    \ee
Then since  
\be 
\bar g_k^{-1}  = g_f^{-1}   +  \sum_{j = k} ^{N-1}   ( \bar g_j^{-1}   - \bar g_{j+1}^{-1} ) 
\ee
we have 
\be 
   g_f^{-1}  +  \frac12 C_-(N-k)   \leq  \bar g_k^{-1}   \leq    g_f^{-1}  +  C_+(N-k)  
\ee
and so 
\be
 \frac{ g_f} { 1  +    C_+  g_f(N-k) } \leq \bar g_k  \leq   \frac{  g_f} { 1  +  \frac12 C_- g_f(N-k) } 
\ee

For the field strength  the solution is 
\be
\bar z_k =  \sum_{\ell=0}^{k-1} \theta_{\ell} \bg_{\ell}^2
\ee
We know  $|\theta_{\ell}|  \leq C$  and 
by  the following remark the sum over    $\bg_{\ell}^2$ is  bounded by $C \bg_k$.    Thus     $|\bar z_k| \leq C \bg_{k}$. 
 \bigskip 

\rem
We will use the following estimates similar to those of  Lemma 2.1 in \cite{BBS15c}.  
One can deduce directly from the equation $\bg_{k+1} = \bg_k + \beta_k \bg_k^2$  that for $0 \leq j \leq k \leq N $ 
\be \label{basicbound} 
\begin{split} 
\sum_{\ell = j}^k \beta_{\ell}   \bar g_{\ell}   \leq   & \  \one |\log \bg_j |   \\
\sum_{\ell = j}^k \beta_{\ell}   \bar g_{ \ell}^n | \log \bg_{\ell}|^m   \leq   &\  \one  \bg_{k+1} ^{n-1}   | \log \bg_{k+1} |^m       \hs n>1, m \geq 0   \\
\end{split} 
\ee
Or since $C_- \leq     \beta_{\ell} \leq C_+$ we have for $C = \one C_-^{-1} $
\be \label{basicbound2} 
\begin{split} 
\sum_{\ell = j}^k \   \bar g_{\ell}   \leq   & \ C |\log \bg_j |   \\
\sum_{\ell = j}^k  \    \bar g_{ \ell}^n | \log \bg_{\ell}|^m   \leq   &\  C \bg_{k+1} ^{n-1}   | \log \bg_{k+1} |^m       \hs n>1, m \geq 0   \\
\end{split} 
\ee

\subsection{Reduction of the problem} 
Now consider the general case. 
We make some changes in notation defining  
 \be
 \begin{split} 
& \si_k =   \cL( \cR E_k +  E_k^*)  \hs \hs 
\rho_k  =  (g_k^*, z_k^*,p_k^*,v_k^*)   \hs \\
\end{split}
\ee
and also the explicit part of the flow
 \be \label{beaver} 
 \begin{split} 
 \bar \phi( g_k, z_k,  p_k,v_k) 
 =&
  \B( g_k +   \beta_k  g^2_k - 2\beta' _kg_kp_k - 4\beta'_k g_kv_k, \    z_k +  \theta_k  g_k^2- \theta^p_k g_k p_k - \theta^v_k g_kv_k,  \\
  & \hs \hs
 p_k -    2 \beta_k' g_kv_k,\   v_k-     \beta'_k g_kp_k\B) \\
 \end{split} 
 \ee
Then  the RG flow is given by  
\be \label{singsong0}
\begin{split} 
&( E_{k+1},  g_{k+1},   z_{k+1},p_{k+1},  v_{k+1}  )  =    \Phi_k ( E_k, g_k,  z_k, p_k,v_k)  \\ 
 \equiv   &  \B( \si_k(E_k, g_k, z_k, p_k,v_k),  \      \bar  \phi_k( g_k, z_k,p_k,v_k )   +    \rho_k(E_k, g_k, z_k, p_k,v_k )  \B)\\
 \end{split} 
 \ee
  We look for solutions with the boundary conditions (\ref{BC}).     
We introduce the notation 
\be
 x_k = (E_k, g_k, z_k,p_k,v_k) 
 \ee
Then (\ref{singsong0})   is also written 
\be \label{singsong1} 
x_{ k +1}  = \Phi_k (x_k)  \equiv    \B( \si_k(x_k ), \bar \phi_k (x_k)   +     \rho_k(x_k) \B)   
\ee

 We introduce an approximate flow  with no feedback from the $E_k$ to the coupling constants.   
So we drop the  $\rho_k $   and study 
\be \label{singsong2} 
x_{k+1}  =  \bar \Phi_{k} (x_k )  \equiv \B(\si_k (x_k),  \bar \phi ( x_k)    \B) 
\ee
 with the boundary conditions  (\ref{BC}).  This has a unique solution  of the form    
\be
\bar x_k  =( \bar E_k ,  \bar   g_k, \bar z_k,  \bar p_k,  \bar v_k       ) 
\ee
with   $\bar p_k=0 ,  \bar v_k=0  $. The $( \bar   g_k, \bar z_k) $ is the solution  of  $( \bar   g_{k+1} , \bar z_{k+1} )   = \bar \phi( \bar   g_k, \bar z_k ,0,0)$
defined in the previous section.
The  $\bar E_k$ 
is the solution of
$ \bar  E_{k+1}   =  \si_k   (\bar  E_k,  \bar   g_k, \bar z_k,0,0) $  starting with $\bar E_0=0$.   (Note that this   involves   $ E^*_k(\bar E_k, \bg_k,  \bar z_k, 0,0)$).        By the remark after theorem \ref{one}, $\bar E_k$    is well-defined for all  $k$ and it satisfies
   \be    \label{entire} 
 \|\bar E_k \|_{h, \Ga_4}   \leq  \one  ( L^{-1}  +  L^2 h^{-\frac12} ) C_E \bar g_k^3 \leq   C_E \bar g_k^3
 \ee

 Next define   for $0 \leq t \leq 1$
  \be \label{etc} 
 x_{k+1} = \Phi^t_{k}(x_k  )  =     \B( \si_k(x_k  ), \bar \phi_k (x_k )   +   t   \rho_k(x_k ) \B)   
 \ee
 This interpolates between $\Phi^0_k(x_k)  = \bar \Phi_k(x_k) $ and $\Phi^1_k (x_k)= \Phi_k(x_k) $. 
 We are looking for a  family of sequences  $x_k(t) $,   each in the domain of theorem \ref{one},     satisfying this recursion with the  boundary condition (\ref{BC}).     At  $t=0$ we have such a sequence namely  $x_k(0) = \bar x_k$,

If $x_k =x_k(t)$ is a such a  solution of (\ref{etc}),  then differentiating 
 $ x_{k+1}(t) = \Phi^t_k(x_k(t)  )  \equiv  \Phi_k(t,x_k(t)  ) $ 
we have a solution to the ODE 
 \be  \label{approx1} 
  \dot x_{k+1}(t)  = \B(  D_x   \Phi^t_{k} (x_k(t) \B) \dot x_k (t)   + \B(0, \rho_k(x_k(t) \B)    \hs \hs x_k(0) = \bar x_k
 \ee
 Conversely a solution of (\ref{approx1})  gives a solution of (\ref{etc}).

As an aid to studying the last equation we study a related equation.   Fix sequences $x_k$ in the domain of theorem 1   and $r_k$ and consider 
the  linear  equation for $y_k=(  E_k,  g_k, z_k, p_k,v_k   )$
\be \label{approx2}
y_{k+1}  =     \B(  D_x   \Phi^t_{k} (x_k ) \B)   y_k    + r_k  
\ee
For  this equation null boundary conditions   
\be \label{bc0} 
g_N =  0  \ \   \textrm{ and }  \ \ E_0 = 0, z_0 = 0,  p_0=0, v_0=0
\ee
are appropriate as we will see.

Furthermore we define $\bar \Phi^0_k$ by setting $t=0$ and also dropping the terms $\sigma_k(x_k) $ so that
\be \label{barphizero}
\bar \Phi^0_k(x_k) = (0, \bar \phi(x_k) )=  \B(0, \bar \phi(g_k,z_k , p_k, v_k ) \B)
\ee  
We replace     (\ref{approx2}) by the simpler   low order equation  
\be \label{approx3}
y_{k+1}  =  \B(    D_x   \bar \Phi^0_{k } (\bar   x_k )  \B)  y_k    + r_k  
\ee
also with null  boundary conditions.  We will  consider solutions of all  these equations in reverse order.

In studying these it will be convenient to introduce the following norms on spaces of sequences $x=(x_0,x_1, \dots,  x_N)$.
\be
\begin{split}
\| x \|_{X_w }  
 =  & \sup_k  \{  \bar g_k^{-3} \|E_k \|_{h,\Ga_4} ,\  \bar g_k^{-2 } |\log \bar  g_k| ^{-1}|g_k|,  \bar g_k^{-2 } |\log \bar  g_k| ^{-1}|z_k|,   \bar g_k^{-2} |p_k|,  \bar g_k^{-2} |v_k|   \}  \\
\| x \|_{X_r}  
 =  & \sup_k  \{   \bar g_k^{-3} \|E_k \|_{h, \Ga_4} ,   \bar g_k^{-3 }|g_k|,  \bar g_k^{-3 } |z_k|,  \bar g_k^{-3} |p_k|, \bar g_k^{-3} |v_k|  \}   \\
\end{split}
\ee
   $X_w, X_r$  are the  Banach spaces for which these norms are finite. 
The norms are chosen to accommodate the next lemma.

\subsection{The first  linear equation}

We study the linear equation (\ref{approx3})  with the notation $L_k =   D_x   \bar \Phi^0_{k } (\bar   x_k )$

\begin{lem} \label{pixie} 
For  $r_k  \in X_r$ there exists 
a unique solution  $y_k =(  E_k, g_k, z_k , p_k, v_k)  $  of  
\be \label{yumyum} 
 y_{k+1}  =   L_k   y_k    + r_k   
 \ee 
 with null boundary conditions (\ref{bc0}) . 
 It   satisfies for some constant $C$
 \be \label{bubbles}
 \| y   \|_{X_w}  \leq   C  \|r\|_{X_r}  
 \ee
  The solution is a linear function of $r$ and  we write it as $y = S^0r$ so
 the estimate  says 
\be
\| S^0\|_{\cL (X_r,X_w) } \leq  C  
\ee 
 \end{lem} 
\bigskip

\pr   Let $r_k = (r_k^E, r_k^g, r_k^z,  r_k^p, r_k^v )$.  The operator $L_k$ is zero on the first component so we just have
 $E_{k+1}  = r^E_{k} $.   Since $ \|E_{k+1} \|_{h, \Ga_4}   = \| r^E_{k}  \|_{h, \Ga_4}   \leq  \bg_k^3 \| r \|_{X_r} $
we have for  the projection $( \pi_E y)_{k+1} = E_{k+1} $
\be
\|\pi_E y \|_{X_w}   = \sup _k   \bar g_{k+1}^{-3} \|E_{k+1} \|_{h, \Ga_4}    
  \leq \| r \|_{X_r} 
\ee

 In the remaining variables $(  g_k, z_k ,p_k, v_k ) $  we compute from  (\ref{beaver}): 
 \be  \label{amos1} 
  D_x   \bar \Phi^0_{k } (   x_k )  = \left( \begin{array}{cccc} 
    1+ 2 \beta_k  g_k -2 \beta_k' p_k  -4\beta'_k v_k & 0    &- 2 \beta_k' g_k & -4\beta_k' g_k \\
 2 \theta_k  g_k  & 1& - \theta^p_k g_k&   - \theta^v_k g_k\\
    -2 \beta_k'v_k & 0 &1&   -2\beta_k' g_k     \\
    -  \beta_k' p_k & 0 & - \beta_k' g_k&1\\
 \end{array} 
 \right) 
 \ee
We evaluate at 
 $\bg_k, \bar z_k$ and $ \bar v_k =0, \bar p_k=0$  and get  
\be       \label{amos2} 
 L_k =  D_x   \bar \Phi^0_{k } (\bar   x_k )  = \left( \begin{array}{cccc} 
    1+ 2 \beta_k \bar g_k& 0    & -2 \beta_k' \bg_k & -4 \beta_k' \bg_k \\
 2 \theta_k \bar g_k  & 1&-  \theta^p_k \bg_k&  -  \theta^v_k \bg_k\\
    0 & 0 &1&  -2 \beta_k' \bg_k     \\
     0 & 0 &-   \beta_k' \bg_k&1\\
 \end{array} 
 \right) 
 \ee
  \bigskip
 
 Our strategy is to first   solve (\ref{yumyum})  for $p_k,v_k$ with zero initial condition.  Then solve for $g_k$  with  final condition $g_N = g_f$.   Then solve for $z_k$ 
with zero initial condition.

 The equations for $p_k,v_k$ are 
 \be
 \begin{split} 
p_{ k+1}  = &  p_k  - 2 \beta_k' \bg_k v_k +r^p_k \\
v_{ k+1}  = &  v_k  - \beta_k' \bg_k v_k + r^v_k  \\
\end{split} 
\ee
The matrix of coefficients on the right side  (i.e. the lower right corner of (\ref{amos2}) ) has eigenvalues $1 \pm \sqrt{2} \beta_k' \bg_k$ and a matrix
of eigenvectors can be taken of the form $I + \cO( \bg_k)$.  Thus we can change  to new variables  $\tilde p_k,\tilde v_k$ and  new $\tilde  r_k^p, \tilde r_k^v  $ which satisfy 
 \be
 \begin{split} 
\tilde p_{ k+1}  = & (1 + \sqrt{2} \beta_k' \bg_k)  \tilde p_k  + \tilde  r_k^p \\
\tilde v_{ k+1}  = &    (1 -  \sqrt{2} \beta_k' \bg_k) \tilde v_k  + \tilde   r_k^v   \\
\end{split} 
\ee
It suffices to prove bounds in these variables. 

 The solution for $\tilde p_k$ with $\tilde p_0 = 0$ is 
 \be
 \tilde p_k =   \sum_{\ell=0} ^{k-1}\B(  \prod_{i = \ell+1} ^{k-1} (1 + \sqrt{2} \beta_i' \bg_i) \B) \tilde  r_{\ell} ^p
 \ee
 In the product we take $\beta'_i \leq \beta_i$ and then 
 by lemma \ref{gamma} to follow with $\gamma = \sqrt 2$ 
\be
  \prod_{i = \ell+1} ^{k-1} (1 + \ga \beta_i' \bg_i)  \leq \one   \left(  \frac{\bg_k} {\bg_{ \ell+1}}  \right)^{\ga} \leq \one   \left(  \frac{\bg_k} {\bg_{ \ell}}  \right)^{\ga} 
  \ee
  We also have $ |\tilde r_{\ell} ^p| \leq  \bg^3_{\ell} \|r\|_{X_r} $.  Then we can estimate by (\ref{basicbound2}) 
 \be \label{amos3} 
| \tilde p_k | \leq    \bg^{\ga} _k \sum_{\ell=0} ^{k-1}\bg^{3- \ga} _{\ell}  \  \|r\|_{X_r}  \leq  C      \bg^2_k  \ \|r\|_{X_r} 
 \ee
 The $\tilde v_k$ trivially satisfies the same bound and hence so do $p_k,v_k$.  Then the projection $ (\pi_p y)_k = p_k$ satisfies
 \be
   \| \pi_py \|_{X_w}   \leq   \sup_k \bg_{k} ^{- 2}    | p_k |  \leq  C   \|r\|_{X_r} 
  \ee
 \bigskip
 
 The equation for  $g_k$ is 
\be
 g_{k+1} =  ( 1 +2  \beta_k  \bg_k ) g_k +\tilde  r^g_ k
\ee
where
\be
\tilde r^g_k  = - 2 \beta'_k \bg_k p_k  - 4  \beta'_k \bg_k p_k + r^g_k
\ee
This is solved starting with $g_N = g_f$  and counting down by 
\be
 g_{N-j}  =  A_{N-j}  ( g_{N-j+1}  - r^g_{N-j}  )  \ \  \textrm{ where } \ \  A_k   = ( 1 + 2 \beta_k  \bar  g_k ) ^{-1}
\ee
The solution is
 \be
g_{N-j} = -    \sum_{\ell =1}^j \B[  \prod_{i=\ell}^{j } A_{N-i} \B] \tilde r^g_{N-\ell }
  \ee 
By lemma \ref{gamma} to follow 
\be
  \prod_{i= \ell}^j  A_{N-i}    \leq    \one    \left(\frac{\bar  g_{N-j} }{\bar g_{N - \ell + 1} }\right) ^2 
\ee   
We also  have     $|r^g_k | \leq  \bg_k^3  \|  r \|_{X_r}   $.  Also  since $|p_k|, |v_k| \leq C \bg_k^2 \|r\|_{X_r} $  both  $  \beta'_k \bg_k |p_k| $   
and  $\beta'_k \bg_k |v_k| $  are bounded by  $C \bg_k^3 \|r\|_{X_r}$.  Therefore
$|\tilde r^g_k| \leq C   \bg_k^3  \|  r \|_{X_r} $ and $|\tilde r^g_{N-\ell} | \leq C   \bg_{N-\ell} ^3  \|  r \|_{X_r} $.  Then again   by (\ref{basicbound2})
\be \label{tiktok2} 
| g_{N-j} |   \leq    C  \bg^2_{N-j } \sum_{\ell =1}^j   \bar g_{N- \ell }   
 \|  r \|_{X_r}    \leq   C \bg^2_{N-j } |\log \bg_{N-j} | \ \|  r \|_{X_r}  
\ee
So for  the projection  $(\pi_g y)_{N-j}  = g_{N-j} $ we have 
\be \label{tiktok3} 
\| \pi_g y\|_{X_w } \ \leq   \sup _j     \bar g_{N-j} ^{-2}| \log \bg_{N-j} |^{-1}  |g_{N-j}| \     \leq C  \|  r \|_{X_r} 
\ee

Finally  consider the $z_k$ equation which is
\be
z_{k+1} = z_k  +2 \theta_k \bar g_k  g_k +\tilde  r^z_k 
\ee
where
\be
\tilde  r^z_k = -  \theta^p_k \bg_k  p_k  -  \theta^v_k \bg_k v_k +   r^z_k 
\ee
With $z_0 =0$  this has the solution
\be
z_k  =  \sum_{\ell =0} ^{k-1} \B( 2\theta_{\ell} \   \bar g_{\ell}  \  g _{\ell}   + \tilde  r^z_{\ell}   \B) 
\ee
Taking account the bounds on $|p_k|, |v_k| $    and $|\theta^p_k|, |\theta^v_k| \leq C$ 
 we have    $  |\tilde r^z_{\ell}|  \leq C  \bar g_{ \ell}  ^3 \|r \|_{X_r}  $.   Also 
 $|g _{\ell}|  
 \leq   C   \bar g^{2}_{\ell}  |\log \bar g_{\ell} | \   \| r \|_{X_r} $ from  (\ref{tiktok2}) 
 and $|\theta_{\ell}| \leq C$. 
Hence again by (\ref{basicbound2}) 
\be
|z_k|   \leq C   \sum_{\ell =0} ^{k-1} \    \bg_{\ell} ^3 \ | \log \bg _{\ell}  |  \| r \|_{X_r} 
\leq  C \bg_{k} ^2 \ | \log \bg _{k} |  \| r \|_{X_r}
\ee
Then we have  for $( \pi_z y)_k  =z_k$
\be
\| \pi_zy  \|_{X_w}   =   \sup_j  \bar g^{-2}_{k}| \log \bar g_{k }|^{-1}   | z_k |  \leq  C  \| r \|_{X_r}
\ee
This completes the proof. 
\bigskip

\begin{lem}  \label{gamma} 
For $g_f$ sufficiently small and $\ga >0$
\be
    \frac12   \left(  \frac{\bg_k} {\bg_{ \ell}}  \right)^{\ga}  \leq \prod_{i=\ell}^{k-1}(1 + \ga \beta_i \bg_i)    \leq     \frac32   \left(  \frac{\bg_k} {\bg_{ \ell}}  \right)^{\ga} 
\ee
\end{lem} 
\bigskip

\pr 
Define $\al_i$ by 
\be
(1 + \ga \beta_i \bg_i)   = (1 + \beta_i \bg_i)^{\ga} ( 1 +  \al_i)  
\ee
Then $\al_i = \cO(\bg^2_i)$ and since  $\sum_i \bg_i^2$ is small we have
\be
\frac12 \leq \prod_{i=\ell}^{k-1} ( 1 +  \al_i)   \leq \frac32
\ee
On the other hand from   $( 1+ \beta_i \bg_i)  = \bg_{i+1} / \bg_i$
\be
\prod_{i=\ell}^{k-1}  (1 + \beta_i \bg_i)^{\ga}   = \prod_{i=\ell}^{k-1}\left(  \frac{\bg_{i+1} }{\bg_i} \right) ^\ga =   \left(  \frac{\bg_k} {\bg_{ \ell}}  \right)^{\ga} 
\ee

\subsection{The second linear equation}

For the next result   we introduce  a new domain. 
Let $\cB_C$ be the   ball of radius $C$ in $X_w$ and we consider the domain   $\bar x + \frac18\cB_{C_E} $.
If  $x \in \bar  x +\frac18 \cB_{C_E} $  then  $x - \bar x \in \frac18  \cB_{C_E}  $
and this says
\be  \label{two} 
\begin{split} 
\| E_k - \bar E_k  \|_{h, \Ga_4}   \leq  &\frac18 C_E \bar g_k^3 \\
|g_k-\bar  g_k|,|z_k-\bar  z_k|       \leq & \frac18  C_E  \bar g_k^2  |\log  \bar g_ k|  \\
|p_k- \bar p_k |,|v_k- \bar v_k| \leq     & \frac18 C_E   \bar g_k^2 \\
\end{split}
\ee
Always assuming $\bg_k \leq g_f <\frac12  g_{\max} $ are sufficiently small,  the bound on $g_k$ implies say 
\be  
\frac45 \bg_k \leq g_k \leq \frac54\bg_k
\ee  

The next result shows that elements of $  \bar x +\frac18  \cB_{C_E} $ satisfy the conditions of theorem \ref{one} and corollary \ref{cor}.  Hence 
 the mapping $ x_{k+1} =  \Phi_k( x_k )  $ is well-defined on  such sequences, as is   $\Phi^t_k$, and we have control over derivatives.

 \begin{lem}  \label{twenty}
 For $x \in   \bar x+ \frac18 \cB_{C_E} $ 
  \be  \label{twotwotwo} 
\begin{split} 
\| E_k  \|_{h, \Ga_4}   \leq  &\frac12 C_E g_k^3 \\
|z_k|       \leq & \frac12 C_Z g_k  \\
|p_k |,|v_k | \leq     &  \frac12  C_E  g_k^2 \\
\end{split}
\ee
\end{lem} 

\pr  
We have     $\|  \bar E_k  \|_{h, \Ga_4}   \leq   \one (L^{-1} + L^2h^{-\frac12} )  C_E \bar g_k^3 $ from (\ref{entire})  which is less than $\frac18   C_E \bar g_k^3 $
for $L,h$ large. 
Together with   (\ref{two}) this gives $ \|  E_k  \|_{h, \Ga_4}   \leq  \frac14 C_E \bar g_k^3 \leq  \frac12 C_E   g_k^3   $.
We have   $|\bar z_k|   \leq  C \bg_k $ from lemma \ref{english} which together with the bound on $|z_k-\bar z_k|$ gives  $  |z_k|       \leq  \frac12 C_Z \bg_k$. 
We have $\bar p_k =0$ so    $|p_k |  \leq   \frac18 C_E   \bar g_k^2  \leq \frac 12 C_E g_k^2$ .   The same bound holds for $v_k$.  This completes the proof.
\bigskip

Now for  $x \in \bar x + \frac18  \cB_{C_E}$ we  study the  equation    $y_{k+1}  =      D_x  \Phi^t_{k}(  x_k )   y_k    + r_k $  from   (\ref{approx2}).  We regard it as  a perturbation 
of the previous equation $  y_{k+1}  =   L_k y_k       + r_k $.  Since $ L_k = D_x \bar  \Phi^0_k  (\bar x_k )$ we    
can write
\be \label{new}
  y_{k+1}  =   L_k y_k  +     \cW_{k}(t,x_k)  y_k     + r_k 
\ee
where 
\be 
 \cW_{k}(t,x_k)   = D_x\Phi^t_{k+1}  (  x_k  )   - D_x\bar  \Phi^0_k  (\bar x_k )   
\ee
We first need estimates on $ \cW_{k}(t,x_k)  $.

 There  are five components  in  
\be
\cW_k= (\cW_E, \cW_V) =  (\cW_E, \cW_g, \cW_z, \cW_p, \cW_v) 
\ee
 and each is a function for the five variables   
 \be x_k =(E_k, V_k)  = (E_k,g_k, z_k,p_k,v_k)  \ee
 We have 
\be
 \cW_k    = \left( \begin{array}{cc} 
   \cW_{EE}  & \cW_{EV}  \\
   \cW_{VE}  & \cW_{VV}  \\
 \end{array} 
 \right) 
 \ee
 Here  the first index indicates the component  we are considering  and  the second index indicates which derivative we are considering
 in the contribution of $D^t_x \Phi_k, D_x  \bar \Phi^0_k$ to $\cW_k$.

\begin{lem}  \label{w} 
\label{www}   For    $x \in \bar x +  \frac18 \cB_{C_E} $
\be
 \begin{split} 
 \|\cW_{EE}  \|_{\cL(X_w, X_r) } \leq  &   \one (L^{-2} + L^2h^{-\frac12})  \\
    \|\cW_{VE}  \|_{\cL(X_w, X_r) } \leq  & \one h^{-2}      \\
     \|\cW_{EV}  \|_{\cL(X_w, X_r) } \leq  &   CC_E h^8   g_f | \log    g_f |   \\
  \|\cW_{VV}  \|_{\cL(X_w, X_r) } \leq  &  C C_E g_f | \log   g_f |^2     \\
\end{split} 
\ee
\end{lem} 
\bigskip

\pr
We  write  
\be   \label{tiny} 
\begin{split} 
 \cW_k(t,x_k)   = &    \cW^{(I)}_k(t,x_k)   +   \cW^{(II)}_k(t,x_k)    \\  
  \cW^{(I)}_k(t,x_k)  =&  D_x \Phi^t_{k}  (  x_k  )    -D_x \bar  \Phi^0_k  ( x_k ) \\
    = &\B( D_x \si_k(x_k),    tD_x\rho_k(x_k) \B) \\
  =& \B( D_x\si_k(x_k),    t D_x g^*_k(E_k)  ,      t D_xz^*_k(E_k), t D_xp_k^* (E_k), t D_xv^*_k(E_k)   \B)  \\
  \cW^{(II )}_k(t,x_k)    = &  D_x  \bar \Phi^0_k ( x_k  )   -   D_x \bar  \Phi^0_k   ( \bar    x_k )  \\
 \end{split} 
   \ee

We begin by studying $\cW_{VV}$ which has the form 
\be
 \cW_{VV}     = \left( \begin{array}{cccc} 
      \cW_{gg}  & \cW_{gz}   & \cW_{gp}  & \cW_{gv}  \\
     \cW_{zg}  &\cW_{zz}   &\cW_{zp}  & \cW_{zv}  \\
       \cW_{pg}  &\cW_{pz}   &\cW_{pp}  & \cW_{pv}  \\
    \cW_{vg}  & \cW_{vz}   & \cW_{vp}  & \cW_{vv}  \\
 \end{array} 
 \right) 
 \ee
There is only a  contribution from  $\cW^{(II)} $. 
 From (\ref{amos1}) and (\ref{amos2}) we have
 \be  \label{wonder} 
 \begin{split} 
\cW_{VV}  =&    D_x   \bar \Phi^0_{k } (   x_k ) - D_x   \bar \Phi^0_{k } ( \bar  x_k ) \\
= &\left( \begin{array}{cccc} 
    2 \beta_k  (g_k-\bg_k)  -2\beta_k'p_k-4 \beta'_kv_k) & 0    & - 2\beta_k'  (g_k-\bg_k)  &- 4\beta_k'  (g_k-\bg_k)  \\
 2 \theta_k   (g_k-\bg_k)   & 0&  \theta^p_k  (g_k-\bg_k) &   2 \theta^v_k  (g_k-\bg_k) \\
    -2 \beta_k'v_k & 0 &0&   -2\beta_k'  (g_k-\bg_k)    \\
    -  \beta_k' p_k & 0 & - \beta_k'  (g_k-\bg_k) &0\\
 \end{array} 
 \right) \\
 \end{split} 
 \ee
 
 Since we are assuming  $|g_k- \bg_k | \leq \frac18 C_E  \bg^2_k | \log \bg_k|$   and $|p_k|,|v_k|   \leq \frac18 C_E \bg_k^2$
 every entry  is bounded by  $  C  C_E \bg^2_k | \log \bg_k| $.  Taking account also that
 $\Phi_k$ is a map from a $k^{th}$ component to $(k+1)^{th}$ component we have for any entry   in $\cW_{VV} $
 \be \label{pingpong} 
\begin{split}
 \|    \cW _{ VV}  f \|_{ X_r  } 
 \leq   & \sup_k     \bar g_{k+1} ^{-3}| ( \cW_{ VV} f)_{k+1}  |\\
 \leq  & \sup_k   \bar g_{k+1} ^{-3}\B[ CC_E  \bg^2_k | \log \bg_k|\B]   | f_k |   \\
\leq  & \sup_k   \bar g_{k+1} ^{-3}\B[ CC_E  \bg^2_k | \log \bg_k|\B]    \bar g_k^2 |\log \bar  g_k| \|f\|_{X_w} \\
 \leq &  \sup_k  CC_E \bar g_k | \log \bar  g_k|^2  \|f\|_{X_w}  \\ 
 \leq  & CC_E    g_f | \log   g_f  | ^2  \|f\|_{X_w} \\
 \end{split} 
\ee
which   says  
 \be \label{pingpong2} 
  \|  \cW _{ VV} \|_{\cL(X_w,X_r) } \leq    CC_E    g_f | \log   g_f  | ^2 \ee

Now consider  $\cW_{EV}$.   Only  $\cW^{(I)} $ contributes  and
 \be
  \cW_{EV}  = D_V \si_k=     D_V \cL ( \cR E_k + E_k^*) =   \cL   D_V E_k^*  
 \ee 
  and
  \be
    D_V E_k^*    =   \B( \pa  E_k^*/ \pa g_k, \      \pa  E_k^* /\pa z_k,\ \pa  E_k^*/ \pa p_k, \      \pa  E_k^* /\pa v_k \B)   
  \ee
  These derivatives are estimated in corollary \ref{cor}.   We have 
 $  \| \pa E^*_k / \pa g_k \|_{\frac12h, \Ga}   \leq     C_E \bg^2_k  $ and
 $  \|\cL \ \pa E^*_k / \pa g_k \|_{h, \Ga_4}   \leq  \one L^2 C_E \bg^2_k  $. The  derivative in  $ z_k $ has   the same bound,  
 and derivatives with respect to $p_k,v_k$ have the same bound but with an extra constant $Ch^8$.
Then  
  \be \label{dd} 
  \begin{split} 
  \|  \cW_{EV}  \|_{\cL(X_w,X_r)}  \leq  & \sup_k    \bg_{k+1} ^{-3} \B[ C C_E h^8\bg_k^2\B]  \bg_k^2 |\log \bg_k| \\
  \leq & \sup_k   CC_E h^8 \bg_k  |\log \bg_k|  \leq   C C _E  h^8   g_f  |\log g_f|  \\
  \end{split} 
  \ee

Next  consider
  \be
  \cW_{VE}  = D_E \rho_ k=    t (D_Eg_k^*,  D_Ez_k^*, D_Ep_k^*, D_Ev_k^*)   
 \ee 
  Since $g^*_k(E_k) $ is a linear function of $E_k$  
 we have  $D_E g^*_k(E_k ) =  g^*_k(E_k ) $.  We have from lemma \ref{local} the estimate    $ | g^*_k(E_k)|  \leq \one  h^{-4} \|E_k\|_{h, \Ga_4}   $ and so 
 $ \| D_E g^*_k   \| \leq  \one  h^{-4} $ (norm in the dual space $\cG'_{h, \Ga_4}$). 
 \be
  \|   D_E g^*_k   \|_{\cL(X_w, X_r) } 
\leq    \sup_k     \bar g_{k+1} ^{-3}\B[\one h^{-4} \B]      \bar g_k^3 \leq  \one  h^{-4}    
   \ee 
 The same bound holds for  $ D_Ep_k^*, D_Ev_k^*$ and for  $  D_Ez_k^*$ it holds with $h^{-2}$ instead of $h^{-4}$ 
Thus we have   
\be
 \| \cW_{VE}  \|_{\cL(X_w, X_r) }  \leq  \one  h^{-2} 
  \ee

 Finally consider
 \be
   \cW_{EE}  = D_E (  \cL \cR E_k  +  \cL  E_k^* )  
   \ee
  From  (\ref{bottle1})    we have $\| \cL \cR E_k \|_{h, \Ga_4}  \leq  \cO(1)L^{-1} \| E_k\|_{h, \Ga_4}  $.       Since it is a linear function it is its own derivative and it follows that
  $\| D_E\cL \cR  \| =  \cO(1) L^{-1}$  (norm in $\cL(\cG_{h, \Ga_4} )$). 
 This yields  
    \be \label{yellow}
 \|D_E \cL \cR \|_{\cL(X_w, X_r) }  =  \sup_k     \bg_{k+1} ^{-3}  [  \cO(1) L^{-1} ]  \bg_{k} ^{3}       \leq \cO(1) L^{-1}
 \ee   
  For the other term   we have   
   $ \| D_E  E^*_k  \|_{\cL(\cG_{h, \Ga_4},  \cG_{h/2, \Ga}) }  \leq \cO(1) h^{-\frac12}  $
  from  corollary \ref{cor},  and $\| \cL \|_{\cL(\cG_{h/2, \Ga},\cG_{h, \Ga_4} ) } \leq \one L^2  $ as in (\ref{bottle2}). 
  Then     
  \be
  \| D_E\cL E^*_k \|_{\cL( \cG_{h, \Ga_4})}   =\| \cL D_E E^*_k \|_{\cL( \cG_{h, \Ga_4})}     \leq  \one L^2 h^{-\frac12} 
   \ee
   and as in  (\ref{yellow})  $ \| D_E\cL E^* \|_{\cL(X_w, X_r) }     \leq  \one L^2 h^{-\frac12} $.
Altogether 
 \be
\|   \cW_{EE}  \|_{\cL(X_w, X_r) }  \leq   \one (L^{-1}   +  L^2  h^{-\frac12} )     
\ee
 This completes the proof.
\bigskip 

\begin{lem} \label{pixie2} 
Let $g_f$ be sufficiently small and suppose   $ x \in \bar x +\frac18 \cB_{C_E}  $.    Then for $r \in X_r$ there exists 
a unique solution of 
\be \label{lumbar} 
y_{k+1}  =      D_x  \Phi_{k}( t, x_k )   y_k    + r_k 
\ee
 in $X_w$  with  null boundary conditions (\ref{bc0}).   There is a  constant $C$ such that the linear solution operator $y  = S(t,x) r $ satisfies
\be  \label{Sbound} 
\begin{split}
\| S(t,x) \|_{\cL(X^r,X^w)} \leq   & \ \ C \\
\end{split}
\ee
\end{lem} 
\bigskip

\pr We claim  (\ref{lumbar})  is equivalent to 
\be \label{verynew} 
  y =    S^0  \cW(t,x) y        +S^0 r 
\ee
This follows since  $y=S^0r$ solves  $y_{k+1}  =   L_k   y_k    + r_k$.   So if  $y$ satsifies   
(\ref{verynew})  then 
\be
\begin{split}
 y _{k+1} = &  ( S^0  \cW(t,x) y)_{k+1}    +  (S^0 r)_{k+1}  \\ 
= & \B(  L_k ( S^0 \cW(t,x) y)_{k}    +    \cW_k(t,x)  y_k   \B)  +   \B(  L_k   (S^0 r) _k    + r_k \B)   \\
= &   L_k \B( ( S^0  \cW(t,x) y)_{k}  + (S^0 r) _k\B)     +    \cW_k(t,x)  y_k   +     r_k     \\
= &   L_k y_k    +    \cW_k(t,x_k)  y_k         + r_k     \\
\end{split} 
\ee
which is (\ref{new}) and hence (\ref{lumbar}).  Conversely a solution of (\ref{lumbar}) gives a solution of (\ref{verynew}).  
 The solution to (\ref{verynew}) is 
 \be
 y = (1 - S^0   \cW(t,x) )^{-1} S^0 r \equiv S(t,x) r 
 \ee
 provided the inverse exists in $\cL(X_w)$.   We argue below that  $ \|  S^0 \cW \|_{\cL(X_w)}   < \frac12$ so the inverse  does exist
 with  $  \| ( I- S^0 \cW )^{-1} \|_{\cL(X_w)}   < 2$.  Then we have the required
 \be
\|  S \|_{\cL(X_r,X_w) }  \leq   \|  S^0 \|_{\cL(X_r,X_w)}  \|   (1 - S^0  \cW(t,x) )^{-1} \|_{\cL(X_w)} \leq C
\ee

 For the estimate we have
   \be
 S^0 \cW   = \left(  \begin{array} {cc}  1 & 0 \\  0 &  S^0  \\ \end{array} \right)
  \left( \begin{array}{cc}   
   \cW_{EE}   &  \cW_{EV}           \\
  \cW_{VE}  & \cW_{VV}    \\
 \end{array} 
 \right) =
 \left( \begin{array}{cc}   
   \cW_{EE}  & \cW_{EV}            \\
 S^0  \cW_{VE}   &S^0 \cW_{VV}    \\
 \end{array} 
 \right) 
\ee
We argue that  each entry is small by  lemma  \ref{pixie} and lemma \ref{w}.
Start with 
 \be
 \| \cW_{EE} \|_{\cL(X_w)} =   \| \cW_{EE} \|_{\cL(X_w,X_r)} \leq \one( L^{-1} + L^2h^{-\frac12}  )  < \frac12 
 \ee
 which holds for first   $L$  and  then $h$  sufficiently large.
Similarly  for $g_f$ sufficiently small  $\| \cW_{EV} \|_{\cL(X_w)}  \leq C C_Eh^8 g_f |\log g_f|  < \frac12$.    
 Next  for  $h$ sufficiently large
  \be 
  \|  S^0 \cW_{VE} \|_{\cL(X_w)}     \leq   \| \cW_{VE} \|_{\cL(X_w,X_r)}  \|S^0\|_{\cL(X_r,X_w) }  \leq Ch^{-2}   < \frac12 
  \ee
  Finally for $g_f$ sufficiently small
  $\|  S^0 \cW_{VV} \|_{\cL(X_w)}   \leq  CC_Eg_f| \log g_f|^2 < \frac12$.   This completes the proof.

 % \newpage

    \subsection{Proof of theorem \ref{another} } 
  
  We  study $x_k = (E_k, g_k,  z_k, p_k, v_k,)$ for $x \in \bar x + \frac18  \cB_{C_E} $.    For any such  sequence $x$  the sequence 
   $y =  S(t,x)  \rho(x) $ is defined to  solve  the equation 
   \be \label{boo} 
 \B( S(t,x)  \rho(x)\B)_{k+1}    =   D_x   \Phi_{k} (t,x_k   ) \B(S(t,x  )  \rho(x) \B)_k    + (0, \rho_k(x))
\ee
with  null boundary conditions (\ref{bc0}). 
  Suppose we can find  a solution $x_k = x_k(t) $  in $\bar x +  \frac18  \cB_{C_E}$ of the system of  ordinary differential equations  
  \be \label{hoo} 
  \dot  x_k  = F_k(t,x) \equiv  (S(t,x)  \rho(x) )_k 
  \ee
  for  $0 \leq t  \leq 1$,  with the initial condition  $x_k(0) =\bar x_k$. 
 The   null  boundary conditions for  $S(t,x)  \rho(x) $ imply  null boundary condition for  $\dot x_k(t)  $.   (That is $\dot g_N(t) =0, \dot E_0(t) =0, \dots $.)  This  means that  
   $x_k(t)$  has  constant boundary conditions, and   since $x_k(0) = \bar x_k $ satisfies the boundary conditions (\ref{BC}), the same is true for  $x(t)$. 
  (That is $g_N(t) = g_N(0) = \bg_N = g_f$, and   $E_0(t) = E_0(0) = \bar E_0   =0$, etc.) 
    
 Now  taking the sequence  $x_k = x_k(t)$ in   (\ref{boo}) 
  we have 
  \be
\dot  x_{k+1} (t) =    D_x   \Phi_{k} \B(t,x_k(t)  \B) \dot x_k(t)     + (0, \rho_k(x_k(t) ) )
\ee
By uniqueness of such solutions   $x_{k+1}(t)  = \Phi_k ^t( x(t))$.   In  particular  $x_k \equiv x_k(1) $, still in  $\bar x +  \frac18  \cB_{C_E}$,  satisfies (\ref{singsong1})   $x_{k+1}  = \Phi_k ( x_k)$
and  hence (\ref{stunning}) with the boundary conditions (\ref{BC}). 

Thus the problem is reduced to  the existence of solutions to (\ref{hoo}) which  is an ordinary differential equation in the  Banach space  $X_w$.   We need a solution $x(t)$ 
 in the domain $  \bar x + \frac18 \cB_{C_E} $with  $x(0) =\bar x$  and defined for at least  $0 \leq t \leq 1$.    By a fundamental theorem of ODE  there is a unique solution provided      $ \| F(t, x)  \|_{X_w}  < \frac18C_E  $  for   $x \in \bar x +  \frac18  \cB$ and $F$   satisfies Lipschitz condition on this domain. 
(See for example \cite{AbMa78}). 

Elements of   $  \bar x + \frac18 \cB_{C_E} $ satisfy the hypotheses of theorem \ref{one}  by lemma \ref{twenty} and so we can conclude  
that $|g_k^*|, |z_k^*|,|p_k^*|,|v_k^*|$ are all bounded by  $C_Eh^{-2} g_k^3 \leq   2 C_Eh^{-2} \bg_k^3 $. 
Therefore
\be
 \| \rho( x)  \|_{X_r}     =   \sup_k  \{   \bar g_k^{-3 }|g^* _k|,   \bar g_k^{-3 } |z^*_k|, \bar g_k^{-3} |v^* _k|,  \bar g_k^{-3} |p^*_k|    \}  \leq 2C_E h^{-2} 
\ee
Then also by the bound on $S$ in lemma \ref{pixie2}  we have for $h$ sufficiently large   the required
\be
  \| F(t,x ) \|_{X_w}  \leq  \| S(t,x) \|_{\cL( X_r,X_w) }  \| \rho( x)  \|_{X_r}  \leq   C C_E h^{-2}   <  \frac18 C_E
\ee

For the Lipschitz continuity recall that  in the proof of corollary \ref{cor}  we observed that $E_k^*$ is analytic in $x_k$.   Keeping this in
mind and inspecting the various terms in $\cW_k$ in the proof of lemma \ref{w} we see that $\cW_k$  is analytic in $x_k$.  Hence the same is
true for $ S(t,x) =  (1 - S^0   \cW(t,x) )^{-1} S^0 $.   Since also $\rho_k(x_k)$ is analytic we have that  $F_k(t,x_k)$ is  analytic and  $F(t,x_k)$ is analytic since every entry is analytic.    The analyticity implies  Lipschitz continuity  in $x_k$ and it is uniform  in $t$ which suffices for the existence theorem.   
Thus theorem \ref{another} is established for all but the vacuum energy $\vep_k$. 
\bigskip

Finally  consider $\vep_k$ which satisfies $\vep_{k+1}  = L^2 ( \vep_k + \vep_k^*)$.  The solution with initial condition $\vep_0$
is then 
\be
\vep_k = L^{2k} \vep_0 + \sum_{j=0}^{k-1} L^{2(k-j) } \vep^*_j
\ee
To get $\vep_N =0$ we choose 
\be
\vep_0 =   - \sum_{j=0}^{N-1} L^{-2j} \vep^*_j
\ee
This completes the proof.

   \section{Ultraviolet stability bound} 
   
   We specialize now to unit volume  $M=0$, so we are studying the partition function $\sZ $ as   on  $\bbT_0 = \bbR^2/ \bbZ^2  $ with momenta cutoff at about $|p| = L^N$  as defined in (\ref{fine}).  This 
  is  scaled up to an integral on the large torus   $\bbT_N =\bbR^2/ L^N \bbZ^2$ with  unit  momenta cutoff at about $|p| =1$.  Successive renormalization group transformations change it to an integral over smaller tori  $\bbT_{N-k} $ still  with unit momentum cutoff.    We choose the coupling constants as in theorem  \ref{another} so we can   continue all the way to $k=N$.  Then we are again on a  unit torus $\bbT_0 = \bbR^2/ \bbZ^2 \equiv \sq$, but now  with unit momentum cutoff. 
   The partition function is given by
   \be
   \sZ    =  \int e^{S_f(\psi) }  d \mu_{G_f} (\psi)  
  \ee
   Here $G_f  \equiv  G_N$  is independent of $N$ and has the kernel 
   \be  \label{kumquat}  
   G_f(x-y) = {\sum}'_{ p \in \bbT_0^* } e^{ip(x-y) } \frac{-i\slp}{p^2}  e^{ -|p|^2 } = \sum_{ p \in 2 \pi \bbZ^2, p \neq 0 } e^{ip(x-y) } \frac{-i\slp}{p^2}  e^{ -|p|^2 } 
   \ee
  Since     $g_N = g_f$ and $\vep_N =0$  the expression for   $S_f \equiv S_N $ from
    (\ref{h1})  is 
   \be
     S_f  =    \int_{\sq}   \B(     - z_N  \bpsi \slpa \psi      +  g_f  (   \bpsi \psi    )^2  
 +  p_N  ( \bpsi  \ga_5 \psi )^2 +  v_N  ( \bpsi  \ga_{\mu} \psi )^2   \B) +   Q _N^{\reg} +  {Q_N'}^{\reg}     +  E_N  
\ee
     
 \begin{thm} (Stability bound) Under the hypotheses of theorem \ref{one} with parameters chosen as in  theorem \ref{another}  we have uniformly in $N$ 
 \be
 \| S_f \|_h  \leq    CC_V g_f   
 \ee
 and  
     \be
\frac12    \leq   \sZ  \leq   \frac32 
\ee
\end{thm} 
\bigskip
 
 \pr   Since our solution has   $x \in \bar x + \frac18 \cB_{C_E} $ the bounds (\ref{twotwotwo}) hold for all $k$ and so for $k=N$.  
 Then as in   lemma \ref{nugget} with $|z_N| \leq C_Zg_f$ 
 \be
\|  z_N  
\int \bpsi \slpa \psi   \|_h \leq  CC_Vg_f  \hs    \|  g_f \int  (    \bpsi \psi    )^2  \|_h \leq Cg_f
 \ee
 All the other terms in $S_N$  are higher order and smaller.  See in particular  (\ref{lunar}), (\ref{lunar1}), and  (\ref{twotwotwo})
 and use  $\| F\|_h < \| F\|_{h, \Ga_4} $ 
The bound on $S_f$ follows.

 For the partition function  note that the Fourier coefficients 
   $\tilde G_f(p) $ are rapidly decreasing and have no small divisors.   Hence $h(G_f)$ defined as in (\ref{sequoia})  is bounded by $\one$
 and we can choose  $h$ so   $ h(G_f )   \leq  h$.
Then $  \| S_f \|_{h(G_f)  } \leq   \| S_f \|_h  \leq  CC_V g_f  $.
  Now we write   
\be
\int   e^{-S_f }  d \mu_{G_f}   = 1  +   \int  (  e^{-S_f} -1  )  d \mu_{G_f} 
\ee
Estimate the last  integral as in lemma \ref{concert} and get for $g_f$ sufficiently small
 \be
 \begin{split} 
| \int   (  e^{-S_f} -1  )  d \mu_{G_f}  |
\leq &   \| e^{-S_f }-1 \|_{ h(G_f ) } 
\leq    e^{  \| S_f  \|_{ h(G_f )  } }  -1  
\leq  e^{  CC_V g_f     } -1 < \frac12 \\
\end{split} 
\ee
This suffices and completes the proof.  
\bigskip

  \begin{appendix} 
  
  \section{Fourier series conventions}
 
 \label{A} 
 We consider functions on the $d$-dimensional  torus $\bbT_j \equiv  \bbR^d/L^j \bbZ^d$.      If  $p \in \bbT_j^* \equiv  2\pi  L^{-j}    \bbZ^d$
then $e^{ipx} $ is invariant under $x_{\mu} \to x_{\mu} + L^j$ and so defines a function on the torus.    We have
$\int  e^{-ipx}  e^{iqx} dx = L^{dj} \de_{p,q} $ so $e_p(x)  = L^{-\frac12 dj} e^{ipx} $ form an orthonormal basis for $L^2(\bbT_j) $
and there is  the expansion 
\be
f(x)  = \sum_{p \in  \bbT_j^*}  (e_p,f )  e_p(x)  = \sum_{p \in  \bbT_j^*}  L^{-dj}    \B[   \int e^{-ipx}   f(x) dx  \B]   e^{ipx} 
\ee
 We write this as 
 \be
f(x)  =  \sum_{p \in  \bbT_j^*}  L^{-dj}   \tilde f( p)   e^{ipx}  \equiv   {\sum}'_{p \in  \bbT_j^*}    \tilde f(p)   e^{ipx} 
\ee
  where
  \be
  \tilde f(p)  =    \int e^{-ipx}   f(x) dx
  \ee
  We also have $\int \overline f(x) g(x)  dx = \sum_p (f,e_p)(e_p, g) $
  which we write as
  \be
  \int \overline{ \bar f(x)}   g(x) dx = \sum_{p \in  \bbT_j^*}  L^{-dj} \overline{ \tilde f(p) } \tilde g(p)  \equiv {\sum}'_{p \in  \bbT_j^*}  \overline{ \tilde f(p) } \tilde g(p)  
 \ee

%\newpage

  \section{Characteristic functions of  Grassmann  integrals}   \label{B}
 
Let $\cG_h$ be the Grassman algebra on some $\bbT_j$  as defined in section \ref{Grassmann}.   We consider integrals   $ \int [ \cdots ] d\mu$       on $\cG_h$ defined 
to be continuous linear functionals on $\cG_h$.    For $F$ of the form (\ref{monkey3}) we have for some constant $c$
\be
|\int F d \mu | \leq   c\|F \|_h 
\ee
 If  $J \in \cC'( \bbT_j \times\bbT_j)$  and  $ <\psi, J \bpsi   >  = \int \psi(x) J(x,y) \bpsi(y) dx dy$
 then 
 \be
  \|   <\psi, J \bpsi   >   \|_h   =  h^2   \|J \|_{\cC'} 
  \ee
and   
 \be \label{pollyanna} 
 \begin{split}
&|  \int    e^ { <\psi, J \bpsi   >  } d \mu(\psi)|    \leq  c  \|   e^ { <\psi, J \bpsi   >  } \|_{h}  
 \leq c \exp \B(  \|<\psi, J \bpsi   >   \|_{h }\B)  \leq  ce^{h^2 \|J\|_{\cC'}   } \\
 \end{split} 
 \ee
This  $\int    e^ { <\psi, J \bpsi   >  } d \mu(\psi)$ is defined to be the characteristic function of the integral.  
  This generalizes   the finite dimensional case,  see \cite{FMRT91}.

 In particular let $   \int [ \cdots ] d\mu_G $
be  a Gaussian integral   with covariance $G$  of the form 
\be
G(x-y) = \int G_1(x-z) G_2(z-y) dz
\ee
with $G_1, G_2 \in \cC^3(\bbT_j) $ so  that $h(G)$ defined in (\ref{sequoia}) is finite.  In
particular we could take  $G=G_k$ defined in (\ref{gk}) or $G=G_k^z$ defined in (\ref{gkz}). 
 Then  as in  lemma \ref{concert}    the integral  $ \int F d \mu_{G} $ is defined on $\cG_{h(C)}$ and
 $| \int F d \mu_{G} | \leq   \| F\|_{h(G)} $.   If $ h(G) < h$ then $\|F\|_{h(G)} \leq \|F\|_h$  and 
 $\cG_h \subset  \cG_{h(G)}$ and so  the Gaussian integral is defined on $\cG_h$.

 \begin{lem}  \label{musket}    Let $G,h $ satisfy the above conditions   
  An integral   $\int [ \cdots ]  d \mu  $ on  the Grassmann algebra $\cG_h$   is Gaussian with covariance  $G$
 if and only if the characteristic function satisfies 
 \be
  \int e^{  <\psi, J \bpsi   > } d \mu =  \det ( I  + J^T   G) 
 \ee
 for all smooth functions $J$. 
 \end{lem} 
 \bigskip
 
 \rems   The determinant is the Fredholm determinant,  see Simon  \cite{Sim79}.   This can also be written
 \be \label{urgent} 
   \int e^{ - <\bpsi, J \psi   > } d \mu =  \det ( I  + J   G) 
 \ee

 \pr  First we note that   $J^TG$ is trace class so the determinant is well defined.  In fact $G$ alone is trace class
 since $Gf = G_1 * G_2* f$ and each of these convolutions is Hilbert-Schmidt.  Indeed the kernel of $f \to G_i*f$
 has Hilbert-Schmidt norm squared
 \be
 \int |G_i(x-y)|^2 dx dy  =  \|G_i\|^2_2 |\bbT_j|  < \infty
 \ee

Now we compute 
\be \label{kingsman} 
\begin{split}  
\int   e^{ <\psi, J \bpsi   >  }  d \mu_{G}   
= & \sum_{n=0} ^{\infty} \frac{1}{n!}  \int    J( x_1, y_1 ) \cdots  J( x_n, y_n )         \det \B \{ G  (x_i,y_i)  \B\}  dx dy     \\ 
= & \sum_{n=0} ^{\infty} \frac{1}{n!}   \int \det \B \{  (J^T G ) (x_i,x_j)  \B\}  dx     \\ 
= & \det ( I  +   J^T G  ) \\ 
\end{split}
\ee
The last step is the Fredholm formula  \cite{Sim79}.  
 
 For the converse suppose  that   $  \int e^{  <\psi, J \bpsi   >}  d \mu =  \det ( I  + J^T   G) $.   We compare the terms $n^{th}$ order in $J$
 and find
  \be 
\begin{split}
&\int    \B[ \int \psi(x_1) \bpsi(y_1)  \cdots \psi(x_n) \bpsi(y_n)  \  d \mu    \B]  \prod_{i=1}^n J(x_i,y_i )   dx dy  \\
=  &  \int  \det \B \{ G  (x_i,y_j) \B\}    \prod_{i=1}^n J(x_i,y_i) dx dy  \\
\end{split} 
\ee
Both sides of this equation are homogeneous functions of $J$ of degree $n$,  that is $T(aJ) = a^n T(J)$.   It follows by the polarization formula that they determine 
multilinear functions of $J$.   Thus in this equation we can replace  $\prod_{i=1}^n J(x_i,y_j )$ by $\prod_{i=1}^n J_i(x_i,y_j )$. Furthermore we can specialize and 
take $  J_i(x_i,y_i) = f_i(x_i) g_i(y_i) $ with $f_i,g_i \in \cC^{\infty} (\bbT) $.  Thus we have an identity between multilinear functions on $\cC^{\infty} (\bbT)$.
Buy the kernel theorem the associated kernels are equal as distributions and hence as functions. 
Thus 
\be  \int \psi(x_1) \bpsi(y_1)  \cdots \psi(x_n) \bpsi(y_n)  \  d \mu   =\det \B \{ G  (x_i,y_j) \B\}  
\ee
which is our result.    

  \begin{lem}  Let $G_k$ on $\bbT_{N+M-k}$ be as in the text.  
  \begin{enumerate}
  \item  There  is     $ \vep'_k $ with   $|\vep'_k| \leq \one  |z_k| $ such that 
\be  \label{esp1} 
\int    e^ {-  z_k \int  \bpsi \slpa \psi } d \mu_{G_k }(\psi)  = \det\B(I + z_k\slpa G_k \B) =   e^{    \vep'_k    | \bbT_{N+M-k}|  }  
\ee
\item  There is an identity between Gaussian integrals
\be \label{polly} 
 \frac{\int  [\cdots] e^ {-  z_k \int  \bpsi \slpa \psi } d \mu_{G_k }(\psi) }{ \int  e^ {-  z_k \int  \bpsi \slpa \psi } d \mu_{G_k }(\psi) }
= \int[ \cdots]   d \mu_{G^z_k } (\psi)  
\ee
where 
 \be 
G^z_k(x,  y)  =   {\sum}'_p    e^{ip(x-y)}  \frac{-i\slp}{p^2} \   \frac{ 1}{ z_k + e^{p^2} }   
\ee
\end{enumerate} 
\end{lem} 
\bigskip

\pr       Start with the estimate 
 \be  
  \|  z_k \int  \bpsi \slpa \psi  \|_h  =     |z_k|  h^2\sup_{\|f \|_{\cC} \leq 1} |\int   \slpa_x f(x,x) | \leq \one |z_k|h^2  | \bbT_{N+M-k}|
    \ee
 Then the   integral in (\ref{esp1})  is finite and by  (\ref{pollyanna}) 
 \be
 |  \int    e^ {-  z_k \int  \bpsi \slpa \psi } d \mu_{G_k }(\psi)|    \leq e^{ \one  |z_k| h^2 | \bbT_{N+M-k}| } 
 \ee
The first identity   in (\ref{esp1})   is (\ref{urgent})  which still holds for the more singular
$J(x,y) = \de(x-y) z_k \slpa_y$.   Note in this case $(JG_k) (x,y) = z_k ( \slpa G_k )(x,y)$ is still the kernel of a  trace class operator. 
For the second identity in (\ref{esp1}) we note that 
\be
z_k \slpa G_k(x,  y)  = z_k  {\sum}'_p    e^{ip(x-y)}    e^{- p^2}    
\ee
The operator has eigenvalues $\{ z_k e^{-p^2} \}$ with multiplicity $2n$ for each $p \in \bbT^*_{M+N-k}$.   Hence
\be
\begin{split} 
  \log  \det ( I  + z_k \slpa G)  = & 2 n \sum_{ p \in \bbT^*_{M+N-k} }  \log( 1 + z_k  e^{-p^2} ) \\
  =&2 n  |\bbT_{N+M-k} |{\sum}'_{ p \in \bbT^*_{M+N-k} }  \log( 1 + z_k  e^{-p^2} ) \\
  \end{split} 
  \ee  
But for $z_k$ small   $|\log( 1 + z_k  e^{-p^2} )| \leq    \one |z_k|  e^{-p^2}     $  which gives    $|\vep'_k| \leq \one  |z_k| $.  
So now part 1 is established.

For part 2  both sides define integrals on $\cG_h$   and  it suffices to check that they  have the same   characteristic function in the form (\ref{urgent}). 
 On the one hand as in lemma \ref{musket}  we have
 the characteristic function of the right side of (\ref{polly}) 
\be
\int   e^{-<  \bpsi, J \psi > } d \mu_{G^z_k } =   \det ( I  + J  G^z_k )
\ee
On the other hand we have similarly 
\be
\int   e^{-< \bpsi,( J+ z_k  \slpa)  \psi> }d \mu_{G_k  } =   \det ( I  + z_k  \slpa  G_k + J G_k ) 
 \ee
The  characteristic function of the left side of (\ref{polly})  is this divided by $ \det ( I  + z_k  \slpa G) $ or multiplied by
$ \det ((  I  + z_k  \slpa  G) ^{-1}) $  which is
\be  
  \det ( I  + JG_k ( I  + z_k  \slpa G_k) ^{-1}  ) 
\ee
But    $(G_k^z)^{-1} = G_k^{-1} + z_k \slpa   =   (  I + z_k \slpa G_k )  G_k^{-1}  $ implies $G_k^z= G_k  ( I  + z_k  \slpa G_k) ^{-1} $.
Thus the  last determinant is  again $  \det ( I  + J  G^z_k )$.   The characteristic functions are equal. 
\bigskip

\section{Gamma matrices} \label{C}

  Let $\ga_0, \ga_1$ be $2 \times 2$ traceless  matrices satisfying $\{ \ga_{\mu} , \ga_{\nu }  \}  = 2 \de_{\mu \nu} $.  If $R$ is a reflection  through a line $x_{\mu} =c$  or a rotation by  a multiple of $\pi/2$
  then there is a matrix $S$ in the group $Pin(2)$  such that  $S^{-1} \ga_{\mu} S = \sum_{\nu} R_{\mu \nu} \ga_{\nu} $. (Actually there are two such.)  We also have
  \be 
  \ga_5 = i \ga_0  \ga_1  = \frac{i}{2} \sum_{\mu \nu}\ep_{\mu \nu} \ga_{\mu} \ga_{\nu}
  \ee
  which satisfies  $S^{-1} \ga_5 S = ( \det R )  \ga_5 $.   For the spinor  group $Pin(n)$ and generalizations see for example   \cite{CBDM89} or \cite{Dim11} .
  
  We relabel  $(  I, \ga_0, \ga_ 1, \ga_5) $ as $( \Ga_1, \Ga_2,  \Ga_3 , \Ga_4) $.     Then the $\Ga_i$ are traceless, self-adjoint matrices satisfying $\Ga_i^2 = I$
  and $\tr( \Ga_i \Ga_j ) =2 \de_{ij} $.  They are linearly independent since if $\sum_i c_i \Ga_i = 0$ then $c_j =  \sum_i c_i \tr( \Ga_i \Ga_j)  =0$.    They form a basis
  for the space $M_2$ of $2 \times 2$ complex matrices.     Any matrix $A$ can be written 
  \be
  A=\sum_i c_i  \Ga_i   \hs \hs c_i = \frac12 \tr( A \Ga_i) 
  \ee
  
  \begin{lem}  { \ } 
  \begin{enumerate}
  \item   If  $\al $ in  $M_2 $ satisfies    $S^{-1}  \al  S = \al $ for all $S$,   then there is a constant $c$ such that $\al = c I $.
 \item    If  $\al_{\mu}  \in M_2$  satisfies      $S^{-1} \al _{\mu} S=\sum_{\nu} R_{\mu \nu} \al _{\nu} $ for all $S$,  then there is a constant $c$ such that $\al_{\mu} = c \ga_{\mu} $.
 \end{enumerate} 
  \end{lem} 
  \bigskip

  \pr For the first item
  we have
  \be
  \begin{split} 
  \tr ( \al \ga_{\mu} ) =  &   \tr (S^{-1}   \al S S^{-1}  \ga_{\mu} S )    = \sum_{\nu} R _{\mu \nu}  \tr (  \al \ga_{\nu} )  \\
 \tr ( \al \ga_5 ) =  &   \tr (S^{-1}   \al  SS^{- 1}  \ga_5 S)  = \det R   \tr (  \al \ga_5)\\
 \end{split} 
  \ee
  In all cases we can pick an $r$ such that the right side is minus the left side and hence the expression is zero.  For example for $ \tr ( \al \ga_0)$  the reflection  $x_0 \to -x_0,x_1 \to x_1$ has this property.   Thus in the expansion in our basis
  only multiples of the identity survive. 
  
  For the second item we have similarly
   \be
  \begin{split} 
   \tr ( \al_{\mu}   ) =  & \sum_{\nu} R_{\mu \nu}  \tr (  \al_{\nu} )  \\
  \tr ( \al_{\mu }  \ga_{\mu' } ) =  &   \tr (S^{-1} \al_{\mu }SS^{-1}   \ga_{\mu' } S)  = \sum_{\mu, \nu} R_{\mu \nu}  R_{\mu' \nu' }  \tr ( \al_{\nu}  \ga_{\nu'} )  \\
    \tr ( \al_{\mu }  \ga_5 ) =  &   \tr (S^{-1} \al_{\mu }S S^{-1}   \ga_5 S )  = \sum_{ \nu} r_{\mu \nu}  \det R  \tr ( \al_{\nu}  \ga_5 )  \\
 \end{split} 
  \ee
In the first and third case we can pick an $R$ such that the right side is minus the left side and hence zero.    This is also true for the second case if $\mu \neq \mu'$.
If $\mu = \mu'$ we can choose $R$ so that  $ \tr(  \al_{0}\ga_{0} ) = \tr( \al_2 \ga_1) $, namely take the rotation $x_0 \to x_1, x_1 \to -x_0$.
 Then  $\tr (\al_{ \mu} \ga_{\nu})  = 2c \de_{\mu \nu} $ for some constant $c$  and
 \be
 \al_{\mu}  =  \sum_{\nu}   \frac12 \tr(  \al_{\mu} \ga_{\nu} )\ga_{\nu} = c \ga_{\mu} 
 \ee

  Now we consider the sixteen linear operators on  $\bbC^2 \otimes \bbC^2$ of the form $\Ga_{ij} \equiv \Ga_i \otimes \Ga_j$.  These are self-adjoint and $\Ga_i^2 =I \otimes I$.  We also have   $ \tr( \Ga_{ij})  = \tr(  \Ga_i \otimes \Ga_j) = ( \tr \Ga_i )(\tr \Ga_j) =0 $  and 
  \be
    \tr( \Ga_{ij}\Ga_{i'j'}   )  = \tr(  \Ga_i\Ga_{i'}  \otimes \Ga_j \Ga_{j'}  ) =  \tr(  \Ga_i\Ga_{i'} ) \tr (  \Ga_j \Ga_{j'}  )  = 4\de_{ij} \de_{i'j'} 
 \ee   
  As before the $\Ga_{ij} $ are a basis for the space  $\cL(\bbC^2 \otimes \bbC^2)$  of linear operators on $\bbC^2 \otimes \bbC^2$,  and  every element  $A$ has an expansion 
  \be \label{expand2} 
  A=\sum_{ij}  c_{ij}   \Ga_{ij}    \hs \hs c_{ij}  = \frac14 \tr( A \Ga_{ij} )  \hs  \hs     
  \ee

  \begin{lem} 
  If   $\al \in \cL(\bbC^2 \otimes \bbC^2)$ satisfies  $( S^{-1}  \otimes S^{-1} )\al ( S  \otimes S ) = \al$ for all $S$,
  then there are constants  $c, c_p, c_v$ such that
  \be 
  \al  =     c  (  I \otimes I ) + c_p(  \ga_5 \otimes \ga_5) 
 + c_v \sum_{\mu}  ( \ga_{\mu} \otimes \ga_{\mu} )    \ee
  \end{lem} 
  \bigskip

  \pr    As in the previous lemma we can argue that   $\tr( \al \Ga_{ij} )  =0$ except for   $\Ga_{ij} = I \otimes I $, $ \ga_{\mu} \otimes  \ga_{\mu}$,
$\ga_5 \otimes \ga_5$.  
For example for any $R$ 
\be 
\tr( \al (\ga_{\mu} \otimes \ga_5 ) ) = \det R  \sum_{\nu} R_{\mu \nu} \tr( \al (\ga_{\nu} \otimes \ga_5 ) )    
\ee
If $\mu =0$ choose  $R$ to be $x_0 \to x_0, x_1 \to -x_1$ and conclude $\tr( \al (\ga_0 \otimes \ga_5 ) )=0$. 
 We can also   argue that   $ \tr( \al  \ga_{\mu} \otimes  \ga_{\mu} ) $ is independent of $\mu$.  Then our basic expansion (\ref{expand2})
gives the result.

  \section{Bounds on propagators}  \label{D}

  \begin{lem} 
  \be
  \begin{split}
 | C_k(x) |  \leq  &\one e^{ - |x|/L}   \\
 |w_k( x) | \leq    & \one |x| ^{- 1} e^{-|x| }  \\
  \end{split}
  \ee
  \end{lem}    
  \bigskip

 \pr   (cf. \cite{BDH98})  We have 
   \be 
C_k(x)  =     {\sum}'_{p  \in \bbT^*_{N+M-k }}    e^{ipx}  \frac{-i\slp}{p^2}  \B(  e^{- p^2 } -  e^{- L^2p^2  } \B)    
\ee
We write
\be \label{sudden1} 
e^{- p^2 } -  e^{- L^2p^2 } =   - \int_1^{L^2}   \frac{d}{d \la} e^{-\la p^2}  d \la  = p^2   \int_1^{L^2}    e^{-\la p^2}  d \la  
\ee
  which gives
\be 
C_k(x)  = -   \int_1^{L^2}  d \la   \  \slpa \   \B[    {\sum}'_{p  \in \bbT^*_{N+M-k }}    e^{ipx}     e^{-\la p^2}        \B] 
\ee    
Our definition of the weighted sum  $ {\sum}'_p$ explicitly excluded the $p=0$ term.   Here we can assume it is restored since
it gives a constant and we are taking a derivative. 
 For the bracketed expression we use the identity  
 \be \label{Poisson} 
   {\sum}'_{p  \in \bbT^*_{N+M-k }}    e^{ipx}     e^{-\la p^2}        =  \frac{1}{4 \pi \la} 
   \sum_{y \in L^{N+M -k  } \bbZ^2} e^{ - |x-y|^2 / 4 \la} 
   \ee
  Both sides are  periodic with period $L^{N+M -k }$ and so can be regarded as a functions on the torus $\bbT_{M+N -k} $.
  This is a special case of the Poisson summation formula.   If we let $\la = kt$ then can be understood as two different
   representations of the fundamental solutions for the heat operator  $\pa/\pa t - k \De$; either by Fourier series on the left or
   by Fourier transform on $\bbR^2$  followed by periodizing on the right.

With this identity we have
\be   \label{inky1} 
\begin{split} 
C_k(x)  =   &  -  \int_1^{L^2} \frac{ d \la }{ 4  \pi \la}    \  \slpa \   \B[    \sum_{y \in L^{N+M -k  } \bbZ^2} e^{ - |x-y|^2 / 4 \la}    \B]  \\
= &  \frac{1}{8 \pi}    \sum_{y \in L^{N+M -k  } \bbZ^2}  ( \slx - \sly ) \B[  \int_1^{L^2} \frac{ d \la }{ \la^2} \   e^{ - |x-y|^2 / 4 \la}    \B]  \\
\end{split} 
\ee   

We work on the bracketed expression in the last equation.   Since  $\al^2 - 4 \al  \geq -4$ we have
\be \label{sound} 
\frac{  |x-y|^2 }{ 4 \la }  - \frac{ 2  |x-y| }{  \sqrt{  \la }}    \geq  - 4
 \ee 
Therefore
\be
  \int_1^{L^2} \frac{ d \la }{ \la^2} \   e^{ - |x-y|^2 / 4 \la}
  \leq     \one  \int_1^{L^2} \frac{ d \la }{ \la^2} \   e^{ -2 |x-y| / \sqrt  \la}  \leq      \one  e^{-2 |x-y|/L} 
  \ee
 and  so  for $- \frac12 L^{N+M-k} \leq x_{\mu} \leq \frac12 L^{N+M-k}$
 \be 
| C_k(x) |  \leq    \one     \sum_{y \in L^{N+M -k  } \bbZ^2}  | x-y |  e^{-2 |x-y|/L}  \leq  \one e^{- |x|/L} 
\ee

Now we turn to the bound on $w_k(x)$  given by 
  \be 
w_k(x)  =     {\sum}'_{p  \in \bbT^*_{N+M-k }}    e^{ipx}  \frac{-i\slp}{p^2}  \B(  e^{- p^2/L^{2k}  } -  e^{- p^2  } \B)    
\ee
Following the same steps as before we have instead of (\ref{inky1}) 
 \be   \label{inky2} 
w_k(x)  =  \frac{1}{8 \pi}    \sum_{y \in L^{N+M -k  } \bbZ^2}  ( \slx - \sly ) \B[  \int_{L^{-2k} }^{1} \frac{ d \la }{ \la^2} \   e^{ - |x-y|^2 / 4 \la}    \B] 
\ee   
Now in the bracketed expression   
\be
\frac{ |x-y|^2 }{4 \la }  \geq \frac32 |x-y|  +    \frac {|x-y|}{ 2\sqrt{ \la} }  - 4
 \ee 
This follows from  $\la\leq 1$ and  $\al^2 - 3 \sqrt{ \la} \al    -\al \geq      \al^2 - 4 \al     \geq - 4$.
Therefore 
\be
\begin{split}
  \int_{L^{-2k} }^{1} \frac{ d \la }{ \la^2} \   e^{ - |x-y|^2 / 4 \la}  
\leq   &   \one   e^{ - \frac32 |x-y|   }   \int_{L^{-2k} }^{1} \frac{ d \la }{ \la^2} \  e^{ -  |x-y|/2\sqrt{ \la} }  \\
\leq   &   \one   e^{ - \frac32 |x-y|   }   \int_0^{\infty } \frac{ d \la }{ \la^2} \  e^{ -  |x-y|/2\sqrt{ \la} }  \\
=  & \one      e^{ - \frac32 |x-y|   } \frac{1}{ |x-y|^2 }  \int_0^{\infty }  \xi e^{-\xi} d \xi \\
=  & \one      e^{ - \frac32 |x-y|   } \frac{1}{ |x-y|^2 } \\
\end{split} 
\ee
This yields
 \be 
|w_k(x)|   \leq \one  \sum_{y \in L^{N+M -k  } \bbZ^2} |x-y|^{-1}  e^{- \frac32 |x-y| }  \leq \one |x|^{-1} e^{-|x|}
\ee

  \begin{lem} For $|z_k| $  sufficiently small
  \be
 | C^z_k(x) - C_k(x) |  \leq     \cO(1) |z_k|   e^{ - |x|/L }   
  \ee
  \end{lem} 
  \bigskip

  \pr
  We have with the sum over $p  \in \bbT^*_{N+M-k}$
  \be  
  \begin{split}
C^z_k (x)   =   &    {\sum}'_p   e^{ipx }   \frac{-i\slp}{p^2} \left(     \frac{1}{  (z_k + e^{p^2} ) } -   \frac{1}{  (z_k + e^{L^2 p^2} ) } \right)
 \\
 =   &    {\sum}'_p     e^{ipx }   \frac{-i\slp}{p^2} \left(     \frac{     e^{L^2p^2}  - e^{p^2}  }{  (z_k + e^{p^2} )   (z_k + e^{L^2 p^2} )  } \right) \\
  =   &    {\sum}'_p   e^{ipx }   \frac{-i\slp}{p^2} \left(     \frac{  e^{-p^2} -   e^{- L^2p^2}   }{  (1 +z_k e^{- p^2} )   (1+ z_k  e^{- L^2 p^2} )  } \right) \\
  = & \sum_{n,m \geq 0}  (-  z_k)^{n+m}   {\sum}'_{p }   e^{ipx }   \frac{-i\slp}{p^2} \B(  e^{-(n+L^2m+1)  p^2} - e^{-(n+L^2m+L^2)  p^2} \B)   \\
\end{split}
\ee
The $C_k(x)$ is the case $n=m=0$ so for the difference we can restriction the sum to $n+m \geq 1$. We use the identity
\be
e^{-(n+L^2m+1)  p^2} - e^{-(n+L^2m+L^2)  p^2}  =p^2    \int_{n+L^2m +1} ^{n + L^2 m + L^2} e^{-\la p^2} d \la
\ee
and the Poisson summation formula (\ref{Poisson}) to obtain
 \be  
  \begin{split}
C^z_k (x) - C_k(x)    
    = &- \sum_{n+ m \geq 1}  (-  z_k)^{n+m}  \   \int_{n+L^2m +1} ^{n + L^2 m + L^2}  d \la  \ \slpa  \ \left[   {\sum}'_{p }   e^{ipx }  \  e^{- \la  p^2}     \right] \\
      = &-  \sum_{n+m \geq 1}  (-  z_k)^{n+m} \  \int_{n+L^2m +1} ^{n + L^2 m + L^2}    \ 
\frac{   d \la } {4 \pi  \la }    \      \slpa  \left[  \sum_y  e^{ - |x-y|^2 / 4 \la}   \right] \\
         = &\frac{1}{8 \pi}  \sum_{n+ m \geq 1}  (-  z_k)^{n+m}  \sum_y   (\slx- \sly) 
  \   \left[    \  \int_{n+L^2m +1} ^{n + L^2 m + L^2}  \frac{d \la}{ \la^2}   \    e^{ - |x-y|^2/4 \la}   \right] \\
\end{split}
\ee
where the sums are over $y \in L^{N+M -k}   \bbZ^2$.   We enlarge the the upper limit of the $\la$ integral to $L^2(n+m+1) $
and then 
\be \label{seesaw} 
| C^z_k (x) - C_k(x)  | \leq    \one   \sum_{n,m \geq 1}  |z_k|^{n+m}  \sum_{y}  |x-y|  
  \    \    e^{ - |x-y|^2 / 4L^2 (n+m+1) }  
  \ee
Now let $\ell = m+n \geq 1$.   If   $\ell =1$ we  use the inequality (\ref{sound})  with $\la = 2L^2$  to get $ e^{ - |x-y|^2 / 8L^2 } \leq  \one  e^{ -\sqrt2  |x-y|/ L } $.    
This gives the required bound   $ \cO(1) |z_k|   e^{ - |x|/L }  $.   For $\ell \geq 2$
 we make the estimate
\be
  |z_k|^{\frac12 \ell}    e^{ - |x-y|^2 / 4L^2 (\ell+1) }   = \exp \B( -  \frac{\ell}{2} | \log|z_k||   - |x-y|^2 / 4L^2 (\ell+1) \B)  \leq e^{-2 |x-y| }  
\ee
The last inequality  requires some explanation.  If $\frac14  \ell| \log |z_k| | \geq |x-y|$ then the first term in exponential  provides the decay. 
On the other hand if  $  |x-y| \geq  \frac14  \ell   | \log |z_k|| $ 
then $ |x-y|^2 / 4 L^2\ell      \geq     |\log |z_k| | |x-y|   / 24L^2 \geq 2 | x-y| $ 
 and the second factor provides the decay.  Inserting the bound   in (\ref{seesaw}) gives the  estimate
 $\one |z_k|  e^{-|x| } $ which suffices. 

   \end{appendix}

\end{document}